\documentclass[]{pasj02} 
\usepackage[switch,mathlines]{lineno} 
\usepackage{threeparttable}
\usepackage{ulem} 
\jyear{2024}
\Received{2024/12/16}
\Accepted{2025/08/27}
\Published{2025/11/08}

\begin{document}

\title{The LMT 2 millimeter receiver system (B4R). II.\par 
Science demonstration observations toward Orion-KL\,/\,OMC-1}

\author{
Teppei \textsc{yonetsu},\altaffilmark{1}\altemailmark\orcid{0000-0003-2386-7427} \email{s\underline{ }t.yonetsu@omu.ac.jp} 
Ryohei \textsc{kawabe},\altaffilmark{2,3}\orcid{0000-0002-8049-7525} 
Yuki \textsc{yoshimura},\altaffilmark{4}\orcid{0000-0002-1413-1963}
Kotomi \textsc{taniguchi},\altaffilmark{2}\orcid{0000-0003-4402-6475}
Yoshito \textsc{shimajiri},\altaffilmark{2,5}\orcid{0000-0001-9368-3143} 
Omar Sergio \textsc{rojas-garc\'{i}a},\altaffilmark{6}\orcid{0000-0003-1054-4637}
Arturo I. \textsc{g\'{o}mez-ruiz},\altaffilmark{6,7}\orcid{0000-0001-9395-1670} 
Takeshi \textsc{sakai},\altaffilmark{8}\orcid{0000-0003-4521-7492}
Kunihiko \textsc{tanaka},\altaffilmark{9}\orcid{0000-0001-8153-1986}
Bunyo \textsc{hatsukade},\altaffilmark{2,3,4}\orcid{0000-0001-6469-8725}
Akio \textsc{taniguchi},\altaffilmark{10,11}\orcid{0000-0002-9695-6183} 
Yoichi \textsc{tamura},\altaffilmark{11}\orcid{0000-0003-4807-8117} 
Tatsuya \textsc{takekoshi},\altaffilmark{10}\orcid{0000-0002-4124-797X} 
Tai \textsc{oshima},\altaffilmark{2,3}\orcid{0009-0005-5915-1035}  
Kotaro \textsc{kohno},\altaffilmark{4,12}\orcid{0000-0002-4052-2394}  
Masato \textsc{hagimoto},\altaffilmark{11}\orcid{0000-0001-8083-5814}
David H. \textsc{hughes},\altaffilmark{6}\orcid{}
F. Peter \textsc{schloerb},\altaffilmark{13}\orcid{}
David \textsc{s\'{a}nchez-arg\"{u}elles},\altaffilmark{6,7}\orcid{0000-0002-7344-9920} 
Kamal \textsc{souccar},\altaffilmark{13}\orcid{0000-0001-7915-5272} 
Gopal \textsc{narayanan},\altaffilmark{13}\orcid{0000-0002-4723-6569}
Min S. \textsc{yun},\altaffilmark{13}\orcid{0000-0001-7095-7543}
V\'{i}ctor \textsc{g\'{o}mez-rivera},\altaffilmark{6,14}\orcid{0009-0003-9025-6121}
Iv\'{a}n \textsc{rodr\'{i}guez-montoya},\altaffilmark{6,7}\orcid{}
Edgar \textsc{col\'{i}n-beltr\'{a}n},\altaffilmark{6,7}\orcid{0000-0002-0758-3160}  
Miguel \textsc{ch\'{a}vez dagostino},\altaffilmark{6}\orcid{}
Javier \textsc{zaragoza-cardiel},\altaffilmark{6,7}\orcid{0000-0001-8216-9800}
Shinji \textsc{fujita},\altaffilmark{15}\orcid{0000-0002-6375-7065}
and 
Hiroyuki \textsc{maezawa}\altaffilmark{1}\orcid{0009-0007-6017-8395}
}

\altaffiltext{1}{Department of Physics, Graduate School of Science, Osaka Metropolitan University, 1-1 Gakuen-cho, Naka-ku, Sakai, Osaka 599-8531, Japan}

\altaffiltext{2}{National Astronomical Observatory of Japan, National Institutes of Natural Sciences, 2-21-1 Osawa, Mitaka, Tokyo 181-8588, Japan}
\altaffiltext{3}{Graduate Institute for Advanced Studies, SOKENDAI, Osawa, Mitaka, Tokyo 181-8588, Japan}

\altaffiltext{4}{Institute of Astronomy, Graduate School of Science, The University of Tokyo, 2-21-1 Osawa, Mitaka, Tokyo 181-0015, Japan}

\altaffiltext{5}{Kyushu Kyoritsu University, 1-8 Jiyugaoka, Yahatanishi-ku, Kitakyushu, Fukuoka, Fukuoka 807-8585, Japan}

\altaffiltext{6}{Instituto Nacional de Astrof\'{i}sica, \'{O}ptica y Electr\'{o}nica, Luis Enrique Erro 1, Tonantzintla C.P. 72840, Puebla, M\'{e}xico}

\altaffiltext{7}{Consejo Nacional de Ciencia y Tecnolog\'{i}a, Av. Insurgentes Sur 1582, Col. Cr\'{e}dito Constructor, Demarcaci\'{o}n Territorial Benito Ju\'{a}rez C.P. 03940, Ciudad de M\'{e}xico, M\'{e}xico}

\altaffiltext{8}{Graduate School of Informatics and Engineering, The University of Electro-Communications, 1-5-1 Chofugaoka, Chofu, Tokyo 182-8585, Japan}

\altaffiltext{9}{Department of Physics, Faculty of Science and Technology, Keio University, 3-14-1 Hiyoshi, Yokohama, Kanagawa 223-8522, Japan}

\altaffiltext{10}{Kitami Institute of Technology, 165, Koen-cho, Kitami, Hokkaido 090-8507, Japan}

\altaffiltext{11}{Department of Physics, Graduate School of Science, Nagoya University, Furocho, Chikusa-ku, Nagoya, Aichi 464-8602, Japan}

\altaffiltext{12}{Research Center for the Early Universe, Graduate School of Science, The University of Tokyo, 7-3-1 Hongo, Bunkyo-ku, Tokyo 113-0033, Japan}

\altaffiltext{13}{Department of Astronomy, University of Massachusetts, Amherst, MA 01003, USA}

\altaffiltext{14}{Corporaci\'{o}n Mexicana de Investigaci\'{o}n en Materiales S.A. de C.V., M\'{e}xico}

\altaffiltext{15}{The Institute of Statistical Mathematics, 10-3 Midori-cho, Tachikawa, Tokyo, 190-8562, Japan}




\maketitle
\begin{abstract}
We present the results of mapping and single-point spectral scans toward Orion-KL/OMC-1 performed as science demonstrations of a 2\,mm superconductor-insulator-superconductor receiver, named the Band 4 Receiver (B4R), installed on the Large Millimeter Telescope (LMT), with a diameter of 50\,m.
To prove the capabilities of mapping and spectral scans with the B4R on the LMT, commissioning observations were conducted employing the On-The-Fly mapping technique toward Orion-KL/OMC-1, which covers a map size of 5$\arcmin\times$5$\arcmin$. 
These mapping observations were performed with two frequency settings providing 10\,GHz in total (131.4--133.9\,GHz and 145.1--147.6\,GHz, 136.2--138.7\,GHz and 149.9--152.4\,GHz) with a frequency resolution of 76.293\,kHz.
In this study, we conducted spectral line identification analysis for the hot core and compact ridge regions in the Orion-KL with a beam size of 11--12$\arcsec$.
We detected nearly 400 emission lines and identified two recombination lines and 29 molecular species, including isotopologues, deuterated molecules, and vibrational excited states, despite the short integration time.
These results of line detection are consistent with those of previous studies.
The 29 molecular species include nitrogen (N)-bearing complex organic molecules (COMs) and oxygen (O)-bearing COMs.
To demonstrate the capability of the B4R in astrochemistry, we conducted detailed analyses of column densities, rotational temperatures, and relative abundances with respect to H$_2$ on two representative COMs, N-bearing C$_2$H$_5$CN and O-bearing CH$_3$OCHO in the central 40$\arcsec\times$40$\arcsec$ area of the map.
The wide bandwidth of 10\,GHz enabled the use of 8 and 34 emission lines, respectively.
The spatial differences in the physical and chemical properties between the above two molecules were derived at a spatial resolution of $\sim$12$\arcsec$. 
The B4R on the LMT was successfully demonstrated to be powerful for mapping and spectral scans and to have high potential for the study of interstellar chemistry.
\end{abstract}

\section{Introduction}
A 2\,mm band receiver plays an important role not only in determining the redshift (z) of dusty star-forming galaxies (DSFGs) or submillimeter galaxies (SMGs) through synergy with a 3\,mm band receiver (e.g., \cite{Bakx2022-fr}) but also in astrochemistry.
The interstellar medium or circumstellar shells host more than 300 molecular species\footnote{$\langle$https://cdms.astro.uni-koeln.de/classic/molecules$\rangle$}, including complex organic molecules (COMs), which are defined as carbon-containing species composed of six or more atoms \citep{Herbst2009}.
The 2\,mm band, which covers molecular species of deuterated molecules (e.g., DCO$^+$, N$_2$D$^+$, DCN, and DNC)\footnote{$\langle$https://splatalogue.online$\rangle$} and many COMs, is a critical frequency band in astrochemistry (e.g., \cite{Ziurys1993-ps,Martin2006-ru,Belloche2014-iy}).

The Band 4 Receiver (B4R), a 2\,mm band superconductor-insulator-superconductor (SIS) receiver, was installed onto the Large Millimeter Telescope (LMT)-50\,m situated at the summit of the Sierra Negra Mountain in Mexico at an altitude of 4600\,m.
We conducted the On-The-Fly (OTF) mapping observations toward the Orion Kleinmann-Low (Orion-KL) region \citep{Kleinmann1967-sm} at a distance of 418\,$\pm$\,6\,pc \citep{Kim2008-yg} in the Orion molecular cloud-1 (OMC-1) to test the observational capabilities of the B4R on the LMT \citep{kawabe_submitted}.
We obtained demonstration maps with a spatial resolution of $\sim11\arcsec$, covering an area of 10$\arcmin\times$10$\arcmin$ in 2018, and 5$\arcmin\times$5$\arcmin$ and 1.5$\arcmin\times$1.5$\arcmin$ in 2019 \citep{kawabe_submitted,Taniguchi2024-pt}.
The capability for the OTF mapping using the B4R on the LMT was successfully demonstrated, and high-quality maps covering large regions of major molecular lines (e.g., CS and H$_2$CO) were obtained \citep{kawabe_submitted}.
Moreover, the B4R is capable of spectral scan observations over a wide frequency range, which makes it valuable and highly competitive in advancing molecular identification efforts and essential for understanding chemical reactions and enrichments.
The spatial variation in the chemical and physical properties of COMs (e.g., column densities and rotational temperatures) is key to understanding interstellar chemistry and its variety.
The Orion-KL region, which is rich in COMs, is one of the most suitable targets for testing the observational capabilities of the B4R on the LMT, such as spectral scans over the 2 mm band and physical and chemical analysis/studies based on mapping spectral scan data.

The Orion-KL region stands as a nearby high-mass star-forming region (HMSFR), harboring protostars, a compact H\,\emissiontype{II} region associated with Source I, infrared sources N and Becklin-Nuegebauer (BN) \citep{Becklin1967-xe}, and dense cores known as the hot core (HC) and compact ridge (CR).
The HC region is located close to the southeast of Source I and is presumed to be externally heated by Source I \citep{Zapata2011-hz}.
The CR region is located at a separate distance southwest of Source I.
The existence of H$_2$O masers and HCOOH in the CR implies the presence of regions of interaction between the Orion-KL outflow and surrounding quiescent gas clouds (e.g., \cite{Genzel1981-nn,Gaume1998-vt,Liu2002-bp}).
The Orion-KL region exhibits distinct chemical properties across its various sub-regions.

The spatial differences between nitrogen (N)-bearing and oxygen (O)-bearing COMs in the HC and CR have also been confirmed by observations toward Orion-KL using single-dish telescopes and interferometers for line surveys and high-resolution imaging (e.g., \cite{Blake1987,Turner1989,Turner1991,Crockett2015-vy,Liu2022,Widicus_Weaver2012-li,Friedel2012-oj,Feng2015-au,Tercero2018-as,yamamoto2016introduction,Peng2013-gz}).
The HC region is rich in N-bearing COMs, and the CR region is characterized by abundant O-bearing COMs.
The presence of saturated O-bearing species suggests interactions between outflows and quiescent ambient molecular material \citep{Blake1987}. 
Meanwhile, among O-bearing COMs, CH$_3$COCH$_3$ has a distribution similar to that of N-bearing COMs, unlike other O-bearing molecules such as CH$_3$OCHO \citep{Widicus_Weaver2012-li,Peng2013-gz}.
\citet{Tercero2018-as} discussed the spatial distribution differences among O-bearing molecules, focusing on C-O-C and C-O-H bonds.

Observations and chemical modeling have been performed to investigate the origins responsible for the chemical differentiation of N- and O-bearing COMs.
\citet{Suzuki2018-ya} evaluated parameters that could influence chemical evolution using the NAUTILUS gas–grain chemical model \citep{Ruaud2016-vw}.
Simulations under different physical conditions (e.g., initial densities, warm-up speed, peak density, timescale of the collapsing phase, and peak temperature) revealed that different temperatures play a significant role, whereas the influence of other parameters is relatively minor.
Under the assumption that the temperature structures and ages of the cores differ, the observed abundances of NGC 6334F (poor in N-bearing species) and G10.47+0.03 (rich in N-bearing species) were consistent with the chemical model.
This suggests that both the different temperature structures inside the core and their evolutionary stages contributed to the observed molecular correlations \citep{Suzuki2018-ya}.
Furthermore, \citet{Garrod2022-pm} incorporated nondiffusive reaction processes occurring on dust grain surfaces and within ice mantles into simulations based on the three-phase astrochemical model MAGICKAL \citep{Garrod2013-yc}.
The observational results of the hot molecular core (HMC) G31.41+0.31, one of the HMSFRs \citep{Mininni2023-vx}, were consistent within a factor of ten with the abundance ratios predicted by the final chemical models of \citet{Garrod2022-pm}.
\citet{Garrod2022-pm} suggested that the observed differences in the abundances of N- and O-bearing species may be related to the efficient formation of nitriles in the gas phase during and after ice mantle desorption. 
They also proposed that the possible duration of chemical reactions in the gas phase after desorption and the thermal history of each source play a significant role in determining its chemical characteristics.

Understanding the chemical differentiation between such N- and O-bearing COMs in the Orion-KL regions and other HMSFRs is also important for comprehending the formation environments and processes of molecules that are considered precursors to prebiotic molecules (e.g., HNCHCN \citep{Zaleski2013-qd,Rivilla_2018}, NH$_2$CH$_2$CN \citep{Li2020-bz,Manna_2022}, NH$_2$CH$_2$CH$_2$OH \citep{Rivilla2021-nr}, and HOCOOH \citep{Sanz-Novo2023-up}).
Therefore, to evaluate the spectral scan capability, we extracted spectra from the OTF mapping data for the HC and CR regions of Orion-KL, where differences in the abundances of N- and O-bearing COMs have been observed.
In addition, we analyzed the spatial distribution of physical quantities to demonstrate the capability of resolving spatial distributions in mapping observations.
We used a 5$\arcmin\times$5$\arcmin$ dataset in 2019 with wide frequency coverage for rotation diagram analysis \citep{Turner1991} to test the observational capabilities.
In addition, observations of distant HMSFRs using interferometers and the nearby Orion-KL region using the B4R on the LMT-50\,m have nearly equivalent spatial scales of angular resolution.
A comparison of observations on the same spatial scales between distant and nearby objects is pivotal for elucidating the universality of the chemical and physical environments conducive to COMs formation.

This paper is organized as follows.
We describe the B4R and report observations and data reductions in Section 2.
Section 3 delves into the details of the line identification and the chemical analysis of C$_2$H$_5$CN (N-bearing) and CH$_3$OCHO (O-bearing).
Section 4 presents a comparative analysis of our results with previous results of observations toward Orion-KL and other distant HMSFRs observed with single-dish telescopes and interferometers and discusses their implications.
Finally, the main conclusions of this study are summarized in Section 5.

\section{Observations and data reductions}
\subsection{Band 4 receiver (B4R)}
The B4R is a 2\,mm single-beam two-polarization heterodyne SIS receiver with 4\,K-cooled two sideband (2SB) mixers designed based on the work of \citet{Asayama2014-ev}.
It was installed in the receiver cabin of the LMT-50\,m, marking the commencement of scientific operations \citep{Hughes2020-by}.
The radio frequency (RF) and intermediate frequency (IF) ranges of the B4R are 125--163\,GHz and 4--8\,GHz, respectively.
The receiver noise temperature ($T_{\rm{RX}}$) is below 60\,K, satisfying the specifications of the Atacama Large Millimeter/submillimeter Array (ALMA) Band 4.
The aperture efficiency was measured to be 48\% to 33\% for 130 to 160\,GHz, which is in good agreement with previous measurements for other receivers installed on the LMT \citep{Hughes2020-by,kawabe_submitted}.
The sideband image rejection ratio surpassed 13\,dB.
Further details can be found in \citet{kawabe_submitted}.

\subsection{Observations}
This paper presents the results of mapping observations in the spectral line OTF mode conducted using the B4R/LMT-50\,m from November 10 to November 29, 2019, encompassing a map size of 5$\arcmin\times$5$\arcmin$ \citep{kawabe_submitted} for testing observational capabilities.
Before the OTF observations, the focus, astigmatism, and pointing were corrected using the SiO maser of Orion-KL.
The pointing drift between calibrations was typically 3--5$\arcsec$ under stable weather conditions (e.g., wind speed $<$ 10\,m/s) \citep{kawabe_submitted}.
Intensity calibration was performed using the chopper-wheel calibration method, and its accuracy is approximately 10\%.
The observations were conducted at two 2SB mode frequency settings (set A: 131.4--133.9\,GHz and 145.1--147.6\,GHz; set B: 136.2--138.7\,GHz and 149.9--152.4\,GHz).
The beam sizes in full width at half maximum (FWHM) were 11.9$\arcsec$ and 10.9$\arcsec$ for set A, and 11.6$\arcsec$ and 10.6$\arcsec$ for set B, as listed in table \ref{tab:obsinfo}.
These beam sizes correspond to approximately 0.02\,pc at the distance of Orion-KL ($\sim418$\,pc; \cite{Kim2008-yg}).
An eXtended bandwidth Fast Fourier Transform Spectrometer (XFFTS, Radiometer Physics GmbH Co.; \cite{Klein2012-ij}) was employed to achieve a frequency resolution of 76.293\,kHz, corresponding to a velocity resolution of $\sim$0.15--0.17\,km/s (2.5\,GHz and 32768 channel).
The center coordinate of the observation field of view was ($\alpha,\delta)_{\rm {J2000.0}}$=$(\timeform{5h35m14s.16}, -\timeform{5D22'21".50}$), with the OFF point located at ($\alpha,\delta)_{\rm {J2000.0}}$=$(\timeform{5h36m15s.50}, -\timeform{5D02'34".40}$).
The total on-source time was approximately 42\,min with 3$\arcsec$ scan separations in each frequency set.
The telescope elevation range was 65.97--58.84 degrees for set A and 66.05--60.42  degrees for set B, whereas the optical depth $\tau_{\rm{225\,GHz}}$ was 0.22 for set A and 0.11 for set B.
The system temperatures ($T_{\rm{SYS}}$) were measured to be 119\,K for set A and 112\,K  for set B on average for both sidebands, respectively.
These observation parameters (observed frequency range, main beam efficiency ($\eta_{\rm{MB}}$), beam size ($\theta_{\rm{beam}}$, FWHM), telescope elevation range, $\tau_{\rm{225\,GHz}}$, and $T_{\rm{SYS}}$) for each frequency set are summarized in table \ref{tab:obsinfo}.

\begin{table*}[h]
  \tbl{Observational parameters.}{%
  \begin{tabular}{llllllllll}
      \hline
      set&sideband&frequency  & $\eta_{\rm{MB}}$ &$\theta$$_{{\rm beam}}$&elevation range\footnotemark[$*$]&$\tau_{\rm{225\,GHz}}$&$T_{\rm{SYS}}$& RMS [K]\footnotemark[$\dagger$]&RMS [K]\footnotemark[$\dagger$]\\ 
        [-0.5mm]&&[GHz]&&FWHM[$\arcsec$]&[degrees]&&[K]&(HC)&(CR)\\
      \hline
       A&lower sideband (LSB)&131.4--133.9&0.69&11.9&65.97--58.84&0.22&119&0.31&0.29\\
       &upper sideband (USB)&145.1--147.6&0.66&10.9&&&&0.37&0.39\\
       B&LSB&136.2--138.7&0.69&11.6&66.05--60.42&0.11&112&0.25&0.25\\
       &USB&149.9--152.4&0.65&10.6&&&&0.28&0.26\\
    \end{tabular}}\label{tab:obsinfo}
\begin{tabnote}
\footnotemark[$*$] telescope elevation range.
\footnotemark[$\dagger$] root mean square (RMS) of noise level calculated from the spectra extracted from the HC and CR regions ($T_{\rm{MB}}$ scale, non-channel binning)
\end{tabnote}

\end{table*}

\subsection{Data reductions}\label{2.3}
We generated measurement sets from the raw B4R data using the B4R pipeline (\texttt{b4rpipe}), developed by our team\footnote{$\langle$https://github.com/b4r-dev/b4rpipe$\rangle$}.
Subsequently, we created FITS datacubes with a grid size of 3$\arcsec$ and a Gaussian grid function using the Common Astronomy Software Applications (CASA) package \citep{CASA2022}.

\begin{figure*}[h]
 \begin{center}
  \includegraphics[width=18cm]{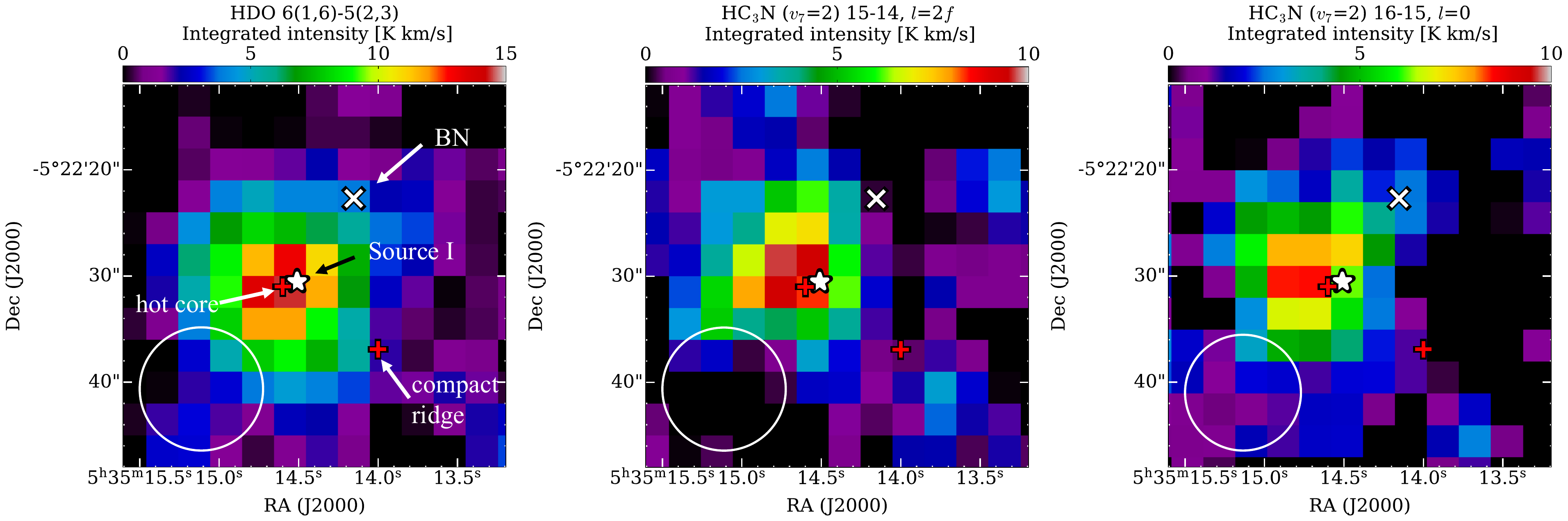}
 \end{center}

\caption{
Integrated intensity maps of HDO (left panel) and HC$_3$N ($v_7=2$) (center) in the 136.2--138.7\,GHz band and HC$_3$N ($v_7=2$) (right) in the 146.1--147.6\,GHz band, after correcting for the main beam efficiency and pointing.
The velocity ranges of integrals are $-$5.8--22 km\,s$^{-1}$, $-$23--11 km\,s$^{-1}$, and $-$2.3--22 km\,s$^{-1}$, respectively.
The red cross (+) denotes the hot core and compact ridge. 
The white cross (x) and star indicate the position of BN and Source I, respectively.
The coordinates of each substructure are summarized in table \ref{tab:cord}.
The lower left circle in each panel denotes the beam size in table \ref{tab:obsinfo}.
{Alt text: The integrated intensity maps of the three panels. The x axis is the right ascension, and the y axis is the declination at the map size of approximately 40$\arcsec\times$40$\arcsec$.}

}\label{fig:point_integmap}
\end{figure*}

In OTF observations, which require a relatively long time for mapping, a slight pointing offset can occur owing to the time interval before the next pointing calibration.
In sets A and B observations, although pointing calibrations were performed using the SiO maser of Orion-KL before OTF observations, a slight pointing offset was identified in sets A and B OTF maps. 
By taking advantage of the wide bandwidth, which enables the simultaneous observation of multiple emission lines, it is possible to select the emission lines that are most suitable for pointing offset correction after observations.
The procedure for pointing correction is as follows.
The HDO and HC$_3$N ($v_7=2$) emission lines were used to calibrate for the pointing offset.
The offset relative to the HC was measured using the HDO line, in which the OTF maps showed a point-like distribution, as shown in figure \ref{fig:point_integmap}.
Because HDO is associated with nearly the HC region (e.g., \cite{Plambeck1987-wz,Neill2013-aj}), the peak coordinates were obtained from the integrated intensity map of the HDO (6(1,6)--5(2,3), upper-level energy $E_{\rm{u}}=$444.03 K\footnote{$\langle$https://splatalogue.online$\rangle$}$^,$\footnote{$\langle$https://spec.jpl.nasa.gov/home.html $\rangle$}) in set B using 2D Gaussian fitting and then calibrated to the HC coordinates ($\alpha,\delta)_{\rm {J2000.0}}$=$(\timeform{5h35m14s.60}, -\timeform{5D22'31".00}$).
The coordinates of each substructure are summarized in table \ref{tab:cord} with the CR, BN, and Source I.
The calibration shift value for set B was ($\Delta \alpha,\Delta  \delta)_{\rm {J2000.0}}$=$(\timeform{0s.455}, \timeform{4".811}$).
Then, by using the HC$_3$N ($v_7=2$) emission lines, which are compact distributions and highly excited lines in both set A (16-15, $l=0$, $E_{\rm{u}}=$701.44 K\footnotemark[4]$^,$\footnote{$\langle$https://cdms.astro.uni-koeln.de/classic/molecules$\rangle$}) and set B (15-14, $l=2f$, $E_{\rm{u}}=$697.69 K\footnotemark[4]$^,$\footnotemark[6]), as shown in figure \ref{fig:point_integmap}, we calibrated set A by matching HC$_3$N ($v_7=2$) peaks coordinate to set B.
The calibration shift value for set A was ($\Delta \alpha,\Delta  \delta)_{\rm {J2000.0}}$=$(\timeform{0s.360}, \timeform{9".73}$).
From the  error of the 2D Gaussian fitting and the slight offset between the peaks of HDO and the position of HC (e.g., \cite{Plambeck1987-wz,Neill2013-aj}), the accuracy of this pointing correction was estimated to be approximately 2--3$\arcsec$, 
This remained within approximately 30\% of the beam size ($\sim$11$\arcsec$).
The baseline calibration over the entire 2.5\,GHz bandwidth was performed using the CASA \texttt{imcontsub} task, with a 5th-order polynomial function fitted to line-free channels selected by intensities below 3$\sigma$.

The HC and CR spectral data were extracted by averaging over circles with diameters corresponding to the each beam size in table \ref{tab:obsinfo} and centered on the HC at ($\alpha,\delta)_{\rm {J2000.0}}$=$(\timeform{5h35m14s.60}, -\timeform{5D22'31".00}$) and the CR at ($\alpha,\delta)_{\rm {J2000.0}}$=$(\timeform{5h35m14s.00}, -\timeform{5D22'36".90}$) \citep{Feng2015-au} from maps corresponding to each frequency set.
The root mean square (RMS) values of the noise level for the HC and CR ($T_{\rm{MB}}$ scale, non-channel binning) were calculated from spectral data, as shown in table \ref{tab:obsinfo}.

\begin{table}[h]
  \tbl{Coordinates of substructures.}{%
  \begin{tabular}{llll}
      \hline
      Source  & RA $\alpha_{\rm{J2000}}$&Dec $\delta_{\rm{J2000}}$&Reference\footnotemark[$*$]\\ 
      \hline
       HC&\timeform{5h35m14s.60}&$-$\timeform{5D22'31".00}&(1)\\
       CR&\timeform{5h35m14s.00}&$-$\timeform{5D22'36".90}&(1)\\
       BN&\timeform{5h35m14s.11}&$-$\timeform{5D22'22".70}&(2)\\
       Source I&\timeform{5h35m14s.51}&$-$\timeform{5D22'30".53}&(3)\\
      \hline
    \end{tabular}}\label{tab:cord}
\begin{tabnote}
\footnotemark[$*$] (1) \citep{Feng2015-au}; (2) \citep{Beckwith1978-zo}; (3) \citep{Menten1995-ph}
\end{tabnote}

\end{table}

\section{Results}
\subsection{Line identification and OTF maps}\label{lineid}
Line identification was performed to test the observational capabilities of spectral scans using the eXtended CASA Line Analysis Software Suite (XCLASS) \citep{xclass2017} and by referring to the splatalogue\footnotemark[4] database and previous studies.
XCLASS refers to two molecular information databases, the Cologne Database for Molecular Spectroscopy (CDMS\footnotemark[6]) \citep{Endres2016-it,Muller2001-ez,Muller2005-kw} and the Jet Propulsion Laboratory molecular databases (JPL\footnotemark[5]) \citep{Pickett1998-ih}.
We performed identifications with a detection threshold of 3$\sigma$ using 0.15--0.17 km/s resolution spectra (non-channel binning).
In this analysis, as described in Section \ref{2.3}, spectra were extracted by averaging each beam size (table \ref{tab:obsinfo}) centered on the HC and CR from maps.
The effective angular resolution is degraded to approximately 1.4 times the original resolution.
As a result, the spectra of HC and CR, which are separated by approximately 11$\arcsec$, are slightly contaminated by each other.
However, to perform line identification and evaluate the spatial resolution capability using spectra with improved signal-to-noise ratio (SNR), we adopted the beam size as the averaging region.

As a result, 404 emission lines are detected in the observed frequency ranges, as shown in figures \ref{fig:all_lines_1}--\ref{fig:all_lines_4} in the Appendix. 
In this study, molecules with more than two emission lines (i.e., two transitions) in the observation frequency ranges are defined as identified.
On the other hand, molecules with only one emission line in the observation frequency ranges, with the exception of the recombination line, are defined as tentatively identified.
Among the 404 emission lines, 337 lines corresponded to identified or tentatively identified lines.
Among the remaining 67 lines, 16 lines were attributed to leakage from the second downconversion or aliasing from the analog-to-digital converter (ADC) owing to the absence of anti-aliasing filters during this observation.
In addition, two lines are ascribed to the leakage of stronger OCS and SO lines from the image sideband to the signal sideband.
These 18 leakage emission lines were labeled as X-lines.
The remaining 49 lines were labeled as unidentified lines (U-lines).
Figures \ref{fig:all_lines_1}--\ref{fig:all_lines_4} and tables \ref{tab:lineid}--\ref{tab:lineid_leak} in the Appendix show detailed information on the spectra of each frequency band, alongside the identified lines, tentatively identified lines, X-lines, and U-lines.

CS and C$^{33}$S were detected for only one line each. 
However, we classified them as identified because these lines exhibit sufficient intensity in figures \ref{fig:all_lines_1}--\ref{fig:all_lines_4} in the Appendix, which is consist with previous studies \citep{Tercero2010-sm}.
In the spectra extracted from the HC and CR, only one transition (150.8519\,GHz) of cyclic (\textit{c})-C$_3$H$_2$ was detected, excluding a transition that is blended with a U-line.
However, the detection of this transition (150.8519\,GHz) is consistent with that of \citet{Ziurys1993-ps}, and other transitions (150.8207\,GHz) buried in noise in the HC and CR were detected in different regions (e.g., regions west and north of the HC and CR) \citep{kawabe_submitted}.
Therefore, \textit{c}-C$_3$H$_2$ was classified as an identified species.
Second excited torsional states of CH$_3$OCHO ($v_{\rm{t}}=2$) were identified through comparison with previous studies \citep{Takano2012-bt,Kobayashi2013-pc}.
Consequently, H35$\alpha$, H51$\gamma$ and 29 molecular species, including isotopologues, deuterated molecules, and vibrational excited states, were identified (NO, CS, C$^{33}$S, SO, $^{33}$SO, SO$_2$ ($v=0, v_2=1$), $^{34}$SO$_2$, $^{33}$SO$_2$, OCS, O$^{13}$CS, HDO, HNCO, H$_2$CO, H$_2$$^{13}$CO, H$_2$CS, cyclic \textit{c}-C$_3$H$_2$, HC$_3$N ($v=0, v_6=1, v_7=1, v_7=2$), t-HCOOH, CH$_3$CN ($v=0, v_8=1$),
CH$_3$OH ($v=0$), $^{13}$CH$_3$OH, CH$_3$CHO, CH$_3$CCH, CH$_3$NCO, C$_2$H$_3$CN,
CH$_3$OCHO ($v=0, v_{18}=1, v_{\rm{t}}=2$), C$_2$H$_5$CN, CH$_3$OCH$_3$, and CH$_3$COCH$_3$).
A total of 12 molecular species, including excited states, were tentatively identified (S$^{18}$O, OC$^{33}$S, $^{18}$OCS, $^{18}$O$^{13}$CS, SiS, H$_2$C$^{34}$S, H$^{13}$CCCN ($v=0$), H(C)OCN, CH$_2$DCN, CH$_3$$^{13}$CN, CH$_3$OH ($v_{12}=1$), and CH$_3$$^{18}$OH).
The results are summarized in table \ref{tab:lines}.

\begin{table}[h]
  \tbl{Identified and tentatively identified species, including exited states.}{%
    {\renewcommand\arraystretch{1.2}
  \begin{tabular}{l}
      \hline
      Identified species and exited states\footnotemark[$*$]\\
      \hline
      H35$\alpha$, H51$\gamma$, NO, CS, C$^{33}$S, SO, $^{33}$SO, SO$_2$ ($v=0, v_2=1$),\\
      $^{34}$SO$_2$, $^{33}$SO$_2$, OCS, O$^{13}$CS, HDO, HNCO, H$_2$CO, \\
      H$_2$$^{13}$CO, H$_2$CS, \textit{c}-C$_3$H$_2$, HC$_3$N ($v=0, v_6=1, v_7=1, v_7=2$), \\
      t-HCOOH, CH$_3$CN ($v=0, v_8=1$), CH$_3$OH ($v=0$),$^{13}$CH$_3$OH,\\
      CH$_3$CHO, CH$_3$NCO, CH$_3$CCH, C$_2$H$_3$CN,\\
      CH$_3$OCHO ($v=0, v_{18}=1, v_{\rm{t}}=2$), C$_2$H$_5$CN, CH$_3$OCH$_3$,\\
      CH$_3$COCH$_3$\\
    \hline
      Tentatively identified species and exited states\footnotemark[$\dagger$]\\
    \hline
    S$^{18}$O, OC$^{33}$S, $^{18}$OCS, $^{18}$O$^{13}$CS, SiS, H$_2$C$^{34}$S, H$^{13}$CCCN ($v=0$),\\
    H(C)OCN, CH$_2$DCN, CH$_3$$^{13}$CN, CH$_3$OH ($v_{12}=1$), CH$_3$$^{18}$OH\\
      \hline
    \end{tabular}}}\label{tab:lines}
\begin{tabnote}
\footnotemark[$*$] Species or exited states with more than two detections of the emission lines in the observation frequency ranges.
\footnotemark[$\dagger$] Species or exited states with only one detection of the emission lines in the observation frequency ranges, except the recombination line.
\end{tabnote}

\end{table}

\begin{figure}[h]
 \begin{center}
  \includegraphics[width=8cm]{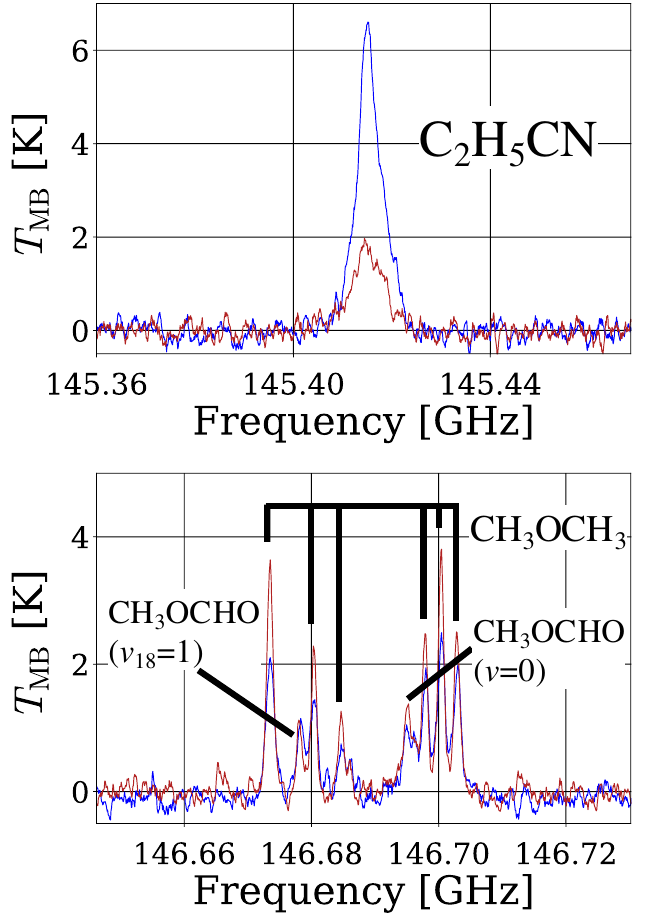}
 \end{center}

\caption{
Spectra of C$_2$H$_5$CN, CH$_3$OCHO ($v=0,v_{18}=1$), and CH$_3$OCH$_3$ in the 145.1--147.6\,GHz band. 
The blue and red lines represent spectra with 10-channel binning, averaging in circles with diameters corresponding to the beam size of 10.9$\arcsec$ centered at the HC and CR \citep{Feng2015-au}, respectively.
{Alt text: The molecular emission line spectra of the two panels. The x axis is the frequency at gigahertz, and the y axis is the intensity at kelvin on the $T_{\rm{MB}}$ scale.}
}\label{fig:lineid_ex_2}
\end{figure}

O-bearing COMs (CH$_3$OH ($v=0$), $^{13}$CH$_3$OH, CH$_3$CHO, CH$_3$OCHO ($v=0, v_{18}=1, v_{\rm{t}}=2$), CH$_3$OCH$_3$, and CH$_3$COCH$_3$) and N-bearing COMs (CH$_3$CN ($v=0, v_8=1$), C$_2$H$_5$CN, and C$_2$H$_3$CN) are detected, as shown in figure \ref{fig:all_lines_1}--\ref{fig:all_lines_4}.
Notably, N-bearing COMs, such as C$_2$H$_5$CN, exhibit a strong spectral intensity in the HC (blue line), whereas O-bearing COMs, such as CH$_3$OCHO ($v=0$) and CH$_3$OCH$_3$, exhibit a strong spectral intensity in the CR (red line), as shown in figure \ref{fig:lineid_ex_2}.
Although the spectra of HC and CR are slightly contaminated by each other, the difference in the spectral intensity indicates that the chemical differences between the HC and CR regions are clearly resolved in these observations.

\begin{figure}[h]
 \begin{center}
  \includegraphics[width=8cm]{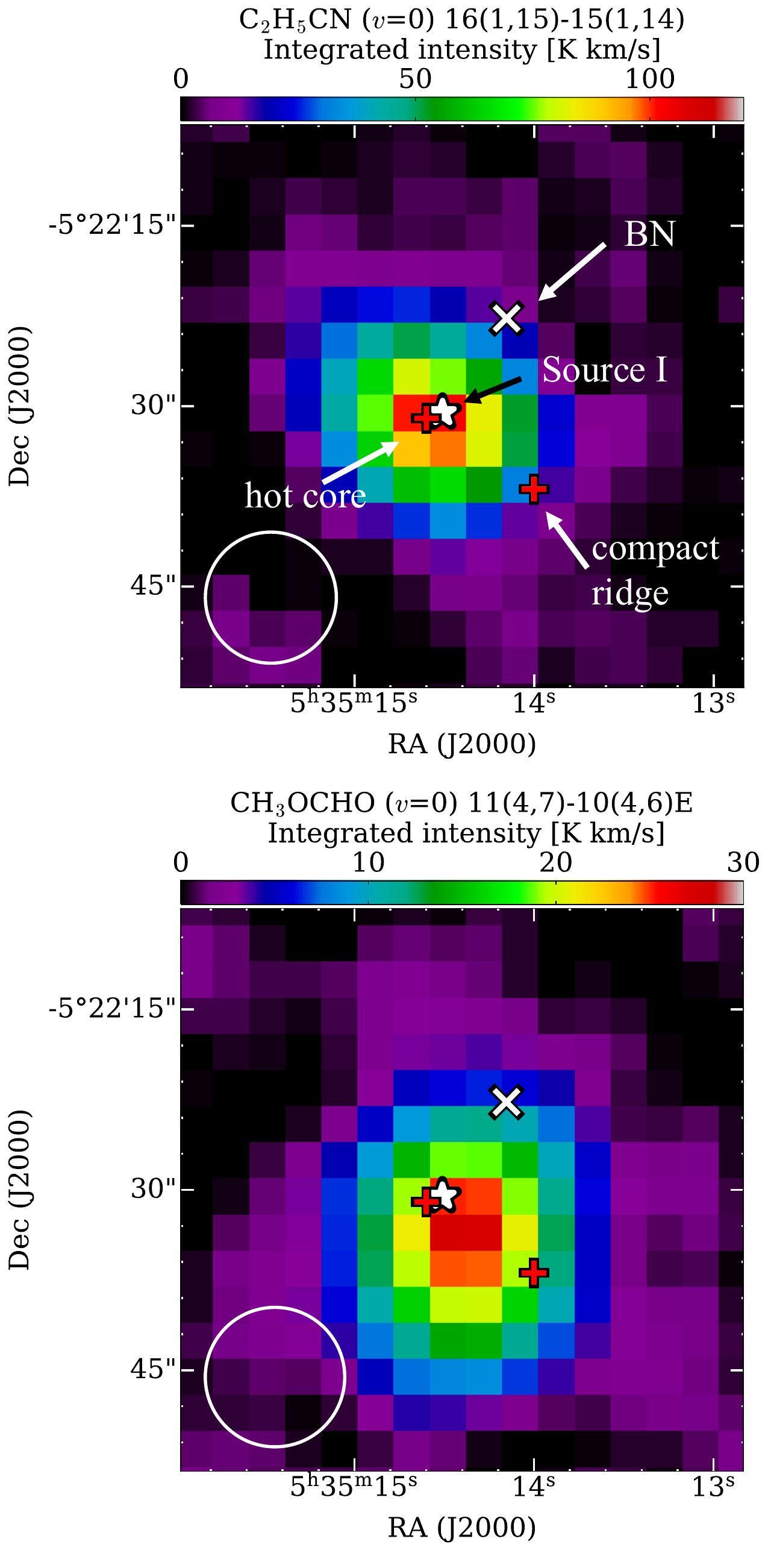}
 \end{center}

\caption{
Integrated intensity maps of C$_2$H$_5$CN ($v=0$) (upper panel) in the 145.1--147.6\,GHz band and CH$_3$OCHO ($v=0$) (lower) in the 136.2--138.7\,GHz band, after correcting for the main beam efficiency and pointing.
The velocity ranges of integrals are $-$40--40 km\,s$^{-1}$ and 0--20 km\,s$^{-1}$, respectively.
The various sources indicated on each map are described in figure \ref{fig:point_integmap} and table \ref{tab:cord}.
The lower left circle in each panel denotes the beam size in table \ref{tab:obsinfo}.
{Alt text: Integrated intensity maps of the two panels. The x axis is the right ascension, and the y axis is the declination at the map size of approximately 40$\arcsec\times$40$\arcsec$.}

}\label{fig:otf_integmap}
\end{figure}

The integrated intensity maps of C$_2$H$_5$CN ($v=0$) and CH$_3$OCHO ($v=0$) are shown in figure \ref{fig:otf_integmap}.
The RMSs of the map noise level are 2.8 K\,km\,s$^{-1}$ and 0.8 K\,km\,s$^{-1}$ for the observation settings A and B, respectively.
The distributions of C$_2$H$_5$CN ($v=0$) and CH$_3$OCHO ($v=0$) are different.
C$_2$H$_5$CN exhibit a peak near the HC, whereas CH$_3$OCHO exhibit a peak between the HC and CR, as shown in figure \ref{fig:otf_integmap}.
The differences in the spectral intensity and the spatial distribution shown in figures \ref{fig:lineid_ex_2} and \ref{fig:otf_integmap} are consistent with those in previous studies (e.g., \cite{Feng2015-au,Friedel2012-oj,Tercero2018-as}) and indicate that the B4R installed on the LMT-50\,m can distinguish between the N-bearing COM-rich HC and O-bearing COM-rich CR.

\subsection{Chemical analysis of C$_2$H$_5$CN and CH$_3$OCHO}\label{RD}
As shown in figure \ref{fig:lineid_ex_2}, the difference in spectral intensity between N- and O-bearing COMs suggests their spatial segregation across our two regions of interest.
C$_2$H$_5$CN and CH$_3$OCHO are representative N- and O-bearing COMs that have been observed not only in the Orion-KL region but also in distant HMSFRs (e.g., \cite{Coletta2020-yo,Bonfand2017-mr,Bonfand2019-fn}).
In addition, these lines do not seem to be strongly blended with other molecular lines in our observational data. 
Therefore, we visualized the spatial distributions as the maps of column densities and rotational temperatures of C$_2$H$_5$CN and CH$_3$OCHO using the rotation diagram method \citep{Turner1991} with the advantage of the B4R on the LMT in terms of spatial resolution (i.e., beam size $\sim$11--12$\arcsec$).
The application of the rotation diagram method to maps has also been reported by \citet{Bell2014-ku}.
The following equations were used in this calculation.

\begin{eqnarray}
\displaystyle \log_{10}\left[\frac{3kW}{(8\rm{\pi}^3\nu S\mu^2\textit{g}_{\rm{I}}\textit{g}_{\rm{K}})}\right] \!\!\!\!\!\!\!\!&=&\!\!\!\!\!\!\!\! \log_{10}\left[\frac{N}{Q(T_{\rm{rot}})}\right]-E_{\rm{u}}\frac{\log_{10}{\rm e}}{T_{ \rm{rot}}}\label{rd} , \\
\displaystyle Q(T_{\rm{rot}}:\rm {C_2H_5CN}) &=& \left[\frac{\pi(\textit{k}\textit{T}_{\rm{rot}})^3}{\textit{h}^3\textit{ABC}}\right]^{1/2}\label{r_n} , \\
\displaystyle Q(T_{\rm{rot}}:\rm {CH_3OCHO}) &=& 2\left[\frac{\pi(\textit{k}\textit{T}_{\rm{rot}})^3}{\textit{h}^3\textit{ABC}}\right]^{1/2} ,\label{r_o} 
\end{eqnarray}
where $W$, $\nu$, $k$, $h$, $\mu$, $S$, $E_{\rm{u}}$, $g_{\rm{I}}$, $g_{\rm{K}}$, $N$, $Q$, and $T_{\rm{rot}}$ represent the integrated intensity, frequency, Boltzmann constant, Planck constant, dipole moment, line strength, upper-level energy, nuclear spin degeneracy, $K$-level degeneracy, column density, partition function, and rotational temperature, respectively.
$A$, $B$, and $C$ in equations (\ref{r_n}) and (\ref{r_o}) indicate the rotational constants for the three principal axes of each molecule.

To apply the rotation diagram method to the maps, we used the spectra of C$_2$H$_5$CN and CH$_3$OCHO extracted by averaging each beam size at the HC and CR (Section \ref{lineid}) as a reference. 
In this reference spectrum, transitions with a spectral intensity of more than 4$\sigma$ in both the HC and CR regions, and that are not blended with other emission lines, were used in the calculations for all pixels.
CH$_3$OCHO has two types of rotational transitions, ground torsional states ($v=0$) and first excited torsional states ($v_{18}=1$), in the emission lines above 4$\sigma$.
Therefore, calculations were performed using only the ground torsional state ($v=0$) and both torsional states ($v=0$ and $v_{18}=1$).
Consequently, 8 C$_2$H$_5$CN, 25 CH$_3$OCHO ($v=0$), and 34 CH$_3$OCHO ($v=0$ and $v_{18}=1$) transitions were used in this calculation.
The degeneracies of C$_2$H$_5$CN are $g_{\rm{I}}=g_{\rm{K}}=1$.
Both A- and E-type transitions were observed for CH$_3$OCHO.
$g_{\rm{I}}=2$ and $g_{\rm{K}}=1$ for A-type transitions, whereas $g_{\rm{I}}=1$ and $g_{\rm{K}}=2$ for E-type transitions \citep{Turner1991}.
In this study, both A- and E-type transitions were used together for fitting; thus, the column density includes contributions from both types.
The formulation of $Q$ (equations (\ref{r_n}) and (\ref{r_o})) was adopted from \citet{Turner1991}.
Because CH$_3$OCHO has A- and E-type transitions, the partition function of CH$_3$OCHO (equation (\ref{r_o})) is multiplied by 2.
The values of $A$, $B$, and $C$ were obtained from \citet{Fukuyama1996-bp} for C$_2$H$_5$CN and from \citet{Karakawa2001-gz} for CH$_3$OCHO.
The values of $\nu$, $S\mu^2$, and $E_{\rm{u}}$ were obtained from the JPL database.
The integrated intensity maps calculated from the OTF observations of 8 C$_2$H$_5$CN lines and 34 CH$_3$OCHO ($v=0$ and $v_{18}=1$) lines, which have intensities above 4$\sigma$ in the spectra extracted by averaging each beam size at the HC and CR, were fitted pixel-by-pixel using equation (\ref{rd}), resulting in maps of $N$ and $T_{\rm{rot}}$.
The beam size of these maps was aligned to 12$\arcsec$ using CASA.

\begin{figure*}
 \begin{center}
      \includegraphics[width=14.5cm]{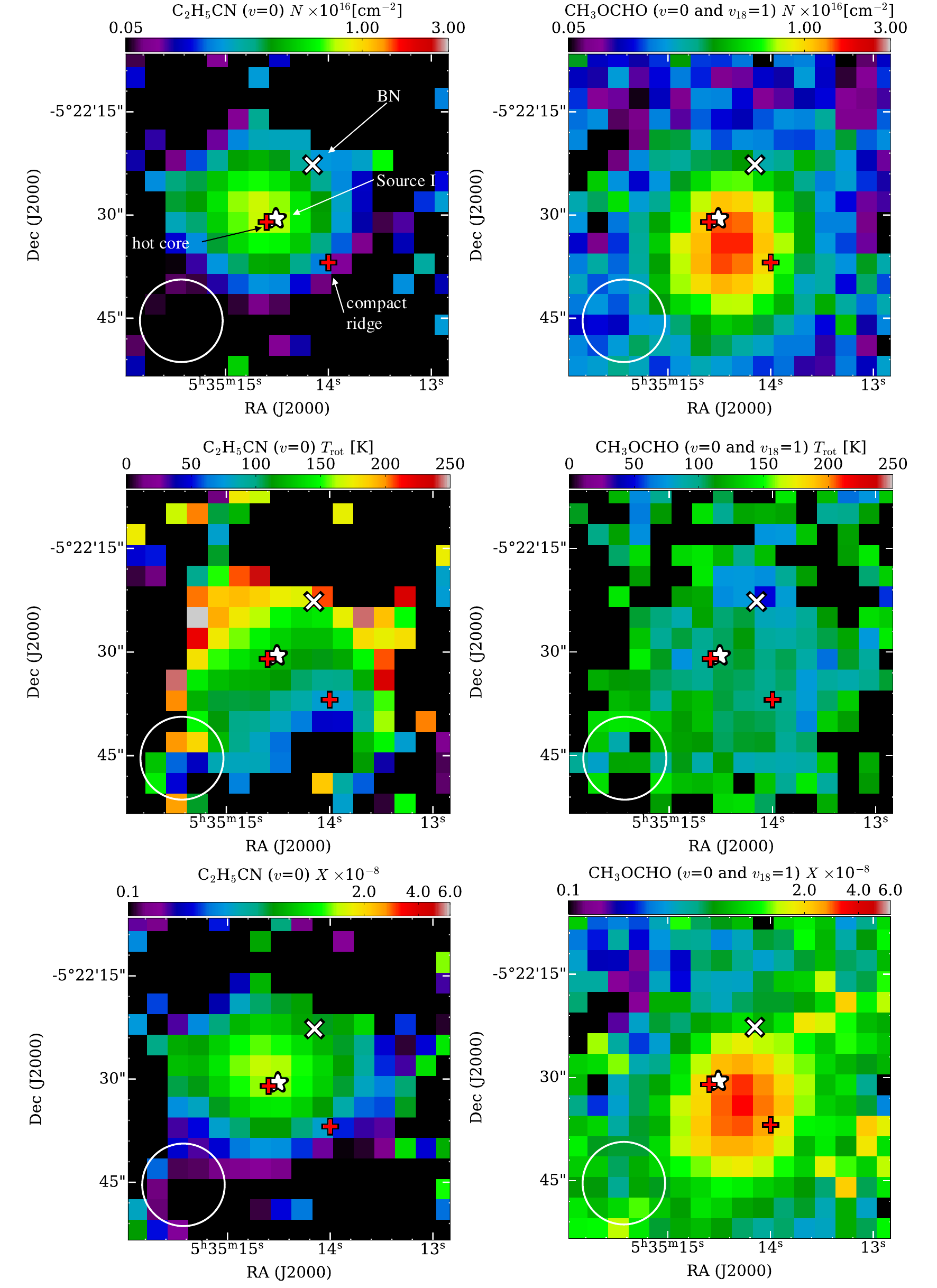}
 \end{center}


\caption{
Column density $N$ (upper row), rotational temperature $T_{\rm{rot}}$ (middle), and relative abundance with respect to $\rm {H_2}$ $X$ (lower) maps of C$_2$H$_5$CN (left column) and CH$_3$OCHO ($v=0$ and $v_{18}=1$) (right). 
The various sources indicated on each map are described in figure \ref{fig:point_integmap} and table \ref{tab:cord}.
The lower left circle in each panel denotes the beam size (12$\arcsec$).
The column density, the rotational temperature, and the abundance maps are masked above the maximum around the HC and CR.
{Alt text: The column density, rotational temperature, and relative abundance with respect to $\rm {H_2}$ maps of the six panels. The x axis is the right ascension, and the y axis is the declination at map size of approximately 40$\arcsec\times$40$\arcsec$.}
}\label{fig:column_trot}

\end{figure*}

\begin{figure*}
 \begin{center}
    \includegraphics[width=16cm]{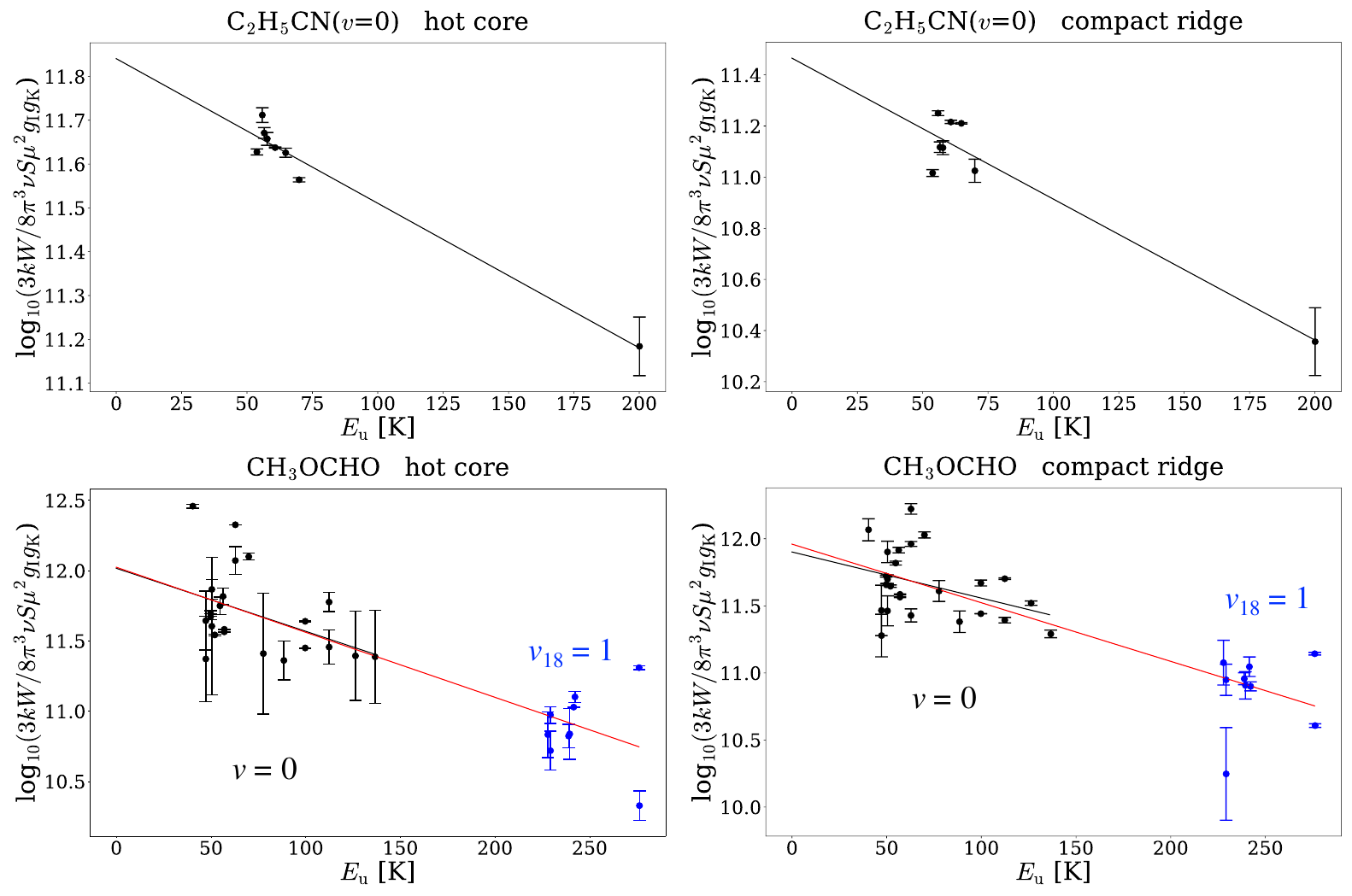}
 \end{center}

\caption{
Rotation diagram plots of C$_2$H$_5$CN and CH$_3$OCHO ($v=0$ and $v_{18}=1$) for the HC and CR (12$\arcsec$). The error bars indicate the 1$\sigma$ noise of $W$. 
The upper panels show the plots of C$_2$H$_5$CN.
In the lower panels, the black plots are CH$_3$OCHO ($v=0$) and the blue plots are CH$_3$OCHO ($v_{18}=1$). 
The black line shows the fitting result using only ($v=0$) states in each panel.
The red line represents the fitting result using both torsional states CH$_3$OCHO ($v=0$ and $v_{18}=1$).
{Alt text: The rotation diagram plots of the four panels. The x axis is the upper-level energy at kelvin, and the y axis is the left side of equation \eqref{rd}.}
}\label{fig:RD1}
\end{figure*}

The relative abundances with respect to $\rm {H_2}$ ($X=N/N(\rm {H_2})$) were calculated.
The values of $N(\rm {H_2})$ are acquired from the FITS maps reported by \citet{Schuller2021-ci}.
The $N(\rm {H_2})$ map has an angular resolution of 8$\arcsec$.
This map was calculated using SPIRE 160, 250, 350, and 500\,$\mu$m data from the Herschel Gould Belt Survey \citep{Andre2010-ao} and ArT\'{e}MiS 350 and 450\,$\mu$m data. 
Previous studies (e.g., \cite{Suzuki2018-ya, Suzuki2016-qw,Comito2005-ru}) have shown that assuming a conservative 10$\arcsec$ source size is an effective method for estimating the relative abundances, because the results agree well with other previous studies within the margin of error.
In this study, we also assumed a 10$\arcsec$ source size to compare with the results of \citet{Suzuki2018-ya} and to evaluate the observational capability of the B4R on the LMT-50\,m.
In addition to $N$, $T_{\rm{rot}}$, and $X$, which were calculated from data with a beam size of 12$\arcsec$, 
$N^{\prime}$, $T^{\prime}_{\rm{rot}}$, and $X^{\prime}$ in the HC and CR were calculated assuming a source size of 10$\arcsec$ based on \citet{Suzuki2018-ya}.
We adjusted the beam size of the $N(\rm {H_2})$ map from 8$\arcsec$ to 12$\arcsec$ and 10$\arcsec$ using CASA and calculated the 1$\sigma$ error from the standard deviations of 12$\arcsec$ and 10$\arcsec$.
The $N(\rm {H_2})$ values in the HC and CR regions are ($5.5\,\pm\,0.1)\,\times\,10^{23}\,$cm$^{-2}$ and ($4.0\,\pm\,0.2)\,\times\,10^{23}\,$cm$^{-2}$, at 12$\arcsec$, respectively.
At 10$\arcsec$, the $N(\rm {H_2})$ values in the HC and CR regions are ($5.8\,\pm\,0.1)\,\times\,10^{23}\,$cm$^{-2}$ and ($4.1\,\pm\,0.3)\,\times\,10^{23}\,$cm$^{-2}$, respectively.
The maps of $N$, $T_{\rm{rot}}$, and $X$ for C$_2$H$_5$CN and CH$_3$OCHO ($v=0$ and $v_{18}=1$) are shown in figure \ref{fig:column_trot}, whereas the rotation diagram plots for the HC and CR are shown in figure \ref{fig:RD1}.
The column density, the rotational temperature, and the relative abundance maps are masked above the maximum around the HC and CR, as shown in figure \ref{fig:column_trot}.
The 1$\sigma$ noise of $W$ is expressed as  error bars in the rotation diagram plots (figure \ref{fig:RD1}).
The $N$, $T_{\rm{rot}}$, $X$, $N^{\prime}$, $T^{\prime}_{\rm{rot}}$, and $X^{\prime}$ values in the HC and CR regions of C$_2$H$_5$CN, CH$_3$OCHO ($v=0$), and CH$_3$OCHO ($v=0$ and $v_{18}=1$) are listed in table \ref{tab:result_all}.
In table \ref{tab:result_all}, we show the fitting results for only the 25 CH$_3$OCHO ($v=0$) transitions as well as the fitting results using data for 34 CH$_3$OCHO ($v=0$ and $v_{18}=1$) transitions.
The errors in $N$, $T_{\rm{rot}}$, $N^{\prime}$, and $T^{\prime}_{\rm{rot}}$ were calculated from the 1$\sigma$ error obtained by fitting, considering error propagation and adding 10\% of the main values as the absolute calibration error.
The errors in $N$ and $N^{\prime}$ also take into account the uncertainty of the rotation temperature in the calculation of the partition function.

Figure \ref{fig:column_trot} shows that the $X$(C$_2$H$_5$CN) and $X$(CH$_3$OCHO) of Orion-KL observed at the 0.02\,pc scale differs between the HC and the CR, indicating that the difference is not due to the total gas abundance but to the chemical properties. 
$X$(C$_2$H$_5$CN) shows a maximum near the HC and is approximately four times stronger than that observed in the CR.
$X$(CH$_3$OCHO) shows a peek in the region between the HC and CR.
In addition, $X$(C$_2$H$_5$CN) has a compact distribution, whereas $X$(CH$_3$OCHO) has a more extended distribution in figure \ref{fig:column_trot}.
Although previous studies (e.g., \cite{Suzuki2018-ya,Feng2015-au}) have determined the temperatures in specific regions, such as the HC and CR, the temperature map in figure \ref{fig:column_trot} shows the continuous temperature distributions in COMs around the HC and CR, as shown in the temperature maps of CH$_3$CN reported by \citet{Bell2014-ku}.
Figure \ref{fig:column_trot} shows a broad distribution at approximately 100\,K for the $T_{\rm{rot}}$ of CH$_3$OCHO; however, it shows that a high-temperature region exists northeast of the HC for the $T_{\rm{rot}}$ of C$_2$H$_5$CN, i.e., CH$_3$OCHO and C$_2$H$_5$CN have different temperature distributions.

\begin{table*}

  \tbl{Column density, rotational temperature, and relative abundance with respect to $\rm {H_2}$ of C$_2$H$_5$CN, CH$_3$OCHO ($v=0$), and CH$_3$OCHO ($v=0$ and $v_{18}=1$).}{%
  {\renewcommand\arraystretch{1.2}
  \begin{tabular}
{l@{\hspace{1mm}}l@{\hspace{1mm}}l@{\hspace{1mm}}l@{\hspace{1mm}}l@{\hspace{1mm}}l@{\hspace{1mm}}l@{\hspace{1mm}}l@{\hspace{1mm}}l@{\hspace{1mm}}}
    \hline
     \multicolumn{1}{c}{no source size assumptions (12$\arcsec$)} & \multicolumn{5}{c}{hot core (HC)}&\multicolumn{3}{c}{compact ridge (CR)}\\
        \cline{3-5}  \cline{7-9}
      & &$N$$\times10^{16}$ & $T_{\rm{rot}}$ &$X$$\times10^{-8}$ &  & $N$$\times10^{16}$ & $T_{\rm{rot}}$ &$X$$\times10^{-8}$  \\
      [-0.5mm]Species&&[cm$^{-2}$]&[K]&&&[cm$^{-2}$]&[K]&\\
      C$_2$H$_5$CN && 0.8($\pm$0.2)&132($\pm$17)&1.4($\pm$0.3)&&0.1($\pm$0.04)&79($\pm$13)&0.4($\pm$0.1)\\
      CH$_3$OCHO ($v=0$)&&1.1($\pm$0.9)&97($\pm$46)&2.0($\pm$1.6)&&1.3($\pm$1.1)&126($\pm$65)&3.1($\pm$2.6)\\
      CH$_3$OCHO ($v=0$ and $v_{18}=1$)&&1.1($\pm$0.4)&94($\pm$15)&1.9($\pm$0.6)&&1.0($\pm$0.3)&99($\pm$16)&2.5($\pm$0.8)\\
      \hline
     \multicolumn{1}{c}{assuming a source size (10$\arcsec$)} & \multicolumn{5}{c}{hot core (HC)}&\multicolumn{3}{c}{compact ridge (CR)}\\
       \cline{3-5}  \cline{7-9}
       & &$N^{\prime}$$\times10^{16}$ & $T^{\prime}_{\rm{rot}}$ &$X^{\prime}$$\times10^{-8}$ &  & $N^{\prime}$$\times10^{16}$ & $T^{\prime}_{\rm{rot}}$ &$X^{\prime}$$\times10^{-8}$  \\
      [-0.5mm]Species&&[cm$^{-2}$]&[K]&&&[cm$^{-2}$]&[K]&\\
      C$_2$H$_5$CN && 1.1($\pm$0.2)&132($\pm$17)&1.9($\pm$0.4)&&0.2($\pm$0.06)&79($\pm$13)&0.5($\pm$0.2)\\
      CH$_3$OCHO ($v=0$)&&1.6($\pm$1.3)&97($\pm$46)&2.7($\pm$2.2)&&1.8($\pm$1.5)&126($\pm$65)&4.4($\pm$3.7)\\
      CH$_3$OCHO ($v=0$ and $v_{18}=1$)&&1.5($\pm$0.5)&94($\pm$15)&2.6($\pm$0.9)&&1.4($\pm$0.5)&99($\pm$16)&3.5($\pm$1.1)\\
      \hline
    \end{tabular}}
}
\label{tab:result_all}
\end{table*}

\section{Discussion}\label{dis}
\subsection{Comparison of line identification with previous studies}
\begin{figure*}[h]
 \begin{center}
  \includegraphics[width=17.5cm]{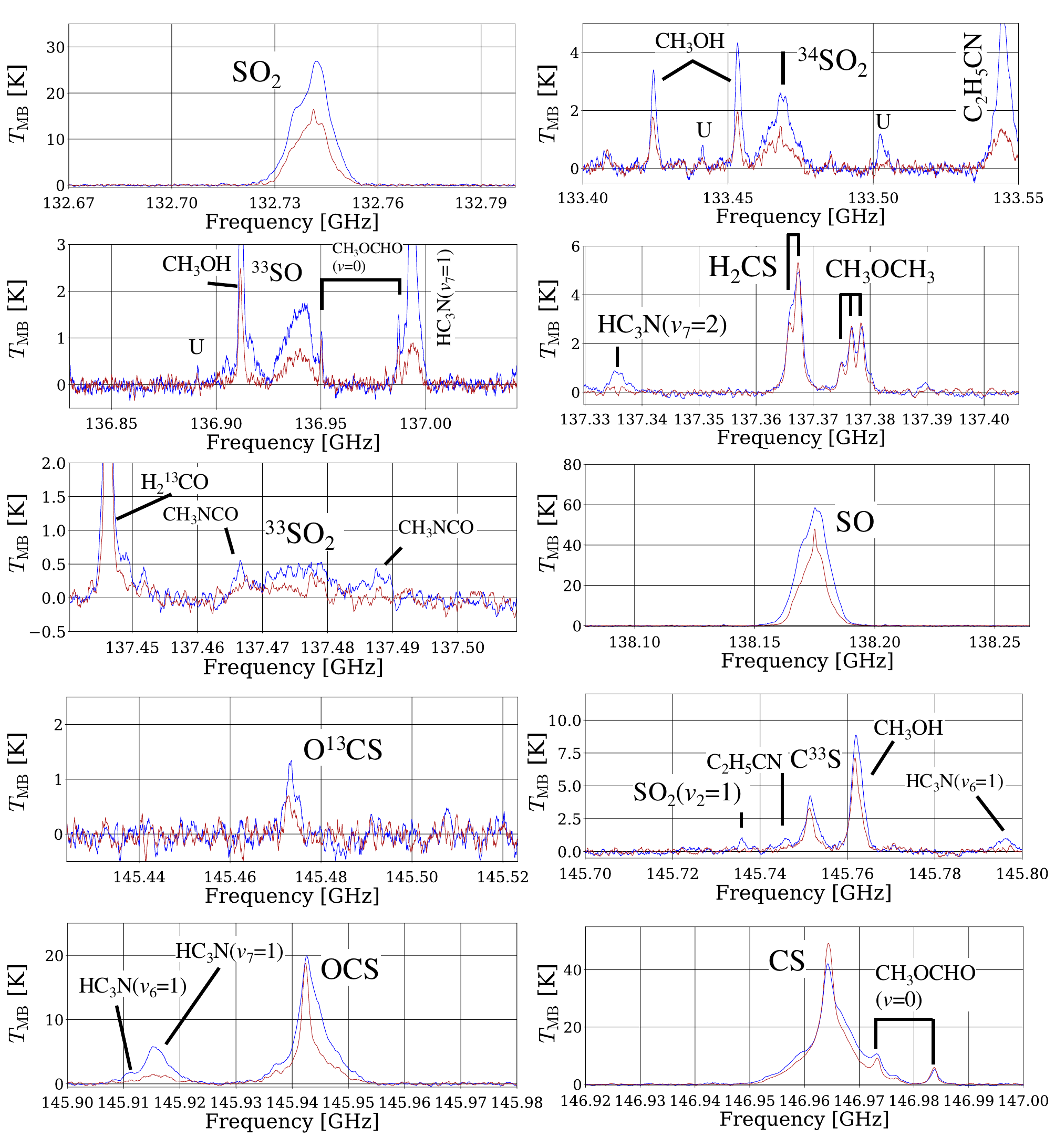}
 \end{center}

\caption{Spectrum of sulfur oxide molecules and sulfur carbon chain molecules (SO$_2$ ($v=0,v_2=1$), $^{34}$SO$_2$, SO, $^{33}$SO, $^{33}$SO$_2$, O$^{13}$CS, OCS, H$_2$CS, CS, and C$^{33}$S) with other lines. 
The blue and red lines represent spectra with 10-channel binning, averaging in circles with diameters corresponding to the beam size for each frequency band (table \ref{tab:obsinfo}, 11--12$\arcsec$) centered at the HC and CR \citep{Feng2015-au}, respectively. Lines labeled with "U" denote undefined lines.
{Alt text: The molecular emission line spectra of the ten panels. The x axis is the frequency at gigahertz, and the y axis is the intensity at kelvin on the $T_{\rm{MB}}$ scale.}
}\label{fig:lineid_ex_1}
\end{figure*}

Sulfur oxide molecules and sulfur carbon chain molecules, such as SO$_2$ ($v=0,v_2=1$), $^{34}$SO$_2$, SO, $^{33}$SO, $^{33}$SO$_2$, O$^{13}$CS, OCS, H$_2$CS, CS, and C$^{33}$S, were identified.
This identification is consistent with previous studies toward Orion-KL conducted at the Institut de radioastronomie millim\'{e}trique (IRAM)-30\,m \citep{Tercero2010-sm,Esplugues2013-yl}.
S$^{18}$O, OC$^{33}$S, $^{18}$OCS, H$_2$C$^{34}$S, and SiS (silicon-bearing molecules) were labeled tentatively identified.
However, their detection in previous studies toward Orion-KL \citep{Tercero2010-sm,Tercero2011-rs,Esplugues2013-yl} substantiates the reliability of our detection.
The spectra of these molecules, excluding the tentative lines, are shown in figure \ref{fig:lineid_ex_1}.

Five lines that were unidentified in the Orion-KL region 2\,mm band line survey by the Taeduk Radio Astronomy Observatory (TRAO)-14\,m \citep{Lee2001-ws} and the Five College Radio Astronomical Observatory (FCRAO)-14\,m \citep{Ziurys1993-ps}, were identified in this study as C$_2$H$_3$CN (138.39516\,GHz), HC$_3$N ($v_7=1$) (146.1276387\,GHz), S$^{18}$O (145.8738078\,GHz, tentatively identified), CH$_3$OCH$_3$ (150.594406\,GHz), and CH$_3$OH ($v_{12}=1$) (150.282794\,GHz, tentatively identified), as shown in figure \ref{fig:all_lines_1}--\ref{fig:all_lines_4}.
In addition, 162 previously undetected lines, with the exception of X-lines \citep{Lee2001-ws,Ziurys1993-ps} were newly detected, and 142 lines were identified in this study.
The remaining 20 lines were labeled U-lines in this study.
The line previously identified as CH$_3$OCH$_3$ at 150.4673\,GHz by \citet{Lee2001-ws} was labeled a U-line in this study.
Despite the short integration time, these results demonstrate the sufficient performance of the B4R receiver.

\subsection{Comparison of chemical properties with results from previous studies}
\subsubsection{Orion-KL}
The $N$, $T_{\rm{rot}}$, $N^{\prime}$, and $T^{\prime}_{\rm{rot}}$ values of CH$_3$OCHO ($v=0$) and C$_2$H$_5$CN in the HC (table \ref{tab:result_all}) are consistent with the results from the Nobeyama Radio Observatory (NRO)-45\,m centered nearly on the HC (assuming a source size of 10$\arcsec$, \cite{Suzuki2018-ya}).
Their $X$ and $X^{\prime}$ values are lower than those reported by \citet{Suzuki2018-ya}.
This is likely because the $N(\rm {H_2})$ values used by \citet{Suzuki2018-ya} were obtained from observations with a large beam size, leading to an underestimation of $N(\rm {H_2})$.
All values of CH$_3$OCHO ($v=0$ and $v_{18}=1$) at the HC and CR are lower than those reported by \citet{Feng2015-au}.
This difference is expected because their observations were performed using an interferometer, which provided a smaller synthetic beam, approximately 6$\arcsec$$\times$4$\arcsec$\citep{Feng2015-au}.

The distribution of C$_2$H$_5$CN centered around the HC resembles the integrated intensity maps obtained using interferometers, such as the Submillimeter Array (SMA), IRAM-30\,m \citep{Feng2015-au}, and the Combined Array for Research in Millimeter-wave Astronomy (CARMA, \cite{Friedel2012-oj}).
The distribution of CH$_3$OCHO extending between the HC and CR is consistent with the integrated intensity distributions obtained from the SMA and IRAM-30\,m \citep{Feng2015-au}, ALMA \citep{Tercero2018-as}, and CARMA \citep{Friedel2012-oj}.
The existence of a high-temperature region northeast of the HC of $T_{\rm{rot}}$ in C$_2$H$_5$CN is similar to that in CH$_3$CN observed at IRAM-30\,m \citep{Bell2014-ku}.
This high-temperature region traced by CH$_3$CN is thought to be caused by shock heating from northeast-southwest outflow gas \citep{Plambeck2009-ai}, which originates from active star formation near IRc2 \citep{Bell2014-ku}.
In addition, the distributions of $N$ and $X$ in C$_2$H$_5$CN, which show a peak near the HC, are similar to the distributions of $N$ and $X$ in the map of CH$_3$CN reported by \citet{Bell2014-ku}.
The formation process of C$_2$H$_5$CN on grains is related to CH$_3$CN (e.g., \cite{Garrod2013-yc,Garrod2017-gw,Garrod2022-pm}), and shocks are considered the most plausible mechanism for releasing C$_2$H$_5$CN from dust grain surfaces (e.g, \cite{Friedel2012-oj}).
The existence of a similar high-temperature region in C$_2$H$_5$CN and CH$_3$CN, as well as the similarity of the distributions of column density $N$ and abundance $X$, which has a peak around the HC, reinforce the strength of the physical/chemical relationship between C$_2$H$_5$CN and CH$_3$CN.

\subsubsection{Other high-mass star-forming regions}
COMs have been detected across various regions, and astrochemical studies, including N- and O-bearing COMs toward distant HMSFRs, such as Sgr B2(N) and G10.47+0.03, have been extended (e.g., \cite{McGuire2022,Coletta2020-yo,Bonfand2017-mr,Bonfand2019-fn,Qin2022-hv,Li2020-bz,Qin2015-vo,Belloche2009-mr,Manna2023-ib,Busch2024-bk,Armijos-Abendano2014-ci,Rolffs2011-jf,Mininni2023-vx,Zeng_2018}).
Spatial distribution differences between N- and O-bearing COMs have been observed in some HMSFRs (e.g., \cite{Wyrowski1999-le,Kalenskii2010-pf,Remijan2004-tw,Zernickel2012-dm,Allen2017-mh,Walt2021-dw,Peng2022-ie,Mininni2023-vx}).

Figure \ref{fig:column_trot} shows the differences in the distributions and abundances of C$_2$H$_5$CN and CH$_3$OCHO in the Orion KL region.
The difference in spatial distributions between C$_2$H$_5$CN and CH$_3$OCHO was more pronounced when the spatial resolution was higher ($\sim$2--6$\arcsec$) as shown in previous interferometric observations \citep{Feng2015-au, Friedel2012-oj, Tercero2018-as}.
These results contrast with previous 2\,mm band observations using the IRAM-30\,m telescope, which revealed that $X$(C$_2$H$_5$CN) and $X$(CH$_3$OCHO) are strongly correlated for eight HMSFRs (18089-1732, G31.41+0.03, G24.78+0.08, W3(OH), G14.33-0.65, W51, G10.47+0.03, and G29.96-0.02) at distances of 2.0--8.9\,kpc \citep{Coletta2020-yo}.
The beam size in the half power beam width (HPBW, $\theta_{\rm{beam}}$) of the IRAM-30\,m in the 2\,mm band was approximately 16--18$\arcsec$ \citep{Coletta2020-yo}.
This beam size corresponds to the spatial scale of the angular resolution ($S_{\rm{beam}}$ [pc]) of approximately 0.16--0.73\,pc at distances of 2.0--8.9\,kpc, and it is an order of magnitude larger than the 0.02\,pc in this study.
The beam size that \citet{Coletta2020-yo} (16--18$\arcsec$) used to calculate the column densities is smaller than that of the observation ($\theta_{\rm{H}_2}$=20--60$\arcsec$) used to calculate $N(\rm{H}_2)$ for calculating $X$.
Therefore, \citet{Coletta2020-yo} rescaled the column densities by $(\theta_{\rm{beam}}/\theta_{\rm{H}_2})^2$ to match the spatial scales of $N$ and $N(\rm{H}_2)$ when $X$ was derived.
Consequently, the apparent $S_{\rm{beam}}$ of $X$(C$_2$H$_5$CN) and $X$(CH$_3$OCHO) became even larger.
We focused on $S_{\rm{beam}}$ to understand the significance of the correlation between $X$(C$_2$H$_5$CN) and $X$(CH$_3$OCHO).
The good correlation between $X$(C$_2$H$_5$CN) and $X$(CH$_3$OCHO), despite the existence of regions with different abundances in the same source, such as Orion-KL, may be due to the observational resolution.

HMSFRs evolve from high-mass starless cores (HMSCs) to high-mass protostellar objects (HMPOs).
As protostars form, these evolve into hot molecular cores (HMCs).
Subsequently, these protostars heat the surroundings, forming hypercompact H\,\emissiontype{II} regions (HCHIIs).
Further evolution results in classification into ultracompact H\,\emissiontype{II} regions (UCHIIs) and eventually H\,\emissiontype{II} regions (e.g., \cite{Van_Dishoeck1998-kz,Tan2014-oz,yamamoto2016introduction,Coletta2020-yo}).
In \citet{Coletta2020-yo}, the phase between HMPOs and UCHIIs, encompassing HCHIIs and phases containing high-mass sources, is termed the intermediate (INT) phase. 
Eight sources studied by \citet{Coletta2020-yo}, 18089-1732(INT), G31.41+0.03 (INT), G24.78+0.08 (INT), W3(OH) (UCHII), G14.33-0.65 (UCHII), W51 (UCHII), G10.47+0.03 (UCHII), and G29.96-0.02 (UCHII), were selected from previous studies \citep{Fontani2011-eh,Fontani2014-eg,Fontani2015-pu,Fontani2015-sv,Fontani2016-tj,Fontani2018-rb,Fontani2019-eb,Colzi2018-cv,Colzi2018-nm,Mininni2018-po}.

A similar linear scale $S_{\rm{beam}}$ with Orion-KL in this study ($S_{\rm{beam}}$ $\approx$ 0.02\,pc) is expected for a target at a distance of several kpcs when it is observed with a spatial resolution of beam sizes of $\sim$1--2$\arcsec$ using interferometers.
In this case, a direct comparison is allowed between Orion KL and more distant sources, which are several kpcs away.
Therefore, we selected Sgr B2 \citep{Bonfand2017-mr,Bonfand2019-fn, Busch2024-bk,Sanchez-Monge2017-uo,Meng2022-df}, which was observed by interferometers ($\theta_{\rm{beam}} \approx$ 1--2$\arcsec$) and has a similar $S_{\rm{beam}}$ to Orion-KL in this study, as a comparison target to compare $X$(C$_2$H$_5$CN) and $X$(CH$_3$OCHO) with the eight sources with $S_{\rm{beam}} > 0.1$ from \citet{Coletta2020-yo}.
The information for each source is presented in table \ref{tab:result_data}.
The classification of the evolutionary stage was based on \citet{Coletta2020-yo, Gerner2014-gl,Gerner2015-vv,Bonfand2017-mr}.
$\theta$$_{{\rm beam}}$ and $\theta$$_{{\rm {H_2}}}$ represent the beam sizes of the telescopes.
$S_{\rm{beam}}$ denotes the spatial scale of the angular resolution calculated from $\theta$$_{{\rm beam}}$ and the distance.

\begin{table*}[h]

  \tbl{High-mass star-forming regions in each observation.}
{
  {\renewcommand\arraystretch{1.2}
  \begin{tabular}{c@{\hspace{2mm}}c@{\hspace{2mm}}c@{\hspace{2mm}}c@{\hspace{2mm}}c@{\hspace{2mm}}c@{\hspace{2mm}}c@{\hspace{2mm}}c@{\hspace{2mm}}c}
      \hline
      Source &$\alpha_{\rm{J2000}}$&$\delta_{\rm{J2000}}$& distance&$\theta$$_{{\rm beam}}$&$S_{\rm{beam}}$&$\theta$$_{{\rm {H_2}}}$&evolutionary stage&References\footnotemark[$\dagger$]\\ 
      [-0.5mm]&[h:m:s]&[$\arcdeg:\arcmin:\arcsec$]&[kpc]&[\arcsec]&[pc]&[\arcsec]&\\
      \hline
      18089-1732&18:11:51.4&-17:31:28&3.6&16-18\footnotemark[(a)]&0.30&28\footnotemark[(g)]&INT\footnotemark[(a)]&(a,b,c,d,e,f,g,h,i,j)\\
      G31.41+0.31&18:47:34.2&-01:12:45&3.8&16-18\footnotemark[(a)]&0.31&36.6\footnotemark[(a)]&INT\footnotemark[(a)]&(a,j,k)\\
      G24.78+0.08&18:36:12.6&-07:12:11&7.7&16-18\footnotemark[(a)]&0.63&36.6\footnotemark[(a)]&INT\footnotemark[(a)]&(a,j,k)\\
      W3(OH)&02:27:04.7&+61:52:25&2.0&16-18\footnotemark[(a)]&0.16&23\footnotemark[(l)]&UCHII\footnotemark[(a)]&(a,j)\\
      G14.33-0.65&18:18:54.8&-16:47:53&2.6&16-18\footnotemark[(a)]&0.21&36.6\footnotemark[(a)]&UCHII\footnotemark[(a)]&(a,j,k)\\
      W51&19:23:43.9&+14:30:32&5.4&16-18\footnotemark[(a)]&0.45&19\footnotemark[(l)]&UCHII\footnotemark[(a)]&(a,j)\\
      G10.47+0.03\footnotemark[(a)]&18:08:38.0&-19:51:50&5.8&16-18\footnotemark[(a)]&0.48&59\footnotemark[(m)]&UCHII\footnotemark[(a)]&(a,j)\\
      G29.96-0.02&18:46:03.0&-02:39:22&8.9&16-18\footnotemark[(a)]&0.73&59\footnotemark[(m)]&UCHII\footnotemark[(a)]&(a,j)\\
      \hline
      Orion-KL(HC)\footnotemark[$*$]&05:35:14.6&-5:22:31.0&0.418&12&0.02&12\footnotemark[(o)]&HMC\footnotemark[(p,q)]&(this work,o,p,q,r)\\
      Orion-KL(CR)\footnotemark[$*$]&05:35:14.0&-5:22:36.9&0.418&12&0.02&12\footnotemark[(o)]&HMC\footnotemark[(p,q)]&(this work,o,p,q,r)\\
      Sgr B2(N1)&17:47:19.87&-28:22:16.0&8.34&0.3-0.75\footnotemark[(t)]&0.01-0.03&0.4-5\footnotemark[(s)]&UCHII\footnotemark[(u)]&(s,t,u)\\
      Sgr B2(N2)&17:47:19.9&-28:22:13.4&8.34&1.6\footnotemark[(u,v)]&0.06&0.4-1.2\footnotemark[(u,v)]&-&(u,v)\\
      Sgr B2(N3)&17:47:19.2&-28:22:14.9&8.34&1.6\footnotemark[(u,v)]&0.06&0.4-1.2\footnotemark[(u,v)]&HMC\footnotemark[(u)]&(u,v)\\
      Sgr B2(N4)&17:47:19.5&-28:22:32.4&8.34&1.6\footnotemark[(u,v)]&0.06&0.4-1.2\footnotemark[(u,v)]&HMC\footnotemark[(u)]&(u,v)\\
      Sgr B2(N5)&17:47:20.0&-28:22:41.3&8.34&1.6\footnotemark[(u,v)]&0.06&0.4-1.2\footnotemark[(u,v)]&UCHII\footnotemark[(u)]&(u,v)\\
      \hline
    \end{tabular}}}\label{tab:result_data}
\begin{tabnote}
\footnotemark[$*$] HC and CR correspond to the hot core and compact ridge, respectively.
\footnotemark[$\dagger$] References.
\footnotemark[(a)]  \citep{Coletta2020-yo}
\footnotemark[(b)]  \citep{Fontani2011-eh}
\footnotemark[(c)]  \citep{Fontani2014-eg}
\footnotemark[(d)]  \citep{Fontani2015-pu}
\footnotemark[(e)]  \citep{Fontani2015-sv}
\footnotemark[(f)]  \citep{Fontani2016-tj}
\footnotemark[(g)]  \citep{Fontani2018-rb}
\footnotemark[(h)]  \citep{Colzi2018-nm}
\footnotemark[(i)]  \citep{Mininni2018-po}
\footnotemark[(j)]  \citep{Fontani2019-eb}
\footnotemark[(k)]  \citep{Colzi2018-cv}
\footnotemark[(l)]  \citep{Rivilla2016-ts}
\footnotemark[(m)]  \citep{Liu2010-ur}
\footnotemark[(n)]  \citep{Rolffs2011-jf}
\footnotemark[(o)]  \citep{Schuller2021-ci}, We adjusted the beam size of from 8$\arcsec$ to 12$\arcsec$ using CASA.
\footnotemark[(p)]  \citep{Gerner2014-gl}
\footnotemark[(q)]  \citep{Gerner2015-vv}
\footnotemark[(r)]  \citep{Taquet2015-br}
\footnotemark[(s)]  \citep{Sanchez-Monge2017-uo}
\footnotemark[(t)]  \citep{Busch2024-bk}
\footnotemark[(u)]  \citep{Bonfand2017-mr}
\footnotemark[(v)]  \citep{Bonfand2019-fn}
\end{tabnote}
\end{table*}

\begin{figure*}

  \includegraphics[width=17cm]{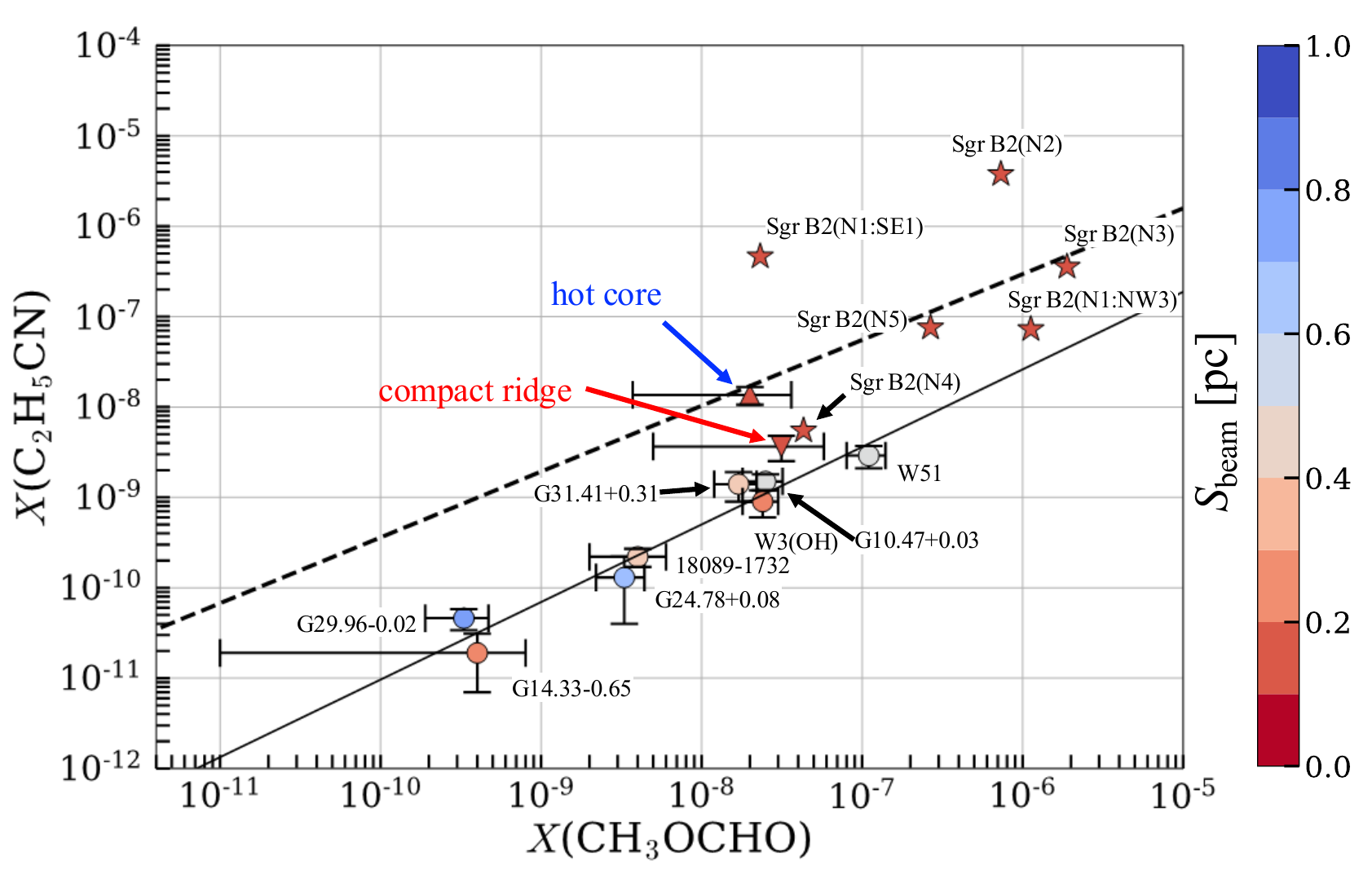} 
\caption{Comparison of $X$(C$_2$H$_5$CN) and $X$(CH$_3$OCHO) among other HMSFRs and the Orion-KL HC, CR.
The triangle and inverted triangle represent $X$ of the HC and CR in this study, respectively (see table \ref{tab:result_all}). 
The circles show the eight sources observed by \citet{Coletta2020-yo}.
The stars denote observations of Sgr B2(N1:SE1, N1:NW3, N2, N3, N4, N5) from \citet{Busch2024-bk,Bonfand2017-mr,Bonfand2019-fn}. 
The solid black line represents the correlation from \citet{Coletta2020-yo}, while the dashed black line represents the fitting results using the plots of the HC, CR (not assuming source size 10$\arcsec$) in this study, and Sgr B2(N) observed with a resolution $S_{\rm{beam}}$ below 0.1\,pc \citep{Bonfand2017-mr,Bonfand2019-fn, Busch2024-bk,Sanchez-Monge2017-uo}.
The color scale corresponds to $S_{\rm{beam}}$ [pc].
{Alt text: The scatter plots and fitting results. The x axis is the relative abundance with respect to $\rm {H_2}$ of CH$_3$OCHO (methyl formate), and the y axis is the the relative abundance with respect to $\rm {H_2}$ of C$_2$H$_5$CN (ethyl cyanide, propionitrile).}
}\label{fig:coletta}
\end{figure*}

Figure \ref{fig:coletta} shows $X$(C$_2$H$_5$CN) and $X$(CH$_3$OCHO) of the sources listed in table \ref{tab:result_data}.
The color scale corresponds to $S_{\rm{beam}}$ [pc].
The results of the eight sources from \citet{Coletta2020-yo} (18089-1732, G31.41+0.03, G24.78+0.08, W3(OH), G14.33-0.65, W51, G10.47+0.03, and G29.96-0.02) are based solely on the transition of CH$_3$OCHO ($v=0$).
Therefore, the plots for the HC and CR in Orion-KL utilized values of $X$(CH$_3$OCHO ($v=0$)) from table \ref{tab:result_all}.
The solid black line represents the fitting and correlation for the eight sources with $S_{\rm{beam}} >$ 0.1 pc (18089-1732, G31.41+0.03, G24.78+0.08, W3(OH), G14.33-0.65, W51, G10.47+0.03, and G29.96-0.02) \citep{Coletta2020-yo}.
The spatial resolutions of the observations for Orion-KL and Sgr B2 ($S_{\rm{beam}} <$ 0.1 pc) are similar to each other but smaller than those of the other eight sources ($S_{\rm{beam}} >$ 0.1 pc).
Therefore, we applied an independent fitting for Orion-KL and Sgr B2. 
The black dashed line in figure \ref{fig:coletta} is the result of a linear fit for Orion-KL (HC and CR, not assuming source size) and Sgr B2 (N1:SE1, N1:NW3, N2, N3, N4, N5) simultaneously.
Sgr B2(N1:SE1) and Sgr B2(N1:NW3) denote observations performed at approximately 0.1\,pc southeast and northwest of Sgr B2 (N1), respectively \citep{Busch2024-bk}.
The results for these sources with $S_{\rm{beam}}$ smaller than 0.1\,pc show a large dispersion, with a correlation coefficient of r = 0.57.
The correlation coefficient remains similar (r = 0.54) even when $X^{\prime}$ (assuming a source size of 10$\arcsec$) is used in place of $X$.

This correlation coefficient is lower than the value reported by \citet{Coletta2020-yo} (r = 0.92). 
Additionally, the solid line \citep{Coletta2020-yo} shifts below the dashed line (in this study) and is closer to the values of $X$(C$_2$H$_5$CN) and $X$(CH$_3$OCHO) in the CR (inverted triangle) than in the HC (triangle) in Orion-KL, as shown in figure \ref{fig:coletta}. This indicates that the eight sources with $S_{\rm{beam}}$ $>$ 0.1 pc tend to have relatively high $X$(CH$_3$OCHO) values.
These considerations highlight the dependence of $X$ on $S_{\rm{beam}}$, indicating that coarser resolution observations ($S_{\rm{beam}}$ $>$ 0.1\,pc) may not adequately resolve regions with distinct chemical properties.
In this case, CH$_3$OCHO (O-bearing COMs), which is more widely distributed and highly abundant, may be predominantly observed, compared with C$_2$H$_5$CN (N-bearing COMs), which is relatively compactly distributed, as shown in figure \ref{fig:column_trot}.

The integrated intensity maps obtained from interferometric observations of G10.47+0.03 \citep{Rolffs2011-jf} and W3(OH) \citep{Wyrowski1999-le}, which are included among the eight sources, support this suggestion.
The integrated intensity distributions of C$_2$H$_5$CN and CH$_3$OCHO obtained with a beam size of approximately 4$\arcsec$$\times$2$\arcsec$ ($S_{\rm{beam}}$ $\approx$ 0.11$\times$0.06\,pc) using SMA are similar \citep{Rolffs2011-jf}.
However, the map observed with a smaller beam size (approximately 0.5$\arcsec$$\times$0.3$\arcsec$, corresponding to $S_{\rm{beam}}$ $\approx$ 0.01$\times$0.008\,pc) shows that the intensity distribution differs, with CH$_3$OCHO having a peak on the west side and C$_2$H$_5$CN having a peak on the east side.
In addition, CH$_3$OCHO is more widely distributed than C$_2$H$_5$CN \citep{Rolffs2011-jf}.
The differences in the distributions of C$_2$H$_5$CN and CH$_3$OCHO and the extended distribution of CH$_3$OCHO at $S_{\rm{beam}} < 0.1$ are similar in this study.
Additionally, these features are also observed in the integrated intensity map of W3(OH) from observations with the Plateau de Bure Interferometer (PdBI) ($\theta$$_{{\rm beam}}$ $\approx$ 0.8$\arcsec$$\times$0.6$\arcsec$, corresponding to $S_{\rm{beam}}$ $\approx$ 0.008$\times$0.006\,pc) \citep{Wyrowski1999-le}.
Therefore, although a precise analysis of the abundances of C$_2$H$_5$CN and CH$_3$OCHO in each region is necessary, these detailed maps observed with the SMA and PdBI support the above suggestions.
The comparison of $X$(C$_2$H$_5$CN) and $X$(CH$_3$OCHO) focusing on $S_{\rm{beam}}$ in this study confirms that CH$_3$OCHO may be predominantly observed in coarser resolution observations ($S_{\rm{beam}} > 0.1$), as suggested by interferometric results.
The good correlation between $X$(C$_2$H$_5$CN) and $X$(CH$_3$OCHO) may be affected by the extent to which the spatial scale of such angular resolution includes each molecular distribution with different spatial coverages due to differences in the physical and chemical environments.
To gain a more detailed understanding of the relationship between $X$(C$_2$H$_5$CN) and $X$(CH$_3$OCHO), it is necessary to observe the differences in their abundances and spatial distributions across sources by observing distant astronomical objects with a spatial resolution sufficient to resolve individual sources.

In the Sgr B2(N) region, where $X$(C$_2$H$_5$CN) appeared higher (positioned upward in figure \ref{fig:coletta}), some prebiotic molecules, such as H$_2$C$_2$N$_2$ (Z-(E-)HNCHCN) and H$_2$C$_4$N$_2$ (NH$_2$CH$_2$CN), have also been observed \citep{Zaleski_2013,Belloche2008-yn,Li2020-bz}.
G10.47+0.03 is expected to contain a region with a high $X$(C$_2$H$_5$CN) \citep{Rolffs2011-jf}, similar to the Sgr B2(N) and Orion-KL HC regions.
NH$_2$CH$_2$CN was also detected in G10.47+0.03 \citep{Manna_2022}.
In addition, chemical model calculations suggest that the Orion-KL region is a probable source for detecting Z-(E-)HNCHCN \citep{Zhang2020-rn}.
The angular resolution of B4R/LMT-50\,m enables us to resolve a 0.02\,pc structure at the distance of Orion-KL, and owing to the wide bandwidth, many molecular emission lines can be observed simultaneously.
These capabilities in nearby star-forming regions such as Orion-KL potentially complement detailed high angular resolution studies on distant sources using interferometers with similar $S_{\rm{beam}}$ at several kpcs and arcseconds.
The B4R/LMT-50\,m capabilities could help us understand the formation of prebiotic molecules and the observed differences in $X$(C$_2$H$_5$CN) and $X$(CH$_3$OCHO) across various sources.

\subsection{Future upgrades and prospects of B4R system}
The B4R and observation system are being developed \citep{kawabe_submitted}.
In May 2024, remote tuning of the LO power at each frequency setting to minimize the receiver noise temperature and monitoring of key parameters (e.g., the temperature of the 4K stage and vacuum gauge output) were integrated and are now ready for future use.
The expansion of the IF bandwidth by adding two XFFTS boards has already been prepared and is expected to be integrated in the near future.
Furthermore, efforts are being made to improve the observational efficiency and SNR through the introduction of data science methods such as GoDec and frequency-modulated local oscillators (FMLO) \citep{Taniguchi2021-ow,Taniguchi2020-mt,Zhou2011-sy}.

These updates are expected to contribute not only to the detection of faint prebiotic molecules and COMs but also to the detection of atomic line emissions from high-z objects.
In addition, the B4R and LMT, a single-dish telescope with a wide bandwidth, high frequency resolution, and high spatial resolution in the 2\,mm band, have the potential to make significant contributions to unbiased survey observations.
Observations of interstellar molecules, including COMs associated with protostars, have also been conducted in nearby low-mass star-forming regions.
Detailed observations with ALMA have enabled studies that approach the issues of material transport and angular momentum transfer from protostellar envelopes to disks (e.g., \cite{Sakai2014-jz}).
The factors behind the chemical diversity observed in astronomical objects during the course of star and planet formation are not yet fully understood, and it is essential to conduct survey observations of the environments surrounding protoplanetary disks using various tracer molecules.
In addition, AGB stars, for which a large number of samples have been obtained through survey observations with infrared space telescopes (e.g., Spitzer; \cite{Blum_2006}), are known to be associated with dense molecular gas. 
Follow-up millimeter-wave survey observations targeting such dense molecular gas are also crucial for deepening our understanding of the chemical and physical environments of protoplanetary disks of high- and low-mass stars that develop from the scale of molecular cloud cores and filaments, which include environments where AGB stars disperse carbon and nitrogen as stellar winds.

\section{Conclusion}
A 2\,mm band SIS receiver, the B4R, was installed on the LMT-50\,m atop the Sierra Negra Mountain in Mexico, at an altitude of 4600\,m.
This study analyzed the results of test observations conducted in November 2019 using the OTF observation technique, with a spatial resolution of 11--12$\arcsec$ and an observed area of 5$\arcmin\times$5$\arcmin$.
The main findings are summarized as follows:

\begin{enumerate}

\item Emission lines from H35$\alpha$, H51$\gamma$ and 29 molecular species, including isotopologues, deuterated molecules, and vibrational states, were identified with a total bandwidth of 10\,GHz using XCLASS \citep{xclass2017}. A total of 337 spectra were identified, and 49 spectra were labeled as unidentified.
These detections and identifications were consistent with previous studies and confirmed the observational capabilities of spectral scans achieved by B4R/LMT-50\,m.
Concurrently, the difference in spectral intensity between N- and O-bearing COMs in the Orion-KL HC and CR regions has demonstrated the competitive spatial resolution achieved by the B4R optics within the LMT facility.
In addition to the IRAM-30\,m with the 2\,mm band receiver Eight MIxer Receiver (EMIR) E150, the B4R/LMT-50\,m offers wide bandwidth, high frequency resolution, and higher spatial resolution observations in the 2\,mm band in the Northern Hemisphere.

\item We computed the maps of column density, rotational temperature, and relative abundance with respect to H$_2$, focusing on representative N- and O-bearing COMs, namely, C$_2$H$_5$CN and CH$_3$OCHO, by a rotation diagram using a two-dimensional method based on OTF observations using the B4R/LMT-50\,m. The spatial distributions and values of the column density and rotational temperature were consistent with previous studies.

\item We compared $X$(C$_2$H$_5$CN) and $X$(CH$_3$OCHO) ($v=0$) in Orion-KL, our targeted observation region, with previously reported HMSFRs. 
The abundance ($X$) was found to be dependent on the spatial scale of the source and, therefore, very sensitive to the angular resolution ($S_{\rm{beam}}$). 
This was confirmed by comparing CH$_3$OCHO (O-bearing COMs), which are more widely distributed and highly abundant, and C$_2$H$_5$CN (N-bearing COMs), which are relatively compactly distributed and could dominate central core observations.
This suggests that the good correlation between $X$(C$_2$H$_5$CN) and $X$(CH$_3$OCHO) reported in a previous study \citep{Coletta2020-yo} may be affected by the extent to which the spatial scale of such angular resolution includes each molecular distribution with different spatial coverages due to differences in the physical and chemical environments.

\par

\end{enumerate}

\begin{ack}
This work was financially supported by JSPS KAKENHI grant Numbers, JP15H02073, 17H06130 (RK, YT, KK), JP19K14754 (TT), 22H04939 (TS, KT, AT, YT, KK), JP19K21885, and JP25K01060 (HM).
TY is supported by JSPS KAKENHI grant No. JP22J22889 and JP22KJ2625.
KT is supported by JSPS KAKENHI Grant Numbers 20K14523 and 24K17096.
TT is supported by MEXT Leading Initiative for Excellent Young Researchers Grant Number JPMXS0320200188.
This paper makes use of data taken by the Large Millimeter Telescope Alfonso Serrano (LMT) in Mexico.
The LMT project is a joint effort of the Instituto Nacional de Astr\'{o}fisica, \'{O}ptica, y Electr\'{o}nica (INAOE) and the University of Massachusetts at Amherst (UMASS).
We also appreciate the support of the technical staff and the support scientists of the LMT during the commissioning campaign of the B4R.
We would like to thank Editage (www.editage.jp) for English language editing.

\end{ack}



\appendix 
\section{Details of observed spectra}\label{ap_id}
\begin{figure*}
  \includegraphics[width=17cm]{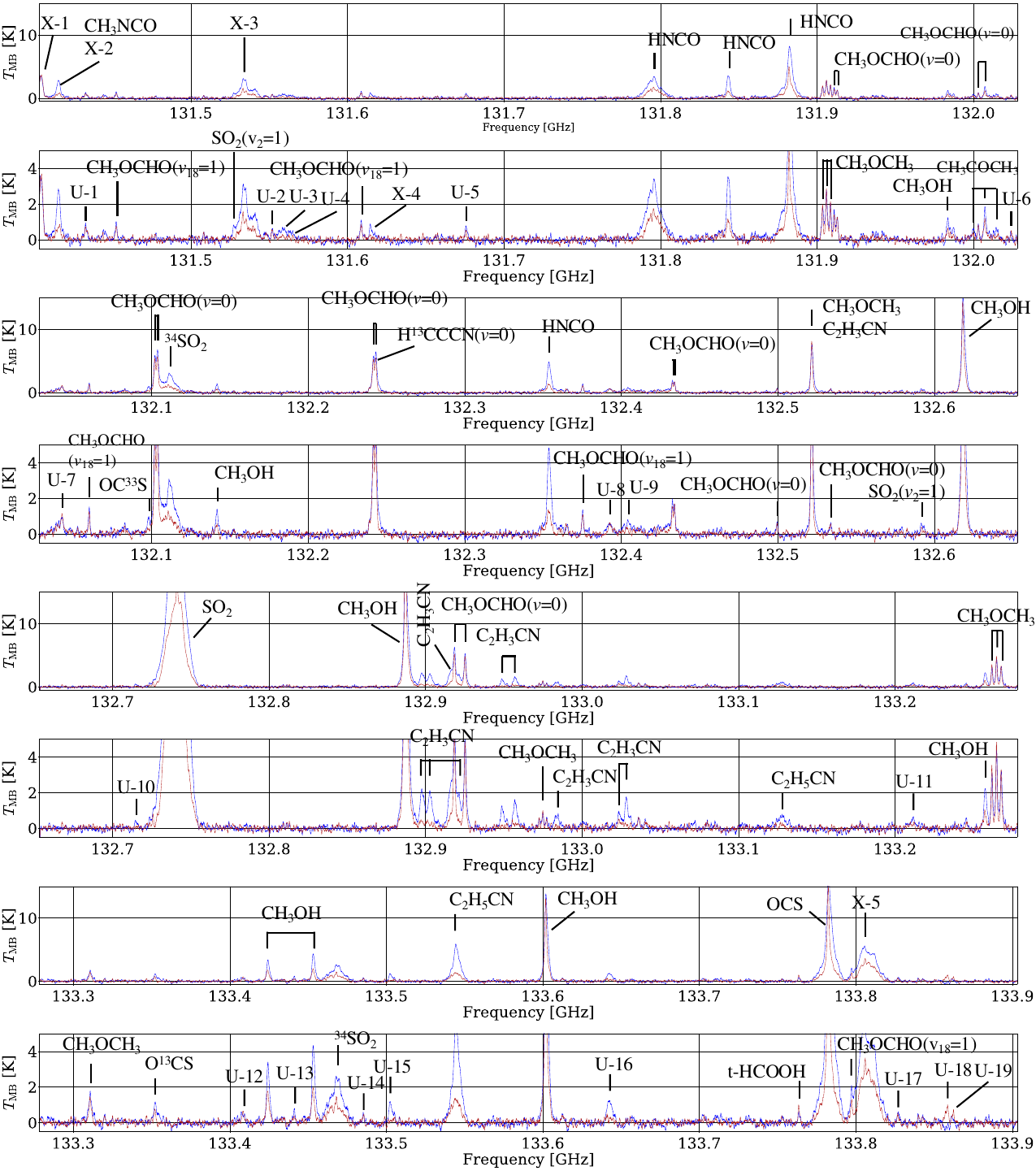} 
\caption{
Observed spectra and identified lines in the range of 131.4--133.9\,GHz are shown in two different vertical scales with 10--channel binning.
The blue lines show the HC, and the red lines show the CR.
The leakage from the second down-conversion or aliasing from the ADC and image sideband leakage are labeled with "X," while unidentified lines are labeled with "U."
{Alt text: The molecular emission line spectra of the eight panels. The x axis is the frequency at gigahertz, and the y axis is the intensity at kelvin on the $T_{\rm{MB}}$ scale.}
}\label{fig:all_lines_1}

\end{figure*}

\begin{figure*}
  \includegraphics[width=17cm]{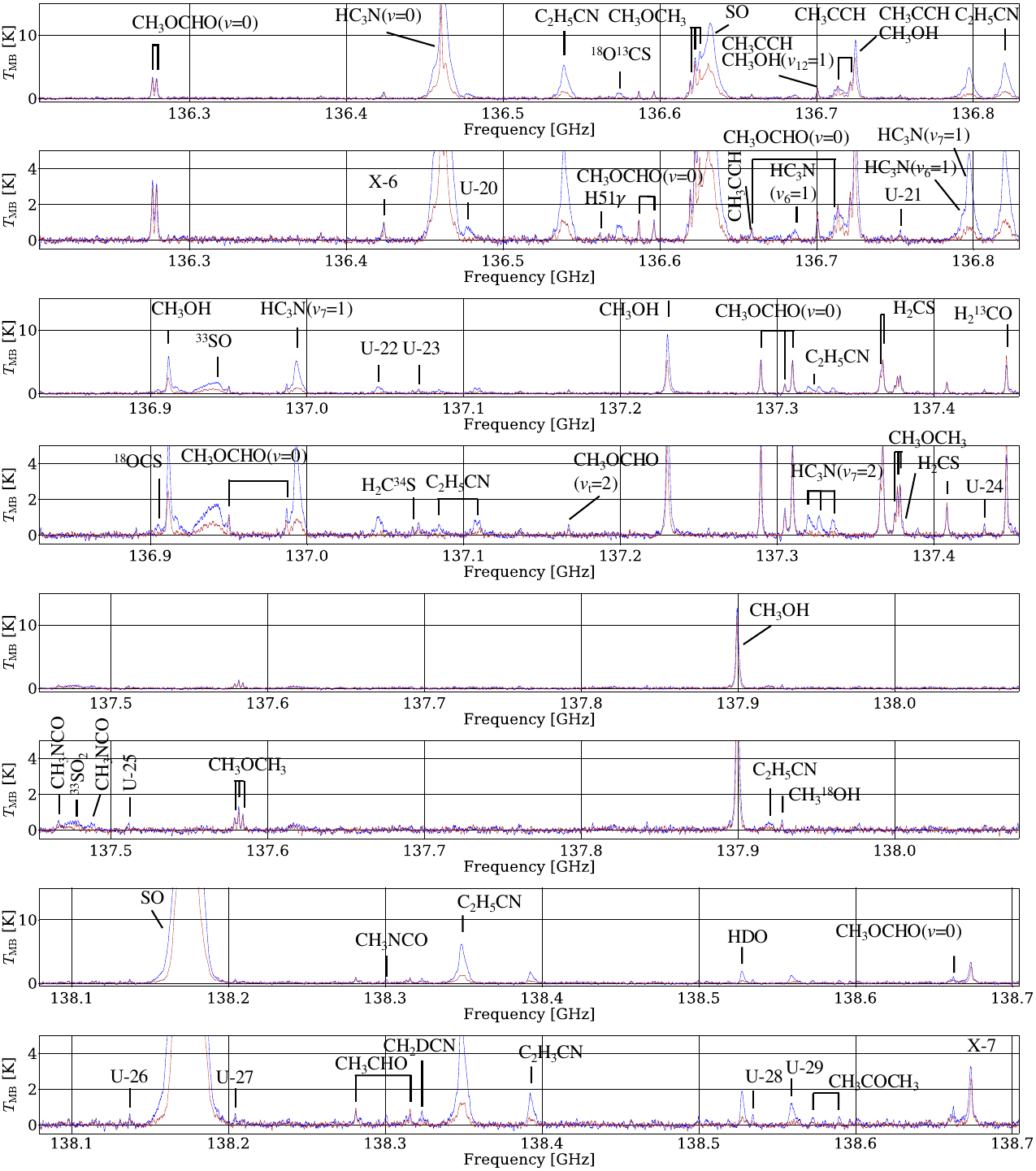} 
\caption{
Observed spectra and identified lines in the range of 136.2--138.7\,GHz are shown in two different vertical scales with 10--channel binning.
The blue lines show the HC, and the red lines show the CR. 
The leakage from the second down-conversion or aliasing from the ADC and image sideband leakage are labeled with "X," while unidentified lines are labeled with "U."
{Alt text: The molecular emission line spectra of the eight panels. The x axis is the frequency at gigahertz, and the y axis is the intensity at kelvin on the $T_{\rm{MB}}$ scale.}

}\label{fig:all_lines_2}

\end{figure*}

\begin{figure*}
  \includegraphics[width=17cm]{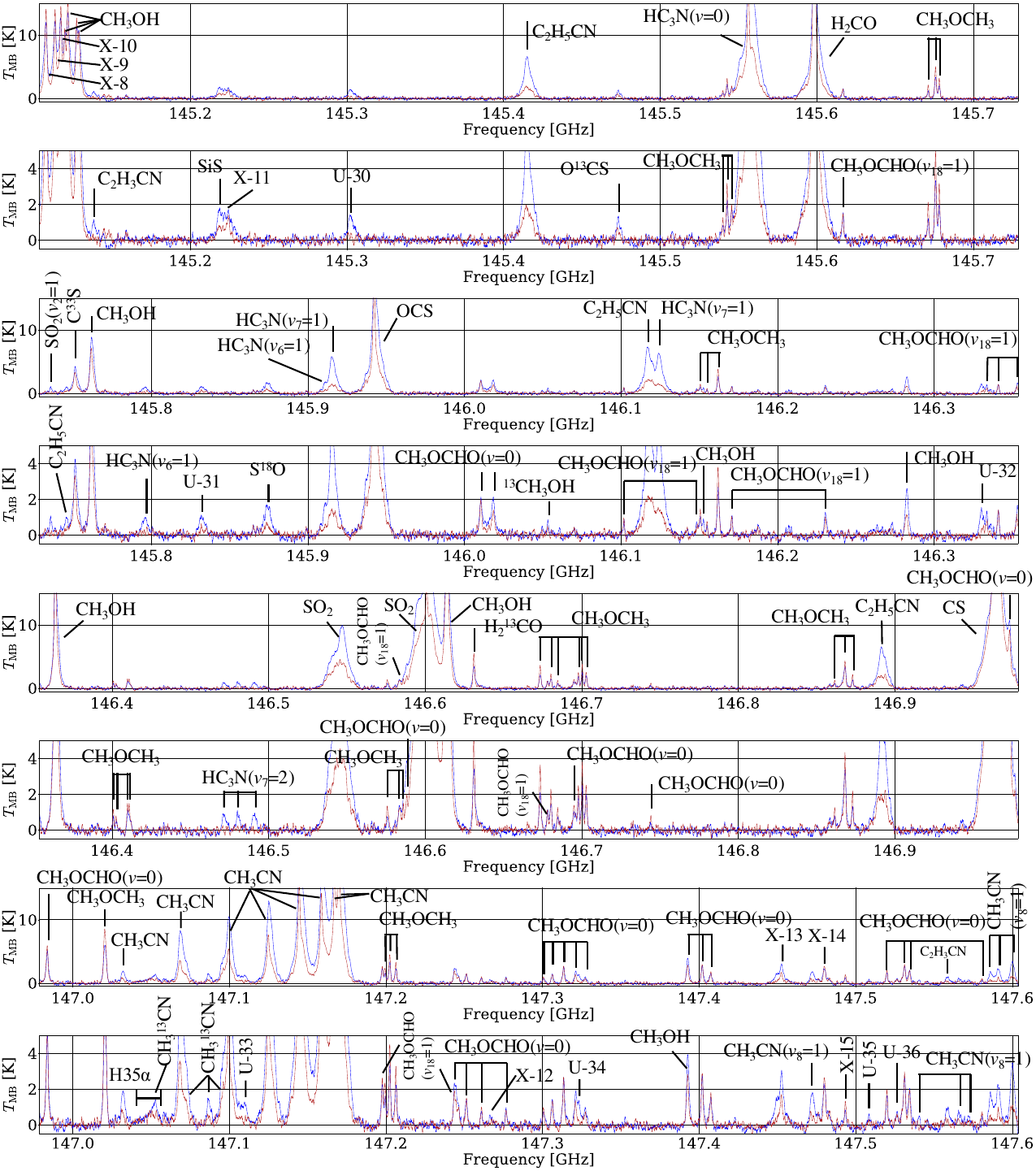} 
\caption{
Observed spectra and identified lines in the range of 145.1--147.6\,GHz are shown in two different vertical scales with 10--channel binning.
The blue lines show the HC, and the red lines show the CR.
The leakage from the second down-conversion or aliasing from the ADC and image sideband leakage are labeled with "X," while unidentified lines are labeled with "U."
{Alt text: The molecular emission line spectra of the eight panels. The x axis is the frequency at gigahertz, and the y axis is the intensity at kelvin on the $T_{\rm{MB}}$ scale.}
}\label{fig:all_lines_3}
\end{figure*}

\begin{figure*}
  \includegraphics[width=17cm]{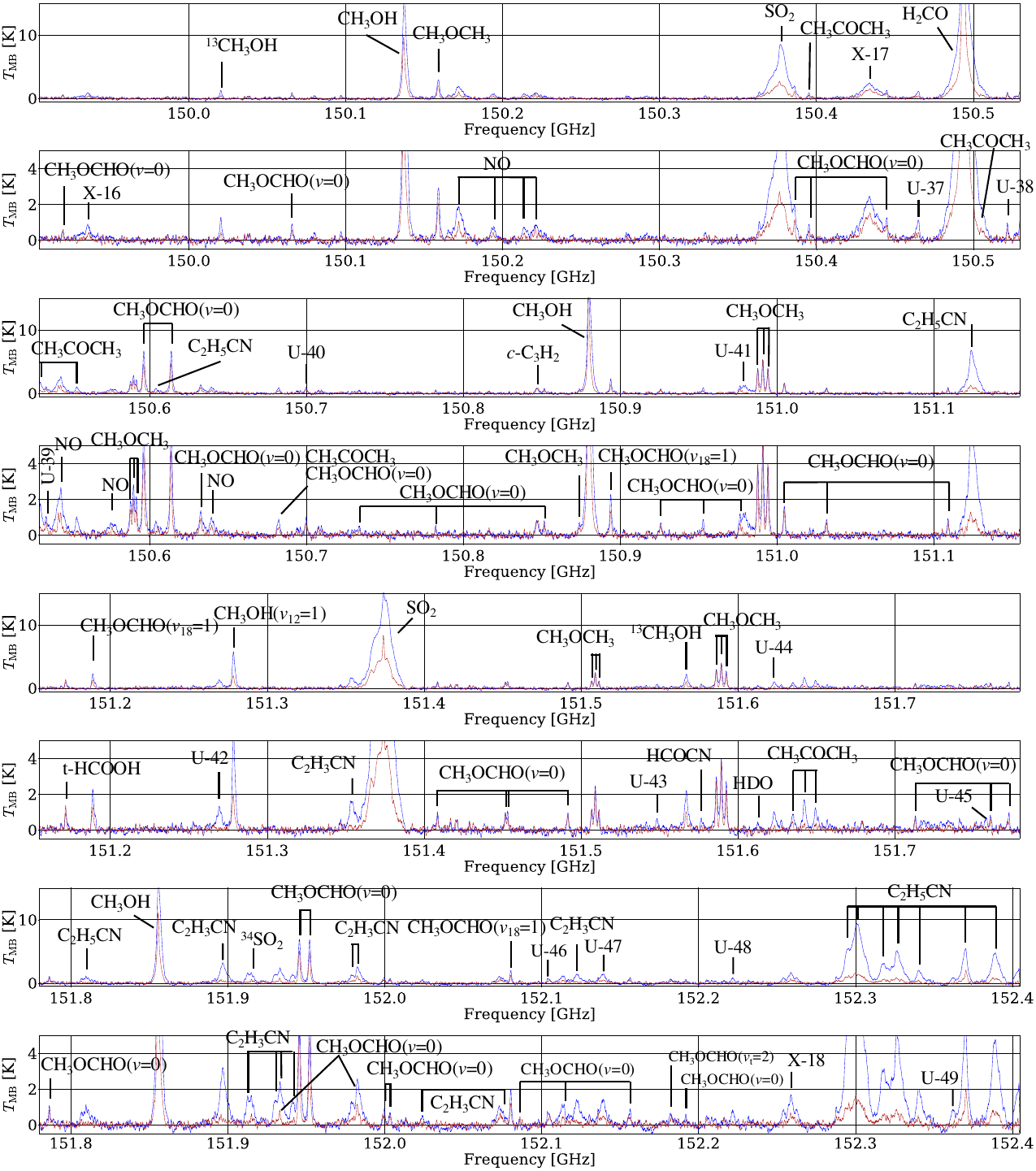} 
\caption{
Observed spectra and identified lines in the range of 149.9--152.4\,GHz are shown in two different vertical scales with 10--channel binning.
The blue lines show the HC, and the red lines show the CR. 
The leakage from the second down-conversion or aliasing from the ADC and image sideband leakage are labeled with "X," while unidentified lines are labeled with "U."
{Alt text: The molecular emission line spectra of the eight panels. The x axis is the frequency at gigahertz, and the y axis is the intensity at kelvin on the $T_{\rm{MB}}$ scale.}

}\label{fig:all_lines_4}

\end{figure*}

This section provides additional details of the observed spectra, as Section \ref{lineid} outlines.
The spectra within the frequency bands of 131.4--133.9\,GHz, 136.2--138.7\,GHz, 145.1--147.6\,GHz, and 149.9--152.4\,GHz, are shown in figures \ref{fig:all_lines_1}--\ref{fig:all_lines_4}. 
Among the 404 lines detected, molecular names are assigned to 337 of the identified and tentatively identified lines, with the identified lines listed in table \ref{tab:lineid} and the tentatively identified lines listed in table \ref{tab:lineid_tent}.
Tables \ref{tab:lineid} and \ref{tab:lineid_tent} present the observed frequency (Obs. freq.), $V_{\rm{LSR}}$, and $T_{\rm{a}}$ obtained from the peak tops, with values provided for both the HC and CR spectra (HC/CR). 
The line widths were determined by fitting using CASSIS \citep{Vastel2015-rn}. 
When fitting with multiple components was appropriate, the widths of each component were listed together.
The transition information and rest frequency (Rest. freq.) predominantly relied on the JPL values sourced from Splatalogue, whereas data exclusively available in the CDMS was obtained from the CDMS. 
The transition information and rest frequency (Rest. freq.) of recombination lines was obtained from Splatalogue.
The 18 X-lines are indicated as (X-1)--(X-18) in figures \ref{fig:all_lines_1}--\ref{fig:all_lines_4}, with details about the candidate molecules summarized in table \ref{tab:lineid_leak}.
Lines resulting from the second downconversion or aliasing from the ADC and image sideband leakage are labeled as DC, ADC, and Image in table \ref{tab:lineid_leak}, respectively.
The 49 U-lines are indicated by (U-1)--(U-49) in figures \ref{fig:all_lines_1}--\ref{fig:all_lines_4} and are listed in table \ref{tab:lineid_und}.
The observed frequency (Obs. freq.) is obtained from the peak tops, with values provided for both the HC and CR spectra (HC/CR) in tables \ref{tab:lineid_leak} and \ref{tab:lineid_und}.
As with tables \ref{tab:lineid} and \ref{tab:lineid_tent}, the transition information and rest frequency (Rest. freq.) listed in table \ref{tab:lineid_leak} are sourced from the JPL and CDMS catalogues.
The values of $T_{\rm{a}}$ and the line widths in table \ref{tab:lineid_und} were obtained using the same method as described for tables \ref{tab:lineid} and \ref{tab:lineid_tent}.

\begin{longtable}{p{.08\textwidth}p{.17\textwidth}p{.08\textwidth}p{.16\textwidth}p{.05\textwidth}p{.06\textwidth}p{.08\textwidth}p{.15\textwidth}}
  \caption{The detected lines in this observations.}\label{tab:lineid}
  \hline              
  Species& Obs. freq. (HC/CR)& Rest. freq.&Transition&$V_{\rm {LSR}}$&$T_{\rm{a}}$&Width&Comments\\ 
    [-0.5mm]&[GHz]&[GHz]&&[km/s]&[K]&[km/s]&\\
\endfirsthead
  \hline
\endhead
  \hline
\endfoot
  \hline
\endlastfoot
\hline
H35$\alpha$&147.052859/147.052938&147.046848&-&-&-&-&blend CH$_3$C$^{13}$N\\
\hline
H51$\gamma$&136.561813&136.559284&-&-5.6&0.58/-&0.97/-&\\
\hline
NO&150.172063/150.171844&150.17648&\textit{J}=3/2-1/2,&8.8/9.3&1.42/0.88&11.0/4.0&\\
&&&\textit{$\Omega$}=1/2$^+$,\textit{F}=5/2-3/2&&&&\\
&150.194500/150.193891&150.19876&\textit{J}=3/2-1/2,&8.5/9.7&0.92/0.57&7.0/3.0&\\
&&&\textit{$\Omega$}=1/2$^+$,\textit{F}=3/2-1/2&&&&\\
&150.214875/150.214266&150.21873&\textit{J}=3/2-1/2,&7.7/8.9&0.75/0.55&7.0/3.5&blend NO\\
&&&\textit{$\Omega$}=1/2$^+$,\textit{F}=3/2-3/2&&&&\\
&150.220281/150.220750&150.22566&\textit{J}=3/2-1/2,&10.7/9.8&0.90/0.58&12.9/12.6&blend NO\\
&&&\textit{$\Omega$}=1/2$^+$,\textit{F}=1/2-1/2&&&&\\
&150.543563/150.543484&150.54652&\textit{J}=3/2-1/2,&5.9/6.0&2.09/1.37&1.2,11.5/1.8&blend U-39\\
&&&\textit{$\Omega$}=1/2$^-$,\textit{F}=5/2-3/2&&&&\\
&150.575234/150.576297&150.58056&\textit{J}=3/2-1/2,&10.6/8.5&0.57/0.58&12.7/2.4&\\
&&&\textit{$\Omega$}=1/2$^-$,\textit{F}=1/2-1/2&&&&\\
&150.639328/150.639859&150.64434&\textit{J}=3/2-1/2,&10.0/8.9&0.96/0.64&6.3/5.7&\\
&&&\textit{$\Omega$}=1/2$^-$,\textit{F}=3/2-1/2&&&&\\
\hline
CS&146.964281/146.964516&146.969033&3-2&9.7/9.2&28.4/33.5&9.8,29.3,3.3&blend\\
&&&&&&/12.4,30.8,3.3&CH$_3$OCHO($v$=0,\\
&&&&&&&$v_{18}$=1)\\
\hline
C$^{33}$S&145.751375/145.751063&145.7557316&3-2&9.0/9.6&3.25/2.72&9.4,2.2/4.9&CDMS, blend\\
&&&&&&&C$_2$H$_5$CN\\
\hline
SO&136.632313/136.631094&136.634799&5(6)-5(5)&5.5/8.1&8.45/3.95&29.7,9.3&blend CH$_3$OCH$_3$,\\
&&&&&&/13.0,1.4,29.0&CH$_3$CCH\\
&138.175156/138.174844&138.1786&4(3)-3(2)&7.5/8.1&40.8/34.6&-&\\
\hline
$^{33}$SO&136.941172/136.942234&136.934082&4(3)-3(2),&-&-&-&CDMS, blend\\
&&&\textit{F}=5/2-3/2&&&&CH$_3$OCHO($v$=0),\\
&&&&&&&$^{33}$SO\\
&&136.939357&4(3)-3(2),&-&-&-&CDMS, blend\\
&&&\textit{F}=7/2-5/2&&&&$^{33}$SO\\
&&136.943672&4(3)-3(2),&-&-&-&CDMS, blend\\
&&&\textit{F}=9/2-7/2&&&&$^{33}$SO\\
&&136.946194&4(3)-3(2),&-&-&-&CDMS, blend\\
&&&\textit{F}=11/2-9/2&&&&$^{33}$SO\\
\hline
SO$_2$&132.741594/132.741141&132.74486&14(2,12)-14(1,13)&7.4/8.4&19.0/12.1&9.7,30.2,6.0&\\
&&&&&&/5.7,2.4,27.3&\\
&146.547547/146.548531&146.55008&10(4,6)-11(3,9)&5.2/3.2&6.95/2.90&7.7,29.2/25.5&\\
&146.601484/146.600953&146.60552&4(2,2)-4(1,3)&8.3/9.3&14.9/11.4&20.1,8.9,13.5&blend CH$_3$OCHO($v$=0,\\
&&&&&&/5.9,2.0,29.6&$v_{18}$=1),CH$_3$OCH$_3$,\\
&&&&&&&CH$_3$OH\\
&150.377156/150.376547&150.3811&15(5,11)-16(4,12)&7.9/9.1&5.76/1.98&7.3,2.9,26.7&\\
&&&&&&/19.9,1.4,9.7&\\
&151.374313/151.374000&151.37863&2(2,0)-2(1,1)&8.5/9.2&10.8/5.81&14.9,10.5,&blend C$_2$H$_3$CN\\
&&&&&&2.5,20.3&\\
&&&&&&/1.6,22.3,6.8&\\
\hline
SO$_2$&131.527141/131.526375&131.5305100&12(2,10)-12(1,11)&-&-&-&blend X-3\\
($v_2$=1)&132.591516/-&132.594390&10(2,8)-10(1,9)&6.5/-&0.83/-&6.1/-&\\
&145.735500/-&145.7399920&16(2,14)-16(1,15)&9.2/-&1.08/-&3.8/-&\\
\hline
$^{33}$SO$_2$&137.478703&137.4735010&6(2,4)-6(1,5),&-&-&-&blend $^{33}$SO$_2$\\
&&&\textit{F}=11/2-9/2&&&&\\
&&137.4744968&6(2,4)-6(1,5),&-&-&-&blend $^{33}$SO$_2$\\
&&&\textit{F}=13/2-15/2&&&&\\
&&137.4754741&6(2,4)-6(1,5),&-&-&-&blend $^{33}$SO$_2$\\
&&&\textit{F}=9/2-9/2&&&&\\
&&137.4764698&6(2,4)-6(1,5),&-&-&-&blend $^{33}$SO$_2$\\
&&&\textit{F}=15/2-15/2&&&&\\
&&137.4792202&6(2,4)-6(1,5),&-&-&-&blend $^{33}$SO$_2$\\
&&&\textit{F}=13/2-11/2&&&&\\
&&137.4796866&6(2,4)-6(1,5),&-&-&-&blend $^{33}$SO$_2$\\
&&&\textit{F}=11/2-11/2&&&&\\
&&137.4806821&6(2,4)-6(1,5),&-&-&-&blend $^{33}$SO$_2$\\
&&&\textit{F}=13/2-13/2&&&&\\
&&137.4811484&6(2,4)-6(1,5),&-&-&-&blend $^{33}$SO$_2$\\
&&&\textit{F}=11/2-13/2&&&&\\
&&137.4816598&6(2,4)-6(1,5),&-&-&-&blend $^{33}$SO$_2$\\
&&&\textit{F}=9/2-11/2&&&&\\
&&137.4826550&6(2,4)-6(1,5),&-&-&-&blend $^{33}$SO$_2$\\
&&&\textit{F}=15/2-13/2&&&&\\
\hline
$^{34}$SO$_2$&132.111359/132.110750&132.11404&12(1,11)-12(0,12)&6.1/7.5&2.48/1.32&11.0/20.4&blend\\
&&&&&&&CH$_3$OCHO($v$=0)\\
&133.469719&133.47147&5(1,5)-4(0,4)&3.9&2.27/0.79&7.8,28.4&\\
&&&&&&/1.5,27.1&\\
&151.915047/151.914969&151.9175597&4(3,1)-5(2,4)&-&-&-&blend C$_2$H$_3$CN\\
\hline
OCS&133.782547/133.782234&133.7859&11-10&7.5/8.2&13.2/11.6&24.9,8.5,2.8&\\
&&&&&&/22.3,7.0,2.0&\\
&145.942578/145.942266&145.946812&12-11&8.7/9.3&13.6/13.8&15.8,8.8,2.2,&blend CH$_3$OCHO\\
&&&&&&8.8/2.6,19.3,&($v$=0,$v_{18}$=1)\\
&&&&&&2.6&\\
\hline
O$^{13}$CS&133.352984/133.352063&133.355416&11-10&5.5/7.5&0.93/0.83&7.8/2.7&\\
&145.473188/145.472125&145.477197&12-11&8.3/10.5&1.25/0.77&6.4/4.5&\\
\hline
HDO&138.528109/138.527734&138.53057&6(1,6)-5(2,3)&5.3/6.1&1.39/0.77&5.2/4.7&\\
&151.612969/-&151.61619&7(3,4)-7(3,5)&6.4/-&0.67/-&3.1/-&\\
\hline
HNCO&131.795719/131.795797&131.7991113&6(3,4)-5(3,3),F=5-5&7.7/7.5&2.82/1.56&4.2,30.0,6.3&\\
&&&&&&/22.5&\\
&&131.7991114&6(3,3)-5(3,2),\textit{F}=5-5&7.7/7.5&&&\\
&&131.7991801&6(3,4)-5(3,3),\textit{F}=5-6&7.9/7.7&&&\\
&&131.7991802&6(3,3)-5(3,2),\textit{F}=5-6&7.9/7.7&&&\\
&&131.7991942&6(3,4)-5(3,3),\textit{F}=5-4&7.9/7.7&&&\\
&&131.7991943&6(3,3)-5(3,2),\textit{F}=5-4&7.9/7.7&&&\\
&&131.7992221&6(3,4)-5(3,3),\textit{F}=7-6&8.0/7.8&&&\\
&&131.7992222&6(3,3)-5(3,2),\textit{F}=7-6&8.0/7.8&&&\\
&&131.799402&6(3,4)-5(3,3),\textit{F}=6-5&8.4/8.2&&&\\
&&131.7994021&6(3,3)-5(3,2),\textit{F}=6-5&8.4/8.2&&&\\
&&131.7994709&6(3,4)-5(3,3),\textit{F}=6-6&8.5/8.4&&&\\
&&131.799471&6(3,3)-5(3,2),\textit{F}=6-6&8.5/8.4&&&\\
&131.843328&131.8453368&6(2,5)-5(2,4),\textit{F}=5-5&4.6&2.88/1.05&5.8/5.9&\\
&&131.8457502&6(2,5)-5(2,4),\textit{F}=5-6&5.5&&&\\
&&131.8458342&6(2,5)-5(2,4),\textit{F}=7-6&5.7&&&\\
&&131.8458344&6(2,5)-5(2,4),\textit{F}=5-4&5.7&&&\\
&&131.8459184&6(2,5)-5(2,4),\textit{F}=6-5&5.9&&&\\
&&131.8460644&6(2,4)-5(2,3),\textit{F}=5-5&6.2&&&\\
&&131.8463318&6(2,5)-5(2,4),\textit{F}=6-6&6.8&&&\\
&&131.8464777&6(2,4)-5(2,3),\textit{F}=5-6&7.2&&&\\
&&131.8465617&6(2,4)-5(2,3),\textit{F}=7-6&7.4&&&\\
&&131.8465619&6(2,4)-5(2,3),\textit{F}=5-4&7.4&&&\\
&&131.8466459&6(2,4)-5(2,3),\textit{F}=6-5&7.5&&&\\
&&131.8470591&6(2,4)-5(2,3),\textit{F}=6-6&8.5&&&\\
&131.882625/131.882156&131.8849274&6(0,6)-5(0,5),\textit{F}=5-5&5.2/6.3&5.96/3.91&6.3,24.0&\\
&&&&&&/14.7,1.9&\\
&&131.8856165&6(0,6)-5(0,5),\textit{F}=5-6&6.8/7.9&&&\\
&&131.8857342&6(0,6)-5(0,5),\textit{F}=7-6&7.1/8.1&&&\\
&&131.8857419&6(0,6)-5(0,5),\textit{F}=6-5&7.1/8.2&&&\\
&&131.8857569&6(0,6)-5(0,5),\textit{F}=5-4&7.1/8.2&&&\\
&&131.886431&6(0,6)-5(0,5),\textit{F}=6-6&8.7/9.7&&&\\
&132.353531/132.353609&132.3561165&6(1,5)-5(1,4),\textit{F}=5-5&5.9/5.7&3.52/1.41&6.9,2.9/7.1&\\
&&132.3566445&6(1,5)-5(1,4),\textit{F}=5-6&7.1/6.9&&&\\
&&132.356738&6(1,5)-5(1,4),\textit{F}=7-6&7.3/7.1&&&\\
&&132.3567521&6(1,5)-5(1,4),\textit{F}=5-4&7.3/7.1&&&\\
&&132.3567638&6(1,5)-5(1,4),\textit{F}=6-5&7.3/7.1&&&\\
&&132.3572919&6(1,5)-5(1,4),\textit{F}=6-6&8.5/8.3&&&\\
\hline
H$_2$CO&145.598250/145.598328&145.602949&2(0,2)-1(0,1)&9.7/9.5&21.0/27.2&28.0,10.1,3.8&\\
&&&&&&/9.0,26.3,2.7&\\
&150.493594/150.493516&150.498334&2(1,1)-1(1,0)&9.4/9.6&30.0/31.0&11.0,27.4,3.8&blend CH$_3$COCH$_3$\\
&&&&&&/8.5,25.4,2.8&\\
\hline
H$_2$$^{13}$CO&137.446500/137.446188&137.4499503&2(1,2)-1(1,1)&7.5/8.2&3.55/4.97&3.1/1.9&\\
&146.631469/146.631016&146.6356717&2(1,1)-1(1,0)&8.6/9.5&2.53/4.25&3.0/2.2&\\
\hline
o-H$_2$CS&137.366000&137.36917&4(3,2)-3(3,1)&-&-&-&blend H$_2$CS($v$=0)\\
p-H$_2$CS&137.367453/137.367375&137.371051&4(0,4)-3(0,3)&7.9/8.0&3.81/4.22&3.5/2.3&blend H$_2$CS($v$=0)\\
&137.378750/137.378359&137.382029&4(2,3)-3(2,2)&-&-&-&blend CH$_3$OCH$_3$\\
&137.408422&137.41177&4(2,2)-3(2,1)&7.3&1.34/1.65&3.0/2.3&\\
\hline
\textit{c}-C$_3$H$_2$&150.847078/150.845938&150.851908&4(1,4)-3(0,3)&9.6/11.9&0.71/0.89&4.4/4.6&\\
\hline
HC$_3$N&136.460563&136.4644013&\textit{J}=15-14&8.4&12.5/11.5&10.5,11.4,3.2,&\\
&&&&&&14.0/5.8,24.0,&\\
&&&&&&2.3&\\
&145.556281&145.560946&\textit{J}=16-15&9.6&14.1/13.8&13.7,13.4,2.7,&blend CH$_3$OCH$_3$\\
&&&&&&9.5/7.6,24.9,&\\
&&&&&&2.5&\\
\hline
HC$_3$N&136.685953/-&136.6882521&\textit{J}=15-14,\textit{l}=1\textit{e}&5.0/-&0.84/-&8.6/-&CDMS\\
($v_6$=1)&136.792547/136.793844&136.7956896&\textit{J}=15-14,\textit{l}=1\textit{f}&-&-&-&CDMS, blend\\
&&&&&&&HC$_3$N($v_7$=1)\\
&145.796234/-&145.799706&\textit{J}=16-15,\textit{l}=1\textit{e}&7.1/-&1.09/-&9.5/-&CDMS\\
&145.911297/145.911219&145.914297&\textit{J}=16-15,\textit{l}=1\textit{f}&-&-&-&CDMS, blend\\
&&&&&&&HC$_3$N($v_7$=1)\\
\hline
HC$_3$N&136.797344/136.797422&136.7997925&\textit{J}=15-14,\textit{l}=1\textit{e}&5.4/5.2&3.61/0.68&6.2,20.4/16.3&CDMS, blend\\
($v_7$=1)&&&&&&&HC$_3$N($v_6$=1)\\
&136.993203/136.993891&136.9957308&\textit{J}=15-14,\textit{l}=1\textit{f}&5.5/4.0&3.69/0.90&14.2,5.2/14.3&CDMS, blend\\
&&&&&&&CH$_3$OCHO($v$=0)\\
&145.914875/145.915641&145.9186679&\textit{J}=16-15,\textit{l}=1\textit{e}&7.8/6.2&4.17/1.36&17.6,6.4,3.3&CDMS, blend\\
&&&&&&/13.2&HC$_3$N($v_6$=1)\\
&146.123859/146.123938&146.1276387&\textit{J}=16-15,\textit{l}=1\textit{f}&7.8/7.6&4.44/1.18&5.6,11.5/13.0&CDMS, blend\\
&&&&&&&C$_2$H$_5$CN\\
\hline
HC$_3$N&137.319766/-&137.3225355&\textit{J}=15-14,\textit{l}=0&6.0/-&0.97/-&7.2/-&CDMS, blend\\
($v_7$=2)&&&&&&&C$_2$H$_5$CN\\
&137.326031/-&137.3295712&\textit{J}=15-14,\textit{l}=2\textit{e}&7.7/-&0.96/-&6.1/-&CDMS\\
&137.336250/-&137.3380387&\textit{J}=15-14,\textit{l}=2\textit{f}&3.9/-&0.81/-&7.1/-&CDMS\\
&146.471703/-&146.47495&\textit{J}=16-15,\textit{l}=0&6.6/-&0.98/-&5.2/-&CDMS\\
&146.480016/-&146.4837002&\textit{J}=16-15,\textit{l}=2\textit{e}&7.5/-&1.03/-&5.6/-&CDMS\\
&146.490859/-&146.4939773&\textit{J}=16-15,\textit{l}=2\textit{f}&6.4/-&0.93/-&6.3/-&CDMS\\
\hline
t-HCOOH&133.763625/133.763391&133.76707&6(0,6)-5(0,5)&7.7/8.2&0.77/1.17&3.2/1.9&\\
&151.171359/151.171438&151.1762809&7(1,7)-6(1,6)&9.8/9.6&0.95/1.23&3.0/1.6&\\
\hline
CH$_3$CN&147.031891/147.031500&147.0358351&8(7)-7(7)&8.0/8.8&1.70/1.17&5.9/4.2&\\
&147.068516&147.0726021&8(6)-7(6)&8.3&5.69/2.83&13.5,4.9&blend CH$_3$$^{13}$CN\\
&&&&&&/13.3,3.7&\\
&147.099719&147.103738&8(5)-7(5)&8.2&7.33/4.14&5.5,18.3&blend U-33, \\
&&&&&&/2.6,15.1&CH$_3$$^{13}$CN\\
&147.125422/147.124813&147.1292302&8(4)-7(4)&7.8/9.0&8.95/5.74&12.5,5.0&\\
&&&&&&/11.7,2.5&\\
&147.144813/147.144656&147.1490683&8(3)-7(3)&8.7/9.0&14.0/10.9&19.4,11.0,2.9,&blend CH$_3$CN($v$=0)\\
&&&&&&7.6/13.9,2.5&\\
&147.159000&147.1632441&8(2)-7(2)&8.6&12.70/9.77&10.9,2.8&blend CH$_3$CN($v$=0)\\
&&&&&&/11.3,2.3&\\
&147.167547/147.167078&147.1717519&8(1)-7(1)&8.6/9.5&16.1/12.5&2.3,17.2,2.3&blend CH$_3$CN($v$=0)\\
&&&&&&/2.1,16.5,2.0&\\
&147.169906/147.170063&147.1745883&8(0)-7(0)&9.5/9.2&18.3/14.1&common with above&blend CH$_3$CN($v$=0)\\
\hline
CH$_3$CN&147.472656/147.473047&147.4749743&\textit{J}=8-7,\textit{K}=$-$6-$-$6&4.7/3.9&1.35/0.86&6.3/2.6&\\
($v_8$=1)&&147.4759962&\textit{J}=8-7,\textit{$^l$K}=$-$1-1&6.8/6.0&1.35/0.86&6.3/2.6&\\
&147.541016/-&147.5439118&\textit{J}=8-7,\textit{K}=$-$4-$-$4&5.9/-&0.79/-&2.9/-&\\
&147.566047/-&147.5698084&\textit{J}=8-7,\textit{K}=$-$3-$-$3&7.6/-&0.85/-&5.4/-&\\
&147.571766/147.571391&147.5755526&\textit{J}=8-7,\textit{K}=5-5&7.7/8.5&0.82/0.80&3.9/-&\\
&147.586422/147.585813&147.5898941&\textit{J}=8-7,\textit{K}=$-$2-$-$2&7.1/8.3&1.36/1.16&5.1/8.8&blend CH$_3$CN($v_8$=1)\\
&147.590844/147.590766&147.5953801&\textit{J}=8-7,\textit{K}=4-4&9.2/9.4&1.69/0.92&8.5/3.2&\\
&147.600234&147.6039578&\textit{J}=8-7,\textit{K}=$-$1-$-$1&7.6&2.62/1.01&6.5/7.8&\\
\hline
CH$_3$OH&131.983797/131.983563&131.986795&21(5)$^+$-22(4)$^+$&6.8/7.3&1.08/0.65&3.8/2.3&\\
&132.141953/132.142031&132.144943&21(5)$^-$-22(4)$^-$&6.8/6.6&1.41/0.89&3.7/3.8&\\
&132.618453/132.618063&132.621824&6(2)$^-$-7(1)$^-$&7.6/8.5&11.3/11.1&7.6,2.5&\\
&&&&&&/7.4,1.9&\\
&132.887016&132.890759&6(-1)-5(0)E2&8.4&21.2/25.5&15.1,7.0,2.3&\\
&&&&&&/1.1,7.9,1.6&\\
&133.257453/133.257375&133.260869&20(4)-19(5)E1&7.7/7.9&1.72/0.93&4.3/3.8&\\
&133.424172&133.427704&20(0)-20(-1)E2&7.9&2.71/1.75&5.4,2.8/4.4&\\
&133.453000/133.453234&133.456518&22(-2)-22(1)E2&7.9/7.4&3.43/1.80&6.2,3.2/4.4&\\
&133.601859/133.601641&133.605439&5(-2)-6(-1)E2&8.0/8.5&10.4/9.8&7.8,2.6&\\
&&&&&&/8.4,2.2&\\
&136.724938/136.724859&136.728343&16(3)$^+$-15(4)$^+$&7.5/7.6&7.04/4.98&8.0,3.0&blend CH$_3$CCH\\
&&&&&&/8.0,2.9&\\
&136.911641&136.915288&19(0)-19(-1)E2&8.0&4.51/2.22&3.8/5.1&\\
&137.229969&137.23338&16(3)$^-$-15(4)$^-$&7.5&7.40/4.53&7.6,2.8&\\
&&&&&&/7.4,2.3&\\
&137.899563/137.899484&137.903064&7(-4)-8(-3)E2&7.6/7.8&9.72/8.62&6.8,2.2&\\
&&&&&&/7.0,1.8&\\
&145.120234/145.119859&145.124332&3(2)$^-$-2(2)$^-$&8.5/9.2&7.62/7.92&1.9,8.3,2.8&blend CH$_3$OH,\\
&&&&&&/2.0,11.2,1.9&X--8,X--9,X--10\\
&145.121688/145.121609&145.126191&3(2)-2(2)E1&9.3/9.5&9.28/11.01&common &blend CH$_3$OH,\\
&&&&&&with above&X--8,X--9,X--10\\
&145.127250&145.131864&3(1)-2(1)E1&9.5&7.90/9.21&3.7,11.0,2.1&blend CH$_3$OH\\
&&&&&&/1.6,7.7,1.8&\\
&145.128703&145.133415&3(2)$^+$-2(2)$^+$&9.7&7.42/7.78&common with above&blend CH$_3$OH\\
&145.761906/145.761672&145.766227&16(0)-16(-1)E2&8.9/9.4&6.33/5.18&6.7,2.8&\\
&&&&&&/1.7,6.8&\\
&146.152938/-&146.156763&25(6)$^-$-24(7)$^-$&7.8/-&0.90/-&0.89/-&blend CH$_3$OCH$_3$\\
&&146.156825&25(6)$^+$-24(7)$^+$&8.0/-&0.90/-&0.89/-&blend CH$_3$OCH$_3$\\
&146.282797&146.286876&20(6)$^+$-21(5)$^+$&8.4&2.04/1.38&4.7/3.8&\\
&146.363891/146.363672&146.368328&3(1)$^-$-2(1)$^-$&9.1/9.5&14.7/18.1&7.8,3.2&\\
&&&&&&/8.3,2.1&\\
&146.614000/146.614078&146.618697&9(0)$^+$-8(1)$^+$&9.6/9.4&21.4/21.5&4.4,0.47,12.5&blend SO$_2$\\
&&&&&&/2.0,8.9,1.9&\\
&147.392844/147.392781&147.396869&24(-2)-24(1)E2&8.2/8.3&3.08/2.39&3.6/3.1&blend\\
&&&&&&&CH$_3$OCHO($v$=0)\\
&150.137203/150.136969&150.141672&14(0)-14(-1)E2&8.9/9.4&10.81/6.83&7.6,2.9&\\
&&&&&&/7.5,2.0&\\
&150.880203/150.879813&150.884543&12(-1)-11(-2)E2&8.6/9.4&15.7/12.4&8.1,0.79,3.1&blend CH$_3$OCH$_3$\\
&&&&&&/6.9,1.7&\\
&151.855828/151.855375&151.860249&13(0)-13(-1)E2&8.7/9.6&11.83/8.17&7.7,2.6/7.2,1.7&\\
\hline
$^{13}$CH$_3$OH&146.053359/-&146.057653&16(0,16)-16(-1,16)&8.8/-&0.85/-&1.9/-&CDMS\\
&150.020391/150.020609&150.024809&14(0,14)-14(-1,14)&8.8/8.4&1.11/0.60&3.1/1.7&CDMS\\
&151.567266/151.567109&151.571517&13(0,13)-13(-1,13)&8.4/8.7&2.18/0.91&0.6,6.5/2.8&CDMS\\
\hline
CH$_3$CHO&138.281063/138.281281&138.28488&7(1,6)-6(1,5)E&8.3/7.8&0.83/1.04&3.1/2.1&\\
&138.316469/138.315844&138.31975&7(1,6)-6(1,5)A&7.1/8.5&0.57/0.98&4.0/1.3&\\
\hline
CH$_3$CCH&136.658781/136.658719&136.66274&8(5)-7(5)&8.7/8.8&0.83/0.90&3.1/1.8&blend\\
&&&&&&&CH$_3$OCHO($v$=0)\\
&136.700531/136.700750&136.7045016&8(3)-7(3)&8.7/8.2&1.19/1.46&3.8/2.2&blend\\
&&&&&&&CH$_3$OH($v_{12}$=1)\\
&136.713641&136.7175596&8(2)-7(2)&8.6&1.70/1.67&-&blend\\
&&&&&&&CH$_3$OCHO($v$=0)\\
&136.721500/136.721656&136.7253966&8(1)-7(1)&8.5/8.2&2.03/2.34&3.5/2.1&blend CH$_3$OH\\
&136.724938/136.724859&136.7280098&8(0)-7(0)&6.7/6.9&7.04/4.98&8.0,3.0&blend CH$_3$OH\\
&&&&&&/8.0,2.9&\\
\hline
CH$_3$NCO&131.415891/131.416281&131.418379&15(-1,0)-14(-1,0),&5.7/4.8&2.34/0.94&6.4/6.3&CDMS,\\
($v$=0)&&&\textit{m}=-3&&&&blend X--2\\
&137.466797/137.467406&137.469287&16(1,16)-15(1,15),&5.4/4.1&0.55/0.57&3.4/-&CDMS\\
&&&\textit{m}=0&&&&\\
&137.489688/-&137.491243&16(1,0)-15(1,0),&3.4/-&0.63/-&10.1/-&CDMS\\
&&&\textit{m}=1&&&&\\
&138.300516/-&138.303496&16(3,14)-15(3,13),&6.5/-&0.64/-&3.4/-&CDMS\\
&&&\textit{m}=0&&&&\\
&&138.304065&16(3,13)-15(3,12),&7.7/7.4&0.64/0.48&3.4/-&CDMS\\
&&&\textit{m}=0&&&&\\
\hline
C$_2$H$_3$CN&132.521859/132.521781&132.524583&14(2,13)-13(2,12)&6.2/6.3&5.83/6.38&7.9,3.0&blend CH$_3$OCH$_3$\\
&&&&&&/8.5,3.0&\\
&132.897703/-&132.90001&14(5,9)-13(5,8)&5.2/-&1.80/-&7.5/-&blend C$_2$H$_3$CN\\
&&132.90001&14(5,10)-13(5,9)&5.2/-&&&\\
&132.902656/-&132.905288&14(6,8)-13(6,7)&5.9/-&1.84/-&7.0/-&blend C$_2$H$_3$CN\\
&&132.905288&14(6,9)-13(6,8)&5.9/-&&&\\
&132.918375/132.918453&132.917752&14(4,11)-13(4,10)&-&-&-&blend CH$_3$OCHO($v$=0)\\
&&132.918991&14(4,10)-13(4,9)&-&-&-&\\
&132.921359/132.920359&132.923739&14(7,7)-13(7,6)&-&-&-&blend CH$_3$OCHO($v$=0)\\
&&132.923739&14(7,8)-13(7,7)&-&-&-&\\
&132.948594/-&132.951274&14(8,6)-13(8,5)&6.0/-&1.33/-&5.0/-&\\
&&132.951274&14(8,7)-13(8,6)&6.0/-&&&\\
&132.957750/-&132.959401&14(3,12)-13(3,11)&3.7/-&1.59/-&6.8/-&\\
&132.984531/-&132.985953&14(9,5)-13(9,4)&3.2/-&0.82/-&9.2/-&\\
&&132.985953&14(9,6)-13(9,5)&3.2/-&&&\\
&133.023672/133.023438&133.0266652&14(10,4)-13(10,3)&6.7/7.3&1.03/0.79&5.5/-&blend C$_2$H$_3$CN\\
&&133.0266652&14(10,5)-13(10,4)&6.7/7.3&&&\\
&133.028328/133.027328&133.030674&14(3,11)-13(3,10)&5.3/7.5&1.68/0.47&4.8/-&blend C$_2$H$_3$CN\\
&138.392609/138.393063&138.39516&15(1,15)-14(1,14)&5.5/4.5&1.38/0.61&9.9,5.1/11.7&\\
&145.138625/145.137703&145.141487&15(1,14)-14(1,13)&5.9/7.8&0.98/0.89&5.6/-&\\
&147.557578/147.557734&147.5617015&16(1,16)-15(1,15)&8.4/8.1&1.25/0.80&5.7/-&\\
&151.353406/-&151.356948&16(2,15)-15(2,14)&7.0/-&1.44/-&9.4/-&blend SO$_2$\\
&151.897344/151.897266&151.900284&16(5,12)-15(5,11)&5.8/6.0&2.37/0.52&17.3,5.8/4.2&\\
&&151.900324&16(5,11)-15(5,10)&5.9/6.0&&&\\
&151.912672/151.911984&151.915868&16(7,9)-15(7,8)&-&-&-&blend $^{34}$SO$_2$\\
&&151.915868&16(7,10)-15(7,9)&&&&\\
&151.930141/151.928391&151.933608&16(4,13)-15(4,12)&6.8/10.3&1.11/0.50&5.1/-&blend C$_2$H$_3$CN\\
&151.933125/151.933203&151.9368513&16(4,12)-15(4,11)&7.4/7.2&1.82/0.94&6.8/1.0&blend CH$_3$OCHO($v$=0),\\
&&&&&&&C$_2$H$_3$CN\\
&151.940906/-&151.94455&16(8,8)-15(8,7)&7.2/-&1.20/-&5.6/-&blend \\
&&&&&&&CH$_3$OCHO($v$=0)\\
&&151.94455&16(8,9)-15(8,8)&7.2/-&&&\\
&151.978594/-&151.9822606&16(9,7)-15(9,6)&7.2/-&1.22/-&5.3/-&blend\\
&&&&&&&CH$_3$OCHO($v$=0)\\
&&151.9822606&16(9,8)-15(9,7)&7.2/-&&&\\
&151.982406&151.986758&16(3,14)-15(3,13)&8.6&2.10/1.10&9.9,3.0/2.8&blend CH$_3$OCHO($v$=0),\\
&&&&&&&C$_2$H$_3$CN\\
&152.024078/152.024609&152.027503&16(10,6)-15(10,5)&6.8/5.7&0.84/0.58&4.7/-&\\
&&152.027503&16(10,7)-15(10,6)&6.8/5.7&&&\\
&152.075797/152.075500&152.0787312&16(11,5)-15(11,4),&5.8/6.4&0.95/0.51&10.8/6.6&\\
&&&\textit{F}=15-15&&&&\\
&&152.0787312&16(11,6)-15(11,5),&5.8/6.4&&&\\
&&&\textit{F}=15-15&&&&\\
&&152.079247&16(11,5)-15(11,4),&6.8/7.4&&&\\
&&&\textit{F}=16-15&&&&\\
&&152.079247&16(11,6)-15(11,5),&6.8/7.4&&&\\
&&&\textit{F}=16-15&&&&\\
&&152.079505&16(11,5)-15(11,4),&7.3/7.9&&&\\
&&&\textit{F}=17-16&&&&\\
&&152.079505&16(11,6)-15(11,5),&7.3/7.9&&&\\
&&&\textit{F}=17-16&&&&\\
&&152.079505&16(11,5)-15(11,4),&7.3/7.9&&&\\
&&&\textit{F}=15-14&&&&\\
&&152.079505&16(11,6)-15(11,5),&7.3/7.9&&&\\
&&&\textit{F}=15-14&&&&\\
&&152.0799279&16(11,5)-15(11,4),&8.1/8.7&&&\\
&&&\textit{F}=16-16&&&&\\
&&152.0799279&16(11,6)-15(11,5),&8.1/8.7&&&\\
&&&\textit{F}=16-16&&&&\\
&152.122422/-&152.1255495&16(3,13)-15(3,12)&6.2/-&1.16/-&8.1/-&blend\\
&&&&&&&CH$_3$OCHO($v$=0)\\
\hline
CH$_3$OCHO&131.911000/131.911078&131.914524&12(0,12)-11(1,11)E&8.0/7.8&1.58/1.60&2.0/2.2&blend CH$_3$OCH$_3$\\
&131.913063&131.916432&12(0,12)-11(1,11)A&7.7&1.44&2.0/1.7&blend CH$_3$OCHO($v$=0)\\
&132.003250&132.006639&10(2,9)-9(1,8)E&7.7&0.97/0.95&2.2/1.9&blend CH$_3$COCH$_3$\\
&132.007375/132.007438&132.010774&10(2,9)-9(1,8)A&7.7/7.6&1.74/1.34&3.6/3.2&blend CH$_3$COCH$_3$\\
&132.102203/132.102063&132.105508&12(1,12)-11(1,11)E&7.5/7.8&4.53/4.26&13.6,2.6,2.8&blend OC$^{33}$S,\\
&&&&&&/2.6,2.5,17.0&$^{34}$SO2\\
&132.103813/132.103891&132.107205&12(1,12)-11(1,11)A&7.7/7.5&5.42/4.47&common with above&blend $^{34}$SO2\\
&132.241766&132.24243&19(6,13)-19(5,14)A&1.5&4.62/4.46&2.7,12.3,2.7&blend\\
&&&&&&/1.9,8.0,1.5&CH$_3$OCHO($v$=0),\\
&&&&&&&H$^{13}$CCCN\\
&&132.244199&19(6,13)-19(5,14)E&5.5&&common with above&\\
&&132.245128&12(0,12)-11(0,11)E&7.6&&common with above&\\
&132.243438/132.243203&132.24673&12(0,12)-11(0,11)A&7.5/8.0&5.01/4.61&common with above&blend H$^{13}$CCCN\\
&132.432656/132.432813&132.436113&12(1,12)-11(0,11)E&7.8/7.5&1.73/1.62&3.1/1.8&blend\\
&&&&&&&CH$_3$OCHO($v$=0)\\
&132.434109&132.437494&12(1,12)-11(0,11)A&7.7&1.55/1.78&1.8/1.8&blend\\
&&&&&&&CH$_3$OCHO($v$=0)\\
&132.499719&132.503091&19(5,15)-19(4,16)E&7.6&0.76/0.74&2.1/1.8&\\
&132.533672/132.533750&132.537122&19(5,15)-19(4,16)A&7.8/7.6&0.80/0.82&2.6/1.0&\\
&132.918609/132.918453&132.921937&11(1,10)-10(1,9)E&7.5/7.9&4.92/4.33&2.2,13.2&blend C$_2$H$_3$CN\\
&&&&&&/2.1,8.0&\\
&132.925391&132.928736&11(1,10)-10(1,9)A&7.5&4.45/4.26&2.0,6.6&blend C$_2$H$_3$CN\\
&&&&&&/1.4,4.1&\\
&136.276531&136.280061&11(4,8)-10(4,7)E&7.8&2.92/2.62&2.5/2.4&blend CH$_3$OCHO($v$=0)\\
&136.278969/136.279047&136.282597&11(4,8)-10(4,7)A&8.0/7.8&2.56/2.75&3.1/2.3&blend CH$_3$OCHO($v$=0)\\
&136.586766/136.586688&136.590238&12(1,11)-11(2,10)E&7.6/7.8&1.30/1.19&2.1/1.3&\\
&136.596453/136.596375&136.599952&12(1,11)-11(2,10)A&7.7/7.9&1.30/1.09&2.8/1.0&\\
&136.658781/136.658719&136.662168&19(4,16)-19(3,17)E&7.4/7.6&0.83/0.90&3.1/1.8&blend CH$_3$CCH\\
&136.711594/136.711656&136.715083&19(4,16)-19(3,17)A&7.7/7.5&1.21/0.73&0.8/2.4&blend CH$_3$CCH\\
&136.950328&136.953766&20(5,16)-20(4,17)E&7.5&1.41/1.26&2.3/2.0&blend $^{33}$SO\\
&136.987031&136.990512&20(5,16)-20(4,17)A&7.6&1.43/1.28&2.5/2.4&blend HC$_3$N($v_7$=1)\\
&137.289703/137.289625&137.293183&11(4,7)-10(4,6)E&7.6/7.8&4.28/4.20&2.0,5.2&\\
&&&&&&/0.88,2.9&\\
&137.304656&137.308132&18(6,12)-18(5,13)E&7.6&1.44/1.16&3.1/2.8&\\
&&137.308622&18(6,12)-18(5,13)A&8.7&&&\\
&137.309844/137.309766&137.31333&11(4,7)-10(4,6)A&7.6/7.8&4.38/4.00&1.1,4.2&\\
&&&&&&/1.9,7.1&\\
&138.662625/138.662859&138.665944&27(5,22)-27(4,23)E&7.2/6.7&1.13/0.56&2.4/2.6&\\
&146.010859/146.010250&146.014526&4(4,0)-3(3,0)E&7.5/8.8&1.64/1.67&4.9/3.1&blend\\
&&146.015063&4(4,1)-3(3,0)A&8.6/9.9&&&CH$_3$OCHO($v$=0)\\
&146.018797/146.018641&146.023024&4(4,0)-3(3,1)A&8.7/9.0&1.69/1.25&4.2/4.8&blend CH$_3$OCHO($v$=0)\\
&146.588141/146.588219&146.59242&31(6,25)-31(5,26)A&8.8/8.6&3.08/1.81&-/2.4&blend SO$_2$,\\
&&&&&&&CH$_3$OCHO($v_{18}$=1),\\
&&&&&&&CH$_3$OCH$_3$\\
&146.695109/146.694953&146.699324&15(6,9)-15(5,10)A&8.6/8.9&0.94/1.29&2.5/2.0&blend\\
&&&&&&&CH$_3$OCHO($v_{18}$=1)\\
&146.744172/146.744234&146.748494&15(6,9)-15(5,10)E&8.8/8.7&0.58/0.98&-/2.1&\\
&146.973281&146.977678&12(3,10)-11(3,9)E&9.0&7.40/6.99&3.1/2.4&blend CS,\\
&&&&&&&CH$_3$OCHO($v_{18}$=1)\\
&146.983891/146.983672&146.988047&12(3,10)-11(3,9)A&8.5/8.9&4.24/4.80&0.90,4.7&\\
&&&&&&/1.1,3.8&\\
&147.243766/147.243609&147.247965&18(6,13)-18(5,14)E&8.5/8.9&1.86/1.82&5.9,3.4&blend \\
&&&&&&/8.8,0.96&CH$_3$OCHO($v_{18}$=1)\\
&147.245297/147.246656&147.250781&12(11,1)-11(11,0)E&-&-&-&blend CH$_3$OCHO($v_{18}$=1)\\
&147.250859/147.251469&147.2556572&12(11,2)-11(11,1)A&9.8/8.5&1.19/1.47&3.5/2.2&\\
&&147.255682&12(11,1)-11(11,0)A&9.8/8.6&&&\\
&147.261078&147.265314&12(11,2)-11(11,1)E&8.6&1.08/1.25&3.8/1.6&\\
&147.276578/147.276875&147.28099&18(6,13)-18(5,14)A&9.0/8.4&0.99/0.98&2.1/1.8&\\
&147.300531/147.300609&147.304789&19(6,14)-19(5,15)E&8.7/8.5&0.98/1.26&2.8/1.7&\\
&147.306250&147.31057&12(10,2)-11(10,1)E&8.8&1.45/1.29&2.5/1.3&\\
&147.313578/147.313500&147.31775&12(10,2)-11(10,1)A&8.5/8.6&2.51/2.70&2.0/1.4&\\
&147.321125/147.321203&147.325392&12(10,3)-11(10,2)E&8.7/8.5&1.84/1.52&3.0/2.5&blend U-34\\
&147.327469/147.327547&147.331633&19(6,14)-19(5,15)A&8.5/8.3&0.97/0.91&5.7/2.2&blend U-34\\
&147.392844/147.392781&147.397073&12(9,3)-11(9,2)E&8.6/8.7&3.08/2.39&3.6/3.1&blend CH$_3$OH\\
&147.402156/147.401938&147.406369&12(9,3)-11(9,2)A&8.6/9.0&2.20/2.24&3.3/2.7&\\
&147.407578/147.407500&147.411819&12(9,4)-11(9,3)E&8.6/8.8&1.45/1.86&4.3/2.3&\\
&147.520188/147.519969&147.524311&12(8,4)-11(8,3)E&8.4/8.8&1.53/1.87&2.3/2.1&\\
&147.531406/147.531328&147.535539&12(8,5)-11(8,4)A&8.4/8.6&2.00/2.40&3.2/2.7&blend\\
&&147.5355507&12(8,4)-11(8,3)A&8.4/8.6&&&CH$_3$OCHO($v$=0)\\
&147.534766/147.534313&147.538644&12(8,5)-11(8,4)E&7.9/8.8&1.75/1.85&3.5/2.5&blend CH$_3$OCHO($v$=0)\\
&&147.539172&17(6,12)-17(5,13)E&9.0/9.9&&&\\
&147.581313/147.581000&147.585206&17(6,12)-17(5,13)A&7.9/8.5&0.80&1.7/2.8&\\
&-/149.919969&149.924426&13(6,8)-13(5,9)E&-/8.9&-/0.69&-/1.3&\\
&150.065625/150.065781&150.070143&13(6,8)-13(5,9)A&9.0/8.7&1.01/0.84&2.3/2.4&\\
&150.386078&150.390454&12(6,6)-12(5,7)A&8.7&1.67/1.15&2.0/1.4&blend SO$_2$\\
&150.395313/150.395234&150.399678&24(7,17)-24(6,18)E&8.7/8.9&0.98/0.89&1.9/0.58&blend CH$_3$COCH$_3$\\
&150.444906&150.449202&12(6,6)-12(5,7)E&8.6&1.21/0.92&2.3/1.8&blend X-17\\
&150.596359&150.600762&12(4,8)-11(4,7)E&8.8&5.14/3.69&5.2,1.6&\\
&&&&&&/1.9,7.4&\\
&150.613844&150.618303&12(4,8)-11(4,7)A&8.9&5.10/3.76&2.1,6.5&\\
&&&&&&/0.96,2.8&\\
&150.632375&150.63674&12(6,7)-12(5,8)A&8.7&1.33&6.1/3.2&\\
&150.682578/150.682438&150.685347&31(8,23)-31(7,24)A&5.5/5.8&0.72/0.53&4.0/3.5&blend CH$_3$COCH$_3$\\
&150.734016/150.733781&150.738141&31(8,23)-31(7,24)E&8.2/8.7&0.68/0.52&3.6/-&\\
&150.782313/150.781922&150.786711&19(2,17)-19(2,18)A&8.7/9.5&0.64/0.57&1.7/1.5&\\
&150.851578/150.851516&150.855977&19(2,17)-19(1,18)E&8.7/8.9&0.77&2.8/1.5&blend \textit{c}-C$_3$H$_2$\\
&150.925828&150.930193&19(2,17)-19(1,18)A&8.7&0.77/0.52&2.4/3.4&\\
&150.952906/-&150.957157&22(6,17)-22(5,18)E&8.4/-&0.79/-&2.1/-&\\
&150.976406&150.980775&22(6,17)-22(5,18)A&8.7&1.42/0.84&0.55/2.7&blend U-41\\
&151.004563&151.009041&11(6,6)-11(5,7)E&8.9&1.38/1.39&2.4/2.0&\\
&151.031656&151.035991&11(6,5)-11(5,6)E&8.6&1.16/0.94&3.2/0.7&\\
&151.108938&151.113186&11(6,6)-11(5,7)A&8.4&1.21/1.04&1.9/1.3&\\
&151.408719&151.413078&10(6,5)-10(5,6)E&8.6&1.27/1.11&2.3/1.2&\\
&151.451906&151.456198&10(6,4)-10(5,5)A&8.5&1.18/0.98&1.1/1.2&blend CH$_3$OCHO($v$=0)\\
&151.453438/151.453578&151.457804&10(6,4)-10(5,5)E&8.6/8.4&1.21/0.74&1.5/1.1&blend CH$_3$OCHO($v$=0)\\
&151.491578/151.491734&151.496105&10(6,5)-10(5,6)A&9.0/8.6&0.77/0.92&2.2/2.4&\\
&151.713313/151.713156&151.717477&9(6,4)-9(5,5)E&8.2/8.5&0.92/0.99&2.3/1.2&\\
&151.761141/151.761063&151.765554&9(6,3)-9(5,4)E&8.7/8.9&0.82/0.63&3.2/1.2&blend U-45\\
&151.772750&151.777093&9(6,3)-9(5,4)A&8.6&0.86/0.53&2.9/3.2&\\
&151.786250/151.786172&151.790613&9(6,4)-9(5,5)A&8.6/8.8&1.09/1.08&1.6/1.2&\\
&151.933125/151.933203&151.937471&8(6,3)-8(5,4)E&8.6/8.4&1.82/0.94&6.8/1.0&blend C$_2$H$_3$CN\\
&151.945719/151.945641&151.9500159&13(2,12)-12(2,11)E&8.5/8.6&5.34/3.93&2.0,5.8&\\
&&&&&&/2.8,0.78&\\
&&151.9500751&17(14,3)-18(13,6)A&8.6/8.7&&&\\
&&151.950079&13(2,12)-12(2,11)E&8.6/8.8&&&\\
&&151.9502357&17(14,3)-18(13,6)A&8.9/9.1&&&\\
&&151.9502357&17(14,4)-18(13,5)A&8.9/9.1&&&\\
&151.952203/151.952047&151.956555&13(2,12)-12(2,11)A&8.6/8.9&5.15/3.58&5.3,2.1&\\
&&&&&&/3.3,1.3&\\
&151.982406&151.9866726&8(6,2)-8(5,3)E&8.4&2.10/1.10&9.9,3.0/2.8&blend C$_2$H$_3$CN\\
&151.999422/151.998891&152.003958&8(6,2)-8(5,3)A&8.9/10.0&1.00/0.76&2.8/4.2&blend CH$_3$OCHO($v$=0)\\
&152.003547&152.007835&8(6,3)-8(5,4)A&8.5&0.77/0.64&3.0/1.1&blend CH$_3$OCHO($v$=0)\\
&-/152.086109&152.090375&7(6,2)-7(5,3)E&-/8.4&-/0.56&-/1.3&\\
&152.112953/-&152.118747&17(1,16)-17(0,17)E&11.4/-&1.04/-&8.1/-&blend C$_2$H$_3$CN\\
&152.156609/152.155609&152.15992&7(6,1)-7(5,2)A&6.5/8.5&0.91/0.56&3.5/4.6&\\
&152.191781/152.192234&152.196627&17(1,16)-17(0,17)A&9.5/8.7&0.67/0.54&2.4/1.4&\\
\hline
CH$_3$OCHO&131.452219/131.452375&131.455648&12(1,12)-11(1,11)E&7.8/7.5&1.00/1.06&2.5/2.4&\\
($v_{18}$=1)&131.608938&131.612329&12(0,12)-11(0,11)E&7.7&1.24/0.98&3.2/2.9&\\
&132.060016&132.063465&11(1,10)-10(1,9)A&7.8&1.58/1.48&2.2/1.0&\\
&132.375203/132.375281&132.378703&11(1,10)-10(1,9)E&7.9/7.7&1.16/1.03&2.4/3.0&\\
&&132.379528&5(3,2)-4(2,2)E&9.8/9.6&1.16/1.03&2.4/3.0&\\
&133.797422/133.797266&133.800739&11(3,9)-10(3,8)A&7.4/7.8&1.91/1.27&1.8/2.5&blend X-5\\
&145.617016/145.616781&145.621125&12(3,10)-11(3,9)A&8.5/8.9&1.71/1.64&2.7/1.7&\\
&146.101813/146.102047&146.106252&12(8,4)-11(8,3)E&9.1/8.6&0.93/1.47&1.3/0.94&\\
&146.148047/146.148734&146.1528653&20(4,17)-20(2,18)E&9.9/8.5&0.75/0.89&1.1/0.83&blend CH$_3$OCH$_3$\\
&&146.152945&12(10,2)-11(10,1)A&10.0/8.6&0.75/0.89&1.1/0.83&blend CH$_3$OCH$_3$\\
&146.170859/146.170781&146.174997&12(9,3)-11(9,2)A&8.5/8.6&1.08/0.99&2.2/1.8&\\
&146.230453/146.230750&146.234988&12(8,5)-11(8,4)A&9.3/8.7&1.07/1.36&3.4/1.8&\\
&146.333766/146.333828&146.337965&12(7,5)-11(7,4)E&8.6/8.5&1.28/0.82&1.7/2.9&blend U-32\\
&146.341156&146.345423&12(3,10)-11(3,9)E&8.7&1.43/1.48&2.4/2.3&\\
&146.353297/146.353141&146.356286&12(11,2)-11(11,1)E&6.1/6.4&2.40/1.16&3.1/3.0&\\
&&146.357492&12(7,6)-11(7,5)A&8.6/8.9&2.40/1.16&3.1/3.0&\\
&146.583484/146.584094&146.587525&12(6,7)-11(6,6)A&8.3/7.0&1.30/1.31&3.7/1.7&blend SO$_2$\\
&&&&&&&CH$_3$OCH$_3$\\
&146.677938/146.678094&146.682383&12(6,6)-11(6,5)E&9.1/8.8&0.97/1.28&2.7/1.5&blend CH$_3$OCH$_3$\\
&147.197688/147.197750&147.201768&12(6,7)-11(6,6)E&8.3/8.2&1.90/2.15&3.1/2.4&blend CH$_3$OCH$_3$\\
&147.243766/147.243609&147.247965&12(5,7)-11(5,6)E&8.5/8.9&1.86/1.82&5.9,3.4&blend\\
&&&&&&/8.8,0.96& CH$_3$OCHO($v$=0)\\
&147.244750/147.244688&147.248992&12(4,9)-11(4,8)A&-&-&-&blend\\
&&&&&&&CH$_3$OCHO($v$=0)\\
&150.893938/150.894000&150.898356&13(2,12)-12(2,11)A&8.8/8.7&2.12/1.69&2.5/1.9&\\
&151.188828/151.188906&151.193305&13(2,12)-12(2,11)E&8.9/8.7&2.01/1.37&3.6/3.0&\\
&152.080297&152.0847&12(2,10)-11(2,9)A&8.7&1.89/1.20&2.2/1.4&blend C$_2$H$_3$CN\\
\hline
CH$_3$OCHO&137.167172/-&137.170769&11(4,8)-10(4,7)A&7.9/-&0.60/-&2.6/-&\\

($v_{\rm t}$=2)&152.182547/152.182922&152.186983&13(2,12)-12(2,11)A&8.7/8.0&0.66/0.59&7.0/3.8&\\
\hline
C$_2$H$_5$CN&133.127750/133.128969&133.129943&16(0,16)-15(1,15)&4.9/2.2&0.72/0.67&15.1/-&\\
&133.544406/133.544031&133.546584&15(2,14)-14(2,13)&4.9/5.7&4.47/1.37&14.9,4.2/16.4&\\
&136.538766/136.538234&136.54127&15(1,14)-14(1,13)&5.5/6.7&3.91/0.99&15.5,4.6/12.6&\\
&136.819938/136.821078&136.8224&15(2,13)-14(2,12)&5.4/2.9&4.15/1.00&15.7,4.7/15.9&\\
&137.084078/-&137.08675&28(4,24)-28(3,25)&5.8/-&0.68/-&8.7/-&\\
&137.108719/-&137.10987&15(1,15)-14(0,14)&2.5/-&0.80/-&11.8/5.4&\\
&137.320984/137.321906&137.324865&23(2,22)-23(1,23)&-&-&-&blend HC$_3$N($v_7$=2)\\
&137.919250/137.919703&137.92213&19(1,18)-18(2,17)&6.3/5.3&0.64/0.60&12.8/10.9&\\
&138.348969/138.350719&138.35105&16(1,16)-15(1,15)&4.5/0.7&4.47/1.08&15.8,4.7/14.2&\\
&145.415203/145.414672&145.41801&16(1,15)-15(1,14)&5.8/6.9&4.71/1.92&15.9,4.8&\\
&&&&&&/17.2,3.1&\\
&145.745813/-&145.74911&25(4,21)-25(3,22)&6.8/-&1.00/-&4.2/-&blend C$^{33}$S\\
&146.116922/146.117984&146.12004&16(2,14)-15(2,13)&6.4/4.2&5.06/1.80&4.3,17.8/14.7&blend HC$_3$N($v_7$=1)\\
&146.891109&146.894524&17(1,17)-16(1,16)&7.0&4.89/1.85&14.1,3.4/17.2&\\
&150.605516/150.605594&150.6068&23(4,19)-23(3,20)&2.6/2.4&0.71/0.56&3.0/1.9&\\
&151.124438/151.123203&151.127264&17(2,16)-16(2,15)&5.6/8.1&4.63/1.10&15.7,3.9/12.8&\\
&151.809594/-&151.81285&18(0,18)-17(1,17)&6.4/-&1.00/-&12.7/-&\\
&152.295766/152.295313&152.297825&17(8,9)-16(8,8)&4.1/4.9&3.88/0.95&5.1/7.2&blend C$_2$H$_5$CN\\
&152.301031/152.302406&152.303789&17(7,11)-16(7,10)&5.4/2.7&6.50/1.11&6.8,23.5/19.4&blend C$_2$H$_5$CN\\
&&152.304641&17(9,8)-16(9,7)&7.1/4.4&6.50/1.11&&\\
&152.317438/152.317906&152.320522&17(10,7)-16(10,6)&6.1/5.1&2.35/0.67&9.6/13.0&blend C$_2$H$_5$CN\\
&152.326219/152.325906&152.329886&17(6,12)-16(6,11)&7.2/7.8&3.87/0.83&18.8,4.4/11.3&blend C$_2$H$_5$CN\\
&152.340250/152.340328&152.343334&17(11,6)-16(11,5)&6.1/5.9&1.80/0.63&8.1/6.7&\\
&152.370469/152.370313&152.371939&17(12,5)-16(12,4)&2.9/3.2&3.71/1.61&3.4,9.4/5.3&blend U-49\\
&152.389391/152.389703&152.3913&17(5,13)-16(5,12)&3.8/3.1&3.40/0.88&14.3,5.0/11.4&\\
&&152.392415&17(5,12)-16(5,11)&5.9/5.3&3.40/0.88&14.3,5.0/11.4&\\

\hline
CH$_3$OCH$_3$&131.903672/131.903609&131.9071431&15(1,14)-15(0,15)AE&7.9/8.0&1.67/2.06&2.5/2.0&\\
&&131.9071435&15(1,14)-15(0,15)EA&7.9/8.0&&&\\
&131.906203&131.9096032&15(1,14)-15(0,15)EE&7.7&2.53/2.71&2.0/2.0&\\
&131.908641&131.9120632&15(1,14)-15(0,15)AA&7.8&1.76/2.06&2.3/1.7&blend CH$_3$OCHO($v$=0)\\
&132.521859/132.521781&132.5247788&8(0,8)-7(1,7)AA&6.6/6.8&5.83/6.38&7.9,3.0&blend C$_2$H$_3$CN\\
&&&&&&/8.5,3.0&\\
&&132.5252393&8(0,8)-7(1,7)EE&7.6/7.8&&&\\
&&132.5256994&8(0,8)-7(1,7)AE&8.7/8.9&&&\\
&&132.5257001&8(0,8)-7(1,7)EA&8.7/8.9&&&\\
&132.975000&132.97761&34(7,28)-35(0,35)EA&5.9&1.19/1.16&1.2/2.0&\\
&&132.9776411&34(7,28)-35(0,35)AE&6.0&&&\\
&&132.9785673&18(3,16)-17(4,13)EE&8.0&&&\\
&133.261953&133.2653105&11(3,8)-11(2,9)AE&7.6&2.82/3.16&1.8/1.8&\\
&&133.2653487&11(3,8)-11(2,9)EA&7.6&&&\\
&133.264859&133.2683191&11(3,8)-11(2,9)EE&7.8&4.04/4.25&2.2/1.8&\\
&133.267906/133.267750&133.2713086&11(3,8)-11(2,9)AA&7.7/8.0&3.00/3.20&3.0/1.9&\\
&133.310781/133.310859&133.3135771&24(3,21)-24(2,22)EE&6.3/6.1&1.58/1.62&6.3/4.5&\\
&&133.3147226&24(3,21)-24(2,22)AA&8.9/8.7&&&\\
&136.619422/136.619500&136.6229215&10(3,7)-10(2,8)AE&7.7/7.5&2.61/2.21&1.4/1.3&blend SO\\
&&136.6229926&10(3,7)-10(2,8)EA&7.8/7.7&&&\\
&136.622625/136.622469&136.6260716&10(3,7)-10(2,8)EE&7.6/7.9&5.43/4.90&2.1/1.8&blend SO\\
&136.625672/136.625594&136.6291861&10(3,7)-10(2,8)AA&7.7/7.9&5.57/3.89&2.6/1.2&blend SO\\
&137.374938/137.375000&137.3784982&20(2,18)-20(1,19)AE&7.8/7.6&1.20/1.15&2.3/2.1&blend H$_2$CS\\
&137.376844/137.376688&137.3804028&20(2,18)-20(1,19)EE&7.8/8.1&2.32/2.56&2.0/1.7&blend H$_2$CS\\
&137.378750/137.378359&137.3823075&20(2,18)-20(1,19)AA&7.8/8.6&2.28/2.32&2.5/2.1&blend H$_2$CS\\
&137.578953&137.5823391&17(3,14)-16(4,13)AA&7.4&0.69/0.76&3.0/1.5&\\
&137.581625&137.5849649&17(3,14)-16(4,13)EE&7.3&1.46/1.17&2.1/1.3&\\
&137.584219/137.584141&137.5875691&17(3,14)-16(4,13)AE&7.3/7.5&0.88/0.85&1.7/1.3&\\
&&137.5876125&17(3,14)-16(4,13)EA&7.4/7.6&&&\\
&145.540031&145.544368&16(1,15)-16(0,16)AE&8.9&0.79/1.23&-/1.5&blend HC$_3$N($v$=0)\\
&&145.5444822&16(1,15)-16(0,16)AE&9.2&&&\\
&&145.5444826&16(1,15)-16(0,16)EA&9.2&&&\\
&145.542859/145.542781&145.547165&16(1,15)-16(0,16)EE&8.9/9.0&1.84/2.65&2.0/1.6&blend HC$_3$N($v$=0)\\
&&145.5472034&16(1,15)-16(0,16)EE&8.9/9.1&&&\\
&145.545984/145.545453&145.5499245&16(1,15)-16(0,16)AA&8.1/9.2&1.39/1.72&6.1/1.9&blend HC$_3$N($v$=0)\\
&&145.549962&16(1,15)-16(0,16)AA&8.2/9.3&&&\\
&145.671484/145.671344&145.675602&5(3,2)-5(2,3)AE&8.5/8.8&1.50/1.93&1.8/1.7&\\
&&145.6756179&5(3,2)-5(2,3)AE&8.5/8.8&&&\\
&145.676000/145.676063&145.67992&5(3,2)-5(2,3)EA&8.1/7.9&3.08/3.98&2.7/2.2&\\
&&145.679936&5(3,3)-5(2,3)EA&8.1/8.0&&&\\
&&145.6803779&5(3,2)-5(2,3)EE&9.0/8.9&&&\\
&&145.680397&5(3,2)-5(2,3)EE&9.0/8.9&&&\\
&145.678359/145.678281&145.6826427&5(3,2)-5(2,3)AA&8.8/9.0&2.02/2.78&2.8/1.9&\\
&&145.682677&5(3,2)-5(2,3)AA&8.9/9.0&&&\\
&146.150875&146.155157&4(3,2)-4(2,2)EE&8.8&1.28/1.43&2.1/2.5&blend CH$_3$OH\\
&&146.1551727&4(3,2)-4(2,2)EE&8.8&&&\\
&-/146.155141&146.15944&4(3,1)-4(2,2)AE&8.5/8.8&-/0.79&-/1.4&blend CH$_3$OH\\
&&146.15946&4(3,1)-4(2,2)AE&8.5/8.9&&&\\
&146.162469/146.161938&146.1662056&4(3,1)-4(2,2)EE&7.7/8.8&2.04/3.32&2.7/2.3&\\
&&146.166246&4(3,1)-4(2,2)EE&7.7/8.8&&&\\
&&146.1665605&4(3,1)-4(2,2)AA&8.4/9.5&&&\\
&&146.166606&4(3,1)-4(2,2)AA&8.5/9.6&&&\\
&&146.1673419&4(3,1)-4(2,2)EA&10.0/11.1&&&\\
&&146.167381&4(3,2)-4(2,2)EA&10.1/11.2&&&\\
&146.401047/146.400750&146.405165&3(3,1)-3(2,1)EE&8.4/9.0&0.99/1.39&1.6/1.0&\\
&&146.4051889&3(3,1)-3(2,1)EE&8.5/9.1&&&\\
&146.402500/-&146.407078&3(3,0)-3(2,1)AE&9.4/-&0.82/-&2.8/-&\\
&&146.4071003&3(3,0)-3(2,1)AE&9.4&&&\\
&146.410359/146.409984&146.4142596&3(3,0)-3(2,1)AA&8.0/8.8&1.19/1.51&5.2/1.5&blend CH$_3$OCH$_3$\\
&&146.414314&3(3,0)-3(2,1)AA&8.1/8.9&&&\\
&146.410359/146.410891&146.4153293&3(3,0)-3(2,1)EE&10.2/9.1&1.19/1.30&0.43,1.4&blend CH$_3$OCH$_3$\\
&&146.415382&3(3,0)-3(2,1)EE&10.3/9.2&&&\\
&&146.4165244&3(3,0)-3(2,1)EA&12.6/11.5&&&\\
&&146.416572&3(3,1)-3(2,1)EA&12.7/11.6&&&\\
&146.575859/146.575703&146.580107&3(3,1)-3(2,2)EE&8.7/9.0&0.97/1.43&2.4/3.5&\\
&&146.5801265&3(3,1)-3(2,2)EE&8.7/9.0&&&\\
&&146.581174&3(3,1)-3(2,2)AE&10.9/11.2&&&\\
&&146.5811959&3(3,1)-3(2,2)AE&10.9/11.2&&&\\
&146.583484/146.584094&146.5883559&3(3,1)-3(2,2)AA&10.0/8.7&1.30/1.31&3.7/1.7&blend SO$_2$,\\
&&&&&&&CH$_3$OCHO($v_{18}$=1)\\
&&146.588412&3(3,1)-3(2,2)AA&10.1/8.8&&&\\
&-/146.585922&146.590267&3(3,0)-3(2,2)EE&-/8.9&-/1.51&-/1.6&blend SO$_2$,\\
&&&&&&&CH$_3$OCHO($v$=0,\\
&&&&&&&$v_{18}$=1)\\
&&146.590323&3(3,0)-3(2,2)EE&5.6/9.0&1.94/1.51&-/1.6&\\
&146.673672/146.673281&146.677588&4(3,2)-4(2,3)AE&8.0/8.8&2.04/2.80&2.9/2.1&\\
&&146.6776074&4(3,2)-4(2,3)AE&8.0/8.8&&&\\
&&146.677951&4(3,2)-4(2,3)EE&8.7/9.5&&&\\
&&146.6779635&4(3,2)-4(2,3)EE&8.8/9.6&&&\\
&&&&&&&\\
&146.680453&146.6847103&4(3,2)-4(2,3)AA&8.7&1.58/2.52&2.8/1.7&blend\\
&&&&&&&CH$_3$OCHO($v_{18}$=1)\\
&&146.684759&4(3,2)-4(2,3)AA&8.8&&&\\
&146.684953&146.6889964&4(3,1)-4(2,3)EE&8.3&0.84/0.98&2.9/1.7&\\
&&146.68904&4(3,1)-4(2,3)EE&8.4&&&\\
&146.697859/146.697703&146.702061&3(2,1)-2(1,2)AE&8.6/8.9&1.52/2.22&2.3/1.6&blend CH$_3$OCHO($v$=0)\\
&&146.7020756&3(2,1)-2(1,2)AE&8.6/8.9&&&\\
&&146.702446&3(2,1)-2(1,2)EA&9.4/9.7&&&\\
&&146.7024544&3(2,1)-2(1,2)EA&9.4/9.7&&&\\
&146.700516/146.700297&146.7047162&3(2,1)-2(1,2)EE&8.6/9.0&2.20/3.12&1.6/2.0&\\
&&146.704743&3(2,1)-2(1,2)EE&8.6/9.1&&&\\
&146.702969/146.702891&146.7071671&3(2,1)-2(1,2)AA&8.6/8.7&1.50/2.02&2.1/2.0&\\
&&146.707233&3(2,1)-2(1,2)AA&8.7/8.9&&&\\
&146.861734/146.861594&146.865928&5(3,2)-5(2,4)EA&8.6/8.8&0.85/1.31&3.3/1.9&\\
&&146.8659748&5(3,3)-5(2,4)EA&8.7/8.9&&&\\
&146.868219&146.8725451&5(3,3)-5(2,4)EE&8.8&2.78/3.60&2.2,5.8&\\
&&&&&&/1.7,7.3&\\
&&146.872547&5(3,3)-5(2,4)EE&8.8&&&\\
&146.873031/146.872953&146.8773085&5(3,3)-5(2,4)AA&8.7/8.9&1.17/1.99&2.7/1.4&\\
&&146.877349&5(3,3)-5(2,4)AA&8.8/9.0&&&\\
&147.020516&147.024205&7(1,7)-6(0,6)EA&7.5&4.63/6.29&4.4/3.8,2.0&\\
&&147.024206&7(1,7)-6(0,6)AE&7.5&&&\\
&&147.0248967&7(1,7)-6(0,6)EE&8.9&&&\\
&&147.024902&7(1,7)-6(0,6)EE&8.9&&&\\
&&147.0255949&7(1,7)-6(0,6)AA&10.4&&&\\
&&147.025599&7(1,7)-6(0,6)AA&10.4&&&\\
&147.199438/147.199359&147.203752&6(3,4)-6(2,5)AE&8.8/8.9&1.47/2.07&1.6/2.0&blend\\
&&147.2037627&6(3,4)-6(2,5)AE&8.8/9.0&&&CH$_3$OCHO($v_{18}$=1)\\
&&147.2038037&46(12,35)-45(13,33)EE&8.9/9.1&&&\\
&147.202484&147.2068019&46(12,34)-45(13,32)EE&8.8&2.83/4.06&2.5/1.5&\\
&&147.2068073&6(3,4)-6(2,5)EE&8.8&&&\\
&&147.206816&6(3,4)-6(2,5)EE&8.8&&&\\
&&147.2074288&46(12,35)-45(13,33)EA&10.1&&&\\
&147.206453/147.206297&147.210399&46(12,35)-45(13,32)AE&8.0/8.4&1.92/2.53&2.6/2.0&\\
&&147.2104546&46(12,34)-&8.1/8.5&&&\\
&&&45(13,33)AE&&&&\\
&&147.2107089&6(3,4)-6(2,5)AA&8.7/9.0&&&\\
&&147.21074&6(3,4)-6(2,5)AA&8.7/9.0&&&\\
&150.159328/150.159094&150.1625706&25(4,21)-25(1,24)AE&6.5/6.9&2.11/1.73&3.8/2.7&\\
&&150.1625707&25(4,21)-25(1,24)EA&6.5/6.9&&&\\
&&150.1628123&25(4,21)-25(1,24)EE&7.0/7.4&&&\\
&&150.1630537&25(4,21)-25(1,24)AA&7.4/7.9&&&\\
&150.587969/150.587750&150.5923227&21(2,19)-21(1,20)AE&8.7/9.1&1.58/1.22&1.7/1.9&blend CH$_3$OCH$_3$\\
&&150.5923228&21(2,19)-21(1,20)EA&8.7/9.1&&&\\
&150.589797&150.594406&21(2,19)-21(1,20)EE&9.2&2.26/1.75&2.3/1.6&blend CH$_3$OCH$_3$\\
&150.591641/150.591563&150.5964893&21(2,19)-21(1,20)AA&9.7/9.8&1.81/1.24&1.5/1.1&blend CH$_3$OCH$_3$\\
&150.873859/150.873719&150.8758548&21(2,20)-20(3,17)AA&4.0/4.2&0.73/0.66&2.1/3.5&blend CH$_3$OH\\
&150.987703/150.987625&150.9920974&10(3,8)-10(2,9)EA&8.7/8.9&2.86/3.19&2.2/1.5&blend U-41\\
&&150.9921683&10(3,8)-10(2,9)AE&8.9/9.0&&&\\
&150.990984/150.990906&150.995391&10(3,8)-10(2,9)EE&8.7/8.9&4.36/4.42&2.1/1.7&\\
&150.994188&150.9986491&10(3,8)-10(2,9)AA&8.9&3.22/3.27&2.1/1.5&\\
&151.507063/151.506688&151.5110971&14(2,12)-13(3,11)AA&8.0/8.7&0.98/1.09&3.6/1.4&\\
&151.509125&151.51348&14(2,12)-13(3,11)EE&8.6&2.24/2.03&1.7/1.6&\\
&151.511500&151.5158562&14(2,12)-13(3,11)AE&8.6&1.20/0.98&1.8/1.4&\\
&&151.5158698&14(2,12)-13(3,11)EA&8.6&&&\\
&151.586266/151.586344&151.5907862&14(2,13)-14(1,14)EA&8.9/8.8&2.54/2.47&2.3/2.0&\\
&151.586266/151.586344&151.5907867&14(2,13)-14(1,14)AE&8.9/8.8&2.54/2.47&2.3/2.0&\\
&151.589625/151.589547&151.5939197&14(2,13)-14(1,14)EE&8.5/8.6&3.54/3.05&2.0/1.7&\\
&151.592672&151.597053&14(2,13)-14(1,14)AA&8.7&2.52/2.25&1.9/1.6&\\
\hline
CH$_3$COCH$_3$&131.999656/131.999359&132.0032346&13(0,13)-12(1,12)AE&8.1/8.8&0.71/0.63&4.0/-&\\
&&132.0032346&13(1,13)-12(0,12)AE&8.1/8.8&&&\\
&&132.0032794&13(0,13)-12(1,12)EA&8.2/8.9&&&\\
&&132.0032794&13(1,13)-12(0,12)EA&8.2/8.9&&&\\
&132.007375/132.007438&132.0105977&13(1,13)-12(0,12)EE&7.3/7.2&1.74/1.34&2.2/1.9&blend CH$_3$OCHO($v$=0)\\
&&132.0105977&13(0,13)-12(1,12)EE&7.3/7.2&&&\\
&132.014391/-&132.0178755&13(1,13)-12(0,12)AA&7.9/-&0.73/-&7.2/-&\\
&&132.0178755&13(0,13)-12(1,12)AA&7.9/5.9&&&\\
&138.572063/-&138.5760885&11(3,8)-10(4,7)EE&8.7/-&0.53/-&4.2/-&\\
&138.589375/-&138.5931203&11(4,8)-10(3,7)EE&8.1/-&0.56/-&3.3/-&\\
&150.395313/150.395234&150.3992893&24(7,17)-24(6,18)AA&7.9/8.1&0.98/0.89&1.9/0.38&blend\\
&&&&&&&CH$_3$OCHO($v$=0)\\
&&150.3992894&24(8,17)-24(7,18)AA&7.9/8.1&&&\\
&150.507328/150.506719&150.5112514&14(1,13)-13(2,12)AE&7.8/9.0&1.05/0.55&-&blend H$_2$CO\\
&&150.5112515&14(2,13)-13(1,12)AE&7.8/9.0&&&\\
&&150.5113151&14(2,13)-13(1,12)EA&7.9/9.2&&&\\
&&150.5113151&14(1,13)-13(2,12)EA&7.9/9.2&&&\\
&150.530219/150.530297&150.5346294&14(2,13)-13(1,12)EE&8.8/8.6&1.62/0.86&3.7/2.6&blend U-39\\
&&150.5346294&14(1,13)-13(2,12)EE&8.8/8.6&&&\\
&150.553172/-&150.5579044&14(1,13)-13(2,12)AA&9.4/-&0.86/-&5.1/-&\\
&&150.5579044&14(2,13)-13(1,12)AA&9.4&&&\\
&150.682578/150.682438&150.6869086&23(7,17)-23(6,18)EE&8.6/8.9&0.72/0.53&4.0/3.5&blend CH$_3$OCHO($v$=0)\\
&&150.6869086&23(6,17)-23(5,18)EE&8.6/8.9&&&\\
&151.634719/151.635094&151.6394264&15(1,15)-14(0,14)AE&9.3/8.6&0.89/0.60&3.7/2.7&\\
&&151.6394264&15(0,15)-14(1,14)AE&9.3/8.6&&&\\
&&151.6394718&15(1,15)-14(0,14)EA&9.4/8.7&&&\\
&&151.6394718&15(0,15)-14(1,14)EA&9.4/8.7&&&\\
&151.642500/151.642344&151.6465978&15(0,15)-14(0,14)EE&8.1/8.4&1.34/0.39&3.9/-&\\
&&151.6465978&15(1,15)-14(0,14)EE&8.1/8.4&&&\\
&&151.6465978&15(0,15)-14(1,14)EE&8.1/8.4&&&\\
&&151.6465978&15(1,15)-14(1,14)EE&8.1/8.4&&&\\
&151.649297/-&151.6536789&15(1,15)-14(1,14)AA&8.7/-&1.06/-&3.1/-&\\
&&151.6536789&15(0,15)-14(0,14)AA&8.7/9.3&&&\\
\hline
\end{longtable}

\begin{longtable}{p{.08\textwidth}p{.17\textwidth}p{.08\textwidth}p{.14\textwidth}p{.06\textwidth}p{.06\textwidth}p{.07\textwidth}p{.15\textwidth}}
  \caption{The tentative lines in this observations.}\label{tab:lineid_tent}
  \hline              
  Species& Obs. freq. (HC/CR)& Rest. freq.&Transition&$V_{\rm {LSR}}$&$T_{\rm{a}}$&Width&Comments\\ 
    [-0.5mm]&[GHz]&[GHz]&&[km/s]&[K]&[km/s]&\\
\endfirsthead
  \hline
\endhead
  \hline
\endfoot
  \hline
\endlastfoot
  \hline
S$^{18}$O&145.873984/145.873609&145.8738078&3(4)-2(3)&-0.4/0.4&1.60/0.82&10.9/9.1&blend\\
&&&&&&&CH$_3$CH$_2$CN\\
&&&&&&&($v_{13}/v_{21}$)\\
&&&&&&&\citep{Esplugues2013-yl}\\
\hline
OC$^{33}$S&132.097781/-&132.101384&11-10&8.2/-&0.99/-&3.4/-&blend\\
&&&&&&& CH$_3$OCHO($v$=0)\\
&&&&&&&CDMS\\
&&&&&&&\citep{Tercero2010-sm}\\
\hline
$^{18}$OCS&136.905078/-&136.908755&12-11&8.1/-&0.57/-&6.5/-&\citep{Tercero2010-sm}\\
\hline
$^{18}$O$^{13}$CS&136.573938/-&136.5777664&12-11&8.4/-&0.96/-&10.1/-&CDMS\\
\hline
SiS&145.218969/145.218125&145.2269975&8-7&16.6/18.3&1.45/0.80&7.1/7.1&\citep{Tercero2011-rs}\\
\hline
o-H$_2$C$^{34}$S&137.067297/137.067453&137.0775425&4(1,3)-3(1,2)&22.4/22.1&0.58/0.63&2.8/1.6&\citep{Tercero2010-sm}\\
\hline
H$^{13}$CCCN&132.243438/132.243203&132.246378&\textit{J}=15-14,\textit{F}=14-13&6.7/7.2&5.01/4.61&-&blend\\
&&&&&&&CH$_3$OCHO($v$=0)\\
\hline
H(C)OCN&151.577031/-&151.581402&16(1,16)-15(1,15)&8.6/-&0.73/-&2.7/-&\\
\hline
CH$_2$DCN&138.323328/-&138.3270249&8(1,8)-7(1,7)&8.0/-&0.71/-&4.6/-&\\
\hline
CH$_3$$^{13}$CN&147.052859/147.052938&147.05656&8(4)-7(4)&7.5/7.4&1.35/0.91&8.0/-&blend H35$\alpha$\\
&147.086063&147.09054&8(2)-7(2)&9.1&1.51/0.94&5.3/4.8&\\
\hline
CH$_3$OH&136.700531/136.700750&136.703454&23(4)$^-$-24(5)$^-$&6.4/5.9&1.19/1.46&3.8/2.2&blend CH$_3$CCH\\
($v_{12}$=1)&&136.704668&23(4)$^+$-24(5)$^+$&9.1/8.6&1.19/1.46&3.8/2.2&\\
&151.278328/151.278547&151.282794&14(-2)-15(-3)E1&8.9/8.4&4.17/1.69&5.1,2.2/3.9&\\
\hline
CH$_3$$^{18}$OH&137.928016/-&137.931708&3(1,3)-2(1,2)A&8.0/-&0.70/-&1.9/-&CDMS\\
\hline

\end{longtable}

\begin{longtable}{p{.04\textwidth}p{.18\textwidth}p{.20\textwidth}p{.25\textwidth}p{.10\textwidth}p{.10\textwidth}}
  \caption{The leakage lines in this observations.\footnotemark[$*$]}\label{tab:lineid_leak}
  \hline              
  Label&Candidate species&Obs. freq. (HC/CR)&Transition&Rest freq.&Comments\\ 
    [-0.5mm]&&[GHz]&&[GHz]&\\
\endfirsthead
  \hline
\endhead
  \hline
\endfoot
  \hline
\endlastfoot
  \hline
X-1&CH$_3$OCH$_3$&131.404219/131.404531&6(1,6)-5(0,5)EA&131.4050391&DC\\
&&&6(1,6)-5(0,5)AE&131.4050401&DC\\
&&&6(1,6)-5(0,5)EE&131.4057961&DC\\
\hline
X-2&HNCO&131.415891/131.415969&6(1,6)-5(1,5),\textit{F}=5-5&131.3934248&DC,blend CH$_3$NCO\\
&&&6(1,6)-5(1,5),\textit{F}=5-6&131.3941370&\\
&&&6(1,6)-5(1,5),\textit{F}=7-6&131.39426190&\\
&&&6(1,6)-5(1,5),\textit{F}=5-4&131.39428210&\\
&&&6(1,6)-5(1,5),\textit{F}=6-5&131.3942897&\\
&&&6(1,6)-5(1,5),\textit{F}=6-6&131.3950019&\\
\hline
X-3&SO$_2$&131.533406/131.533250&16(5,11)-17(4,14)&131.27493&DC\\
\hline
X-4&$^{33}$SO$_2$&131.614656&14(2,12)-14(1,13),F=29/2-31/2&131.1863884&DC\\
&&&14(2,12)-14(1,13),\textit{F}=27/2-25/2&131.18660810&DC\\
&&&14(2,12)-14(1,13),\textit{F}=29/2-27/2&131.1895759&DC\\
&&&14(2,12)-14(1,13),\textit{F}=29/2-29/2&131.1899456&DC\\
&&&14(2,12)-14(1,13),\textit{F}=27/2-27/2&131.1901653&DC\\
&&&14(2,12)-14(1,13),\textit{F}=27/2-29/2&131.1905351&DC\\
&&&14(2,12)-14(1,13),\textit{F}=31/2-31/2&131.19205980&DC\\
&&&14(2,12)-14(1,13),\textit{F}=25/2-25/2&131.1922796&DC\\
&&&14(2,12)-14(1,13),\textit{F}=31/2-29/2&131.1956171&DC\\
&&&14(2,12)-14(1,13),\textit{F}=25/2-27/2&131.1958369&DC\\
\hline
X-5&SO$_2$&133.805359/133.805813&8(2,6)-8(1,7)&134.004860&ADC\\
\hline
X-6&CH$_3$OCHO&136.423781/136.424094&11(5,6)-10(5,5)A&135.9885&DC\\
\hline
X-7&$^{13}$CS&138.673609/138.674000&3-2&138.739335&ADC\\
\hline
X-8&CH$_3$OH&145.107797/145.107875&3(0)$^+$-2(0)$^+$,vt=0&145.103185&ADC\\
&&&3(1)-0(0)E1,vt=1&145.103185&ADC\\
\hline
X-9&CH$_3$OH&145.113375/145.113594&3(-1)-2(-1)E2,vt=0&145.097435&ADC\\
&&&3(1)-\,-(0)E1,vt=1&145.097435&ADC\\
\hline
X-10&CH$_3$OH&145.117188/145.117344&3(0)-2(0)E1,vt=0&145.093754&ADC\\
&&&3(0)-0(0)E1,vt=1&145.093754&ADC\\
\hline
X-11&OCS&145.224688/145.224469&11-10&133.7859&Image\\
\hline
X-12&CH$_3$OH($v_{12}=1$)&147.266344/147.265734&7(1)-6(0)E1,vt=1&147.9436730&DC\\
&&&35(-9)-35(10)E1,vt=1&147.944291&DC\\
\hline
X-13&C$_2$H$_5$CN&147.452891/147.452516&17(0,17)-16(0,16)&147.7567110&DC\\
&CH$_3$CN($v_8=1$)&147.452891/147.452516&8(1)-7(-1),\textit{F}=8-8,\textit{l}=1&147.7591845&DC\\
\hline
X-14&CH$_3$OCHO&147.479594/147.479984&28(5,23)-28(5,24)A&147.7291239&DC\\
&&&28(5,23)-28(5,24)A&147.729318&DC\\
&&&12(7,6)-11(7,5)A&147.7306066&DC\\
&&&12(7,6)-11(7,5)A&147.730607&DC\\
&&&12(7,5)-11(7,4)A&147.730751&DC\\
&&&12(7,5)-11(7,4)A&147.7307717&DC\\
&&&12(7,6)-11(7,5)E&147.731209&DC\\
&&&12(7,6)-11(7,5)E&147.731334&DC\\
&CH$_3$OCHO($v_{18}=1$)&147.479594/147.479984&12(5,8)-11(5,7)E&147.728224&DC\\
&CH$_3$OCH$_3$&147.479594/147.479984&7(3,5)-7(2,6)EA&147.7274357&DC\\
&&&7(3,5)-7(2,6)AE&147.7280999&DC\\
&&&7(3,5)-7(2,6)EE&147.731357&DC\\
&&&7(3,5)-7(2,6)AA&147.7349473&DC\\
\hline
X-15&CH$_3$OCHO&147.493406/147.493188&12(7,5)-11(7,4)E&147.717696&DC\\
\hline
X-16&CH$_3$OCH$_3$&149.935469/149.935922&5(5,0)-6(4,2)EA&149.8772306&ADC\\
&&&5(5,1)-6(4,2)AE&149.8787933&ADC\\
&H$^{13}$CCCN&149.935469/149.935922&\textit{J}=17-16,\textit{F}=16-15&149.877008&ADC\\
&&&\textit{J}=17-16,\textit{F}=17-16&149.87700940&ADC\\
\hline
X-17&SO&150.433844/150.434234&4(3)-3(2)&138.1786&Image\\
\hline
X-18&C$_2$H$_5$CN&152.258531/152.258688&17(4,13)-16(4,12)&152.552899&DC\\
\hline

\end{longtable}

\begin{table*}[h]
  \tbl{The leakage lines in this observations.\footnotemark[$*$]}{%
  \begin{tabular}{p{.05\textwidth}p{.16\textwidth}p{.08\textwidth}p{.08\textwidth}p{.05\textwidth}p{.16\textwidth}p{.08\textwidth}p{.08\textwidth}}
      \hline          
  Label&Obs. freq. (HC/CR)&$T_{\rm{a}}$&Width&Label&Obs. freq. (HC/CR)&$T_{\rm{a}}$&Width \\ 
    [-0.5mm]&[GHz]&[K]&[km/s]&&[GHz]&[K]&[km/s]\\
      \hline
U-1&131.432688/131.432453&0.93/0.84&3.3/5.5&U-26&138.136859/138.137609&0.65/0.57&1.9/2.4\\
U-2&131.552094/131.552172&0.89/0.65&2.7/2.6&U-27&138.204453/138.204219&0.66/0.55&1.5/1.7\\
U-3&131.558953/-&0.81/-&16.2/-&U-28&138.534453/-&0.60/-&2.5/-\\
U-4&131.563234/-&0.67/-&7.9/-&U-29&138.559547/-&1.05/-&7.5/-\\
U-5&131.675625/131.675234&0.74&4.0/2.0&U-30&145.302750/-&1.25/-&10.0/-\\
U-6&-/132.023922&-/0.62&-/1.6&U-31&145.832250/145.831797&1.03/0.92&9.8/3.1\\
U-7&132.042844/132.042844&0.91/1.15&4.4,9.2/4.0,4.1&U-32&146.330016/-&1.44/-&4.9/-\\
U-8&132.392453/132.392906&0.68/0.73&7.8/8.2&U-33\footnotemark[$*$]&147.109703/-&1.24/-&8.4/-\\
U-9&132.403891/-&0.81/-&6.2/-&U-34&147.322969/-&1.22/-&3.3/-\\
U-10&132.715500/-&0.72/-&4.7/-&U-35&147.508516/-&0.81/-&3.3/-\\
U-11&133.211438/-&0.79/-&7.6/-&U-36&147.526531/147.526141&0.78/-&3.2/3.7\\
U-12&133.407156/133.408219&0.78/0.70&7.5/3.1&U-37&150.464828/150.465203&0.98/0.50&4.9/2.9\\
U-13&133.441172/-&0.90/-&2.9/-&U-38&150.521891/150.521672&0.98/0.61&2.4/1.8\\
U-14&133.485438/133.485438&0.89/0.84&1.9/2.4&U-39&150.534188/150.534109&1.05/0.75&7.2/3.2\\
U-15&133.502375/-&1.05/-&7.2/-&U-40&150.699750/150.699672&0.96/0.64&2.6/3.8\\
U-16&133.642453/-&1.16/-&9.0/-&U-41&150.979922/-&1.10/-&11.3/-\\
U-17&133.827859/-&0.81/-&3.8/1.8&U-42&151.269094/-&1.28/-&7.5/-\\
U-18&-/133.858453&-/0.93&-/4.4&U-43&151.548188/151.548953&0.76/0.56&3.0/-\\
U-19&133.862656/133.862344&0.65/0.85&2.5/3.3&U-44&151.623047/-&0.96/-&4.8/-\\
U-20&136.478109/-&0.73/-&11.2/-&U-45&151.757859/-&1.01/-&4.9/-\\
U-21&136.753469/136.752859&0.76/0.62&3.1/5.4&U-46&152.103953/152.105250&0.87/0.51&4.2/5.3\\
U-22&137.044563/-&1.02/-&9.7/-&U-47\footnotemark[$\dagger$]&152.139203/152.139813&1.51/0.61&10.3/11.5\\
U-23&137.071344/137.070875&0.69/0.54&3.3/2.7&U-48&152.222375/-&1.04/-&4.8/-\\
U-24&137.432234/137.432234&0.86/0.53&2.6/0.87&U-49\footnotemark[$\dagger$]&152.362984/-&0.87/-&6.4/-\\
U-25&137.511891/-&0.56/-&2.7/-&&&&\\
      \hline
    \end{tabular}}\label{tab:lineid_und}
\begin{tabnote}
\footnotemark[$*$] This line is also reported as U-line in \citet{Lee2001-ws}.
\footnotemark[$\dagger$] This line is also reported as U-line in \citet{Ziurys1993-ps}.
\end{tabnote}

\end{table*}

\bibliography{refs}
\bibliographystyle{pasjbib2020}

\end{document}